\documentclass[nohyper,12pt,letterpaper]{JHEP3}
\usepackage{amsmath}
\usepackage{amsthm}
\usepackage{amsfonts}

\DeclareMathOperator{\tr}{tr} 

\date{\today}

\title{A Deformation of Twistor Space
and a Chiral Mass Term in $N=4$ Super Yang-Mills Theory}

\author{
Dah-Wei Chiou, Ori J. Ganor and Bom Soo Kim\\

Department of Physics,
University of California, Berkeley, CA 94720 \\
  and\\
Theoretical Physics Group,
Lawrence Berkeley National Laboratory,\\
Berkeley, CA 94720\\ \\

Emails:\\
       \email{dwchiou,origa,bskim@socrates.berkeley.edu}
}

\abstract{
Super twistor space admits a certain 
(super) complex structure deformation that
preserves the Poincar\'e subgroup of the symmetry group
$PSL(4|4)$ and depends on 10 parameters.
In a previous paper [hep-th/0502076], it was proposed
that in twistor string theory this deformation corresponds
to augmenting $N=4$ super Yang-Mills theory by a mass
term for the left-chirality spinors.
In this paper we analyze this proposal in more detail.
We calculate 4-particle scattering amplitudes of fermions,
gluons and scalars and show that they are supported on holomorphic
curves in the deformed twistor space.
}


\received{December 19, 2005}

\preprint{\hepth{0512242}\\ UCB-PTH-05/39 \\ LBNL-54227}
\begin{document}

\def\be{\begin{equation}}
\def\ee{\end{equation}}
\def\bear{\begin{eqnarray}}
\def\eear{\end{eqnarray}}
\def\nn{\nonumber}


\def\defineas{{:=}}

\def\a{\alpha}
\def\b{\beta}
\def\g{\gamma}
\def\d{\delta}
\def\r{\rho}
\def\th{{\theta}}
\def\lam{{\lambda}}

\def\dta{{\dot{\a}}}
\def\dtb{{\dot{\b}}}
\def\dtg{{\dot{\g}}}
\def\dtd{{\dot{\d}}}

\def\bth{{\overline{\theta}}}
\def\blam{{\overline{\lambda}}}
\def\bpsi{{\overline{\psi}}}
\def\bsig{{\overline{\sigma}}}
\def\Dslash{{\relax{\not\kern-.18em \partial}}} 
\def\SL{{{\mbox{\rm SL}}}} 
\def\GL{{{\mbox{\rm GL}}}} 
\def\rt{{\rightarrow}}
\def\cc{{\mbox{c.c.}}}

\newcommand\SUSY[1]{{${\cal N}={#1}$}}  
\newcommand\px[1]{{\partial_{#1}}}
\newcommand\qx[1]{{\partial^{#1}}}

\newcommand\ppx[1]{{\frac{\partial}{\partial {#1}}}}
\newcommand\pspxs[1]{{\frac{\partial^2}{\partial {#1}^2}}}
\newcommand\pspxpx[2]{{\frac{\partial^2}{\partial {#1}\partial {#2}}}}
\newcommand\pypx[2]{{\frac{\partial {#1}}{\partial {#2}}}}

\newcommand{\field}[1]{\mathbb{#1}}
\newcommand{\ring}[1]{\mathbb{#1}}
\newcommand{\C}{\field{C}}
\newcommand{\R}{\field{R}}
\newcommand{\Z}{\ring{Z}}
\newcommand{\N}{\ring{N}}
\newcommand{\CP}{\field{C}\mathbb{P}} 
\newcommand{\RP}{\field{R}\mathbb{P}} 


\providecommand{\abs}[1]{{\lvert#1\rvert}}
\providecommand{\norm}[1]{{\lVert#1\rVert}}
\providecommand{\divides}{{\vert}}
\providecommand{\suchthat}{{:\quad}}


\def\gYM{{g_{\textit{YM}}}} 
\newcommand\rep[1]{{\bf {#1}}} 

\newcommand\inner[2]{{\langle {#1}, {#2} \rangle}} 


\def\Horava{{Ho\v{r}ava\ }}
\def\Cech{{\v{C}ech\ }}


\newtheorem{thm}{Theorem}[section]
\newtheorem{lem}[thm]{Lemma}

\numberwithin{equation}{section}

\newcommand{\thmref}[1]{Theorem~\ref{#1}}
\newcommand{\secref}[1]{\S\ref{#1}}
\newcommand{\lemref}[1]{Lemma~\ref{#1}}

\newcommand{\figref}[1]{Figure~\ref{#1}}
\newcommand{\appref}[1]{Appendix~\ref{#1}}


\def\tlam{{\tilde{\lam}}} 
\def\bQ{{\overline{Q}}} 

\newcommand\rsprod[2]{{\langle {#1}, {#2}\rangle}} 
\def\Lag{{\cal L}} 

\def\cA{{{\cal A}}} 
\def\wchi{{\tilde{\chi}}} 
\def\wvrho{{\widetilde{\varrho}}} 
\def\hwvrho{{\widehat{\widetilde{\varrho}}}} 
\def\hvarrho{{\widehat{\varrho}}} 
\def\hzeta{{\widehat{\zeta}}}

\def\wchipr{{{\chi^\prime}}} 

\def\bW{{\overline{W}}} 
\def\bPhi{{\overline{\Phi}}} 
\def\spr{{\zeta}} 
\def\bspr{{\overline{\spr}}} 

\def\cV{{\cal V}} 

\centerline{\bf DISCLAIMER}
This document was prepared as an account of work sponsored by the United States Government. While this document is believed to contain correct information, neither the United States Government nor any agency thereof, nor The Regents of the University of California, nor any of their employees, makes any warranty, express or implied, or assumes any legal responsibility for the accuracy, completeness, or usefulness of any information, apparatus, product, or process disclosed, or represents that its use would not infringe privately owned rights. Reference herein to any specific commercial product, process, or service by its trade name, trademark, manufacturer, or otherwise, does not necessarily constitute or imply its endorsement, recommendation, or favoring by the United States Government or any agency thereof, or The Regents of the University of California. The views and opinions of authors expressed herein do not necessarily state or reflect those of the United States Government or any agency thereof or The Regents of the University of California.

\newpage
\tableofcontents

\section{Introduction}\label{sec:intro}
Twistor \cite{Penrose:1967wn}
string theory \cite{Witten:2003nn}
studies perturbative scattering amplitudes of massless
particles in $N=4$ Super-Yang-Mills theory in terms
of a topological B-model with target space $\CP^{3|4}.$
This target space is a Calabi-Yau supermanifold
\cite{Sethi:1994ch}\cite{Aganagic:2004yh}.
(For alternative formulations of twistor string theory
see 
\cite{Berkovits:2004hg}\cite{Berkovits:2004tx}\cite{Siegel:2004dj}.)
Twistor techniques are, in general, useful
for dealing with massless particles.
They have recently been used to derive simple
expressions for scattering amplitudes that have previously
never been written in closed form.
(See \cite{Cachazo:2005ga} for a recent review.)

In this paper we would like to describe a theoretical
extension of twistor string theory that includes
a mass term for the fermions of the vector
multiplet of Super-Yang-Mills theory.
Such a mass term, of course, breaks supersymmetry and
conformal invariance as well.
In general, a mass term precludes the use of
the twistor transform
which requires that external particles
have lightlike momenta.
(But see \cite{Badger:2005zh} for recent developments
that use twistor techniques indirectly to calculate
scattering amplitudes of massive particles.)
However, if the mass term only involves
spinors of one chirality and does not include the
spinors of the opposite chirality,
the plane-wave solutions of the free Dirac equation are still
lightlike. Of course, such a model breaks CPT symmetry,
but it is consistent mathematically,
and we can calculate scattering amplitudes in this model.
The amplitudes
are holomorphic functions of the chiral mass parameters.

The physical relevance of the scattering amplitudes that we get
in such a model can be described as follows.
The scattering amplitudes of a model with a CPT-invariant
fermion mass term depend on the complex mass parameter $M$
and its complex conjugate $M^*.$
It can be written as an analytic expression in
two formally independent variables $M$ and $M^*.$
The amplitudes of the chiral-mass theory can be defined
as the expressions that we get when we formally
set $M^*=0$ in the physical amplitudes.

In this work we study the twistor approach to
$N=4$ Super-Yang-Mills theory with an extra chiral mass term,
and we expand on ideas presented in \cite{Chiou:2005jn}.
There, it was argued that the free-field equations
of motion of the augmented theory still have a twistor description;
the relevant twistor space is a certain super complex structure
deformation of $\CP^{3|4}.$
In this paper we calculate 4-particle scattering amplitudes
and extend the notion of {\it maximally helicity violating}
(MHV) amplitudes to include the chiral mass term.
In the massless theory, Witten discovered that MHV scattering
amplitudes vanish, when expressed in twistor variables,
unless certain algebraic conditions hold.
The amplitude does not vanish only if there exists an
algebraic curve of degree $d=1$ in supertwistor space, $\CP^{3|4}$,
such that all the twistors that label
the external particles lie on this curve \cite{Witten:2003nn}.
Does a similar assertion
hold for the theory with the chiral mass term?

In this paper we will explore this question for 4-particle
amplitudes.
In \secref{sec:scamp} we extend the definition
of helicity to the fermions with a chiral mass term,
and we calculate 4-particle
(extended) MHV scattering amplitudes.
In addition to the chiral mass term,
we also include in the calculations
a possible 3-scalar interaction,
which has the same dimension ($\Delta=3$) and
R-symmetry quantum numbers as the fermion mass term.
In \secref{sec:chbmt} we describe the  deformation
of super twistor space that corresponds to
adding the chiral mass term, and
we look for algebraic curves
in the deformed space.
There, we define a natural extension of the notion of
degree $d=1$ curves for the deformed case;
the equations describing these curves
contain quadratic terms.
In \secref{sec:twmass} we
show that if we set the 3-scalar
coupling correctly, 4-particle (extended) MHV
amplitudes are indeed supported on these $d=1$
algebraic curves.
Furthermore, we find that the amplitudes are given by
an integral over the moduli space of $d=1$ curves
that is essentially the same as the one for the massless case
\cite{Witten:2003nn}; the only modification is the expression
for the curve itself.
We conclude with a discussion in \secref{sec:disc}.
The appendices contain more technical details about
the Feynman rules in the presence of the
unusual CPT-violating chiral mass terms.

\section{Chiral and anti-chiral fermion mass terms}\label{sec:cfmt}
We denote the negative helicity fermions by
\textit{$\psi_\a^A$},
where $\a$ is a spinor index
($\a=1,2$) and $A$ is an $SU(4)$ R-symmetry index ($A=1,\dots,4$).
We denote the positive helicity fermions by
\textit{$\bpsi^\dta_A$}.
The full $N=4$ Super Yang-Mills Lagrangian is
presented in \appref{app:Feynman rules}, for completeness.
An anti-chiral mass term is
\textit{$M_{AB} \psi_\a^A\psi^{\a B}$}
and a chiral mass term is
\textit{$M^{AB}\bpsi_{\dta A}\bpsi^\dta_B$}.
Here \textit{$M_{AB}=M_{BA}$} and \textit{$M^{AB}=M^{BA}$}
are the corresponding mass matrices, with
10 independent complex parameters each.
We are going to add a chiral mass term
to the $N=4$ SYM Lagrangian. This, of course, breaks
CPT invariance, but the perturbative Feynman diagrams
are well-defined.

\subsection{Free field equations of motion}\label{subsec:eom}
In the presence of a chiral mass term,
the negative helicity fermions acquire a left-chirality
($\dta$) component. To see this, we write down the
Dirac equations:
\be\label{eqn:DiracEq}
p_{\a\dta}\bpsi^\dta_A = 0,
\qquad
p^{\a\dta}\psi_\a^A = M^{AB}\bpsi^\dta_B.
\ee
These equations imply that the momentum $p_{\a\dta}$
is lightlike. It can therefore be written as a product
of two spinors,
\be\label{eqn:lamtlam}
p_{\a\dta} = \lam_\a\tlam_\dta,
\ee
as in the \textit{$M^{AB}=0$} case.
A basis for the solutions of \eqref{eqn:DiracEq} is given by
\be\label{eqn:DiracSol}
\bpsi^\dta_{A} = \tlam^\dta\wvrho_A,
\qquad
\psi_\a^A = \lam_\a\varrho^A +M^{AB}\eta_\a\wvrho_B,
\ee
where $\wvrho_A, \varrho^A$ are arbitrary parameters
(scalars in the fundamental representation of
the R-symmetry group)
and $\eta^\a$ is an arbitrary chiral spinor that
is only required to satisfy
\be\label{eqn:etalam}
\lam^\a\eta_\a = 1.
\ee
Once \textit{$\eta_\a$} is fixed,
we can define {\it helicity} as follows:
A solution with helicity $(-)$ has
\textit{$\bpsi^\dta_A=0$} and
\textit{$\psi^{\a A}=\lam^\a\varrho^A$};
a solution with helicity $(+)$ has
\textit{$\bpsi^\dta_A=\tlam^\dta\wvrho_A$}
and \textit{$\psi^{\a A}=M^{AB}\eta^\a\wvrho_B$}.\footnote{If we treat
$\varrho$, $\wvrho$ and $\eta$ as continuous functions of $\lam$ and $\tlam$,
the helicity can be alternatively defined as \eqref{eqn:helicity}.
These two definitions
turn out to be equivalent, as discussed in \appref{app:helicity}.}
In Feynman diagrams, external lines of negative helicity fermions
only have left-moving \textit{$\psi^{\a A}$} components,
but external lines of positive helicity fermions have
both left-moving and right-moving components.
This is depicted in \figref{fig:extFermion} and
\figref{fig:extantiFermion}.
%
%
%
%
%

\subsection{3-scalar interaction}\label{subsec:3phi}
The fermion mass term
that we added in \secref{subsec:eom}
is a linear combination of operators
\be\label{eqn:MassOp}
\cV'_{AB} \defineas \tr\bigl\{\bpsi_{\dta A}\bpsi^\dta_B\bigr\}
\ee
of conformal dimension $\Delta=3$,
at lowest order in perturbation theory.
These operators are in the $SU(4)$
(R-symmetry) irreducible representation $\rep{10}$
(i.e., a symmetric covariant 2-tensor).
There is another set of operators of $N=4$ super Yang-Mills
with the same quantum numbers, at lowest order in perturbation
theory. They are cubic in the scalar fields.
Let us denote these scalar fields by
\be\label{eqn:scalars}
\phi_{\mathcal{I}},\qquad
\mathcal{I}=1,\dots,6.
\ee
Here $\mathcal{I}$ is an R-symmetry index in
the fundamental representation of $so(6)\simeq su(4).$
(For convenience,
we present some relevant identities
in \appref{app:susy-indices}.)

The second set of operators of conformal dimension
$\Delta =3$ (at $0^{th}$ order of perturbation theory)
and $so(6)\simeq su(4)$ representation $\rep{10}$
can now be written as
\be\label{eqn:cVppAB}
\cV''_{AB}\,\defineas\,
\Gamma^{\mathcal{I}\mathcal{J}\mathcal{K}}_{AB}
\tr\bigl\{
\phi_{\mathcal{I}}
\phi_{\mathcal{J}}
\phi_{\mathcal{K}}
\bigr\},
\ee
using the $SU(4)$-invariant symbol
\textit{$\Gamma^{\mathcal{I}\mathcal{J}\mathcal{K}}_{AB}$},
defined at the end of \appref{app:susy-indices}.
This symbol is anti-symmetric  in
the $so(6)$ indices $\mathcal{I}\mathcal{J}\mathcal{K}$
and symmetric in the $su(4)$ indices $AB$,
and it connects the representation $\rep{10}$ of $so(6)$
(self-dual 3-tensors) to the representation $\rep{10}$ of $su(4).$

There is a linear combination of $\cV'_{AB}$
and $\cV''_{AB}$ that lies in a short supermultiplet.
This is the combination
\be\label{eqn:cVAB}
\cV_{AB}\defineas \cV'_{AB}+\frac{1}{4}\cV''_{AB},
\ee
and its conformal dimension $\Delta=3$ is exact.
These operators can be obtained by acting with
two supersymmetry transformations on the
chiral primary operators
\textit{$\cV_{\mathcal{I}\mathcal{J}}\defineas
\tr\{\phi_{\mathcal{I}\mathcal{J}}\}$}.
(See \cite{Howe:1981xy}\cite{Intriligator:1998ig}
for more details.)

In the next section we will calculate 4-point tree level
scattering amplitudes in the presence of the perturbations
discussed above.
We will include both the 2-fermion and the 3-scalar
perturbations in the combination
\be\label{eqn:defk}
g^2 \delta\Lag
=\frac{1}{2} M^{AB}\cV_{AB},
\ee
where $g$ is the Yang-Mills coupling constant
and the unperturbed Lagrangian is presented in
\eqref{eqn:susy4Lag}.

In \cite{Berkovits:2004jj} it was shown that
twistor string theory contains a sector that
is described by $N=4$ conformal supergravity (CSUGRA).
Furthermore, tree-level amplitudes in CSUGRA
have been calculated in \cite{Ahn:2005es}
using twistor string theory.
The fields of CSUGRA couple to the
fields of $N=4$ Super Yang-Mills (SYM).
To linear order, each CSUGRA
field couples to an
$N=4$ SYM operator from the
short supermultiplet of the chiral primary field
\textit{$\cV_{\mathcal{I}\mathcal{J}}$}.
For example, CSUGRA contains an $SU(4)$ gauge field that
couples to the R-symmetry current of $N=4$ SYM.
CSUGRA also contain scalar fields in the representation
$\overline{\rep{10}}$ of $SU(4)$, which were denoted
by \textit{$\overline{E}^{AB}$} in \cite{Berkovits:2004jj}.
To linear order,
these fields couple to the $N=4$ SYM operators
\textit{$\cV_{AB}$},
and the mass terms that we are considering here can
be interpreted as VEVs,
\be\label{eqn:VEVEAB}
M^{AB} \sim \left\langle\overline{E}^{AB}\right\rangle,
\ee
as suggested in \cite{Chiou:2005jn}.

\section{Extended MHV amplitudes}\label{sec:scamp}

%
%
%


In this section we will calculate several scattering
amplitudes with a chiral mass term for tree-level planar diagrams.
The mass term mainly changes the
Feynman diagram rules for the fermions and 3-scalar interaction.
We present the fermion
propagators and external wavefunctions in
\figref{fig:FermionProp} -- \figref{fig:extantiFermion},
and the 3-scalar vertex in \figref{fig:scalar-vertex}.
All the relevant Feynman rules are given in
\appref{app:Feynman rules}.

When we label the helicity of the amplitude, we  use the
convention that all external particles are incoming.
For example $A(+1,+1,-1,-1)$ represents
 the amplitude with two incoming helicity
$+1$ and two incoming helicity $-1$ gluons.
On the other hand, for convenience,
the convention depicted in the Figures (and discussed in
\appref{app:Feynman rules}) will be that the helicity and
momentum  are all physical
(2 incoming and 2 outgoing particles
with their physical momenta and helicities).\footnote{This
difference in conventions corresponds to replacing
$p^\mu\rightarrow-p^\mu$, or $\lam\rightarrow i\lam$ and
$\tlam\rightarrow i\tlam$ for the outgoing particles. This does not
affect our result because we scale $(\lam_{i1},\lam_{i2})$ to
$(1,Z_i=\lam_{i2}/\lam_{i1})$ in the end.
However, the form of the momentum
conservation condition depends on the
convention:
$p_1+\dots+p_4=0$ for
``incoming'' momenta while $p_1+p_2=p_3+p_4$ for ``physical''
momenta. The former leads to
$\frac{[1,3]}{[1,4]}=-\frac{\langle2,4\rangle}{\langle2,3\rangle}$,
$\frac{[2,1]}{[2,3]}=-\frac{\langle4,3\rangle}{\langle4,1\rangle}$
and so on, while the latter gives
$\frac{[1,3]}{[1,4]}=-\frac{\langle2,4\rangle}{\langle2,3\rangle}$
and
$\frac{[2,1]}{[2,3]}=\frac{\langle4,3\rangle}{\langle4,1\rangle}$
(an extra minus sign may arise).}
For the planar diagrams with
external particle indices $i$ cyclically attached, all amplitudes
include an overall group theory factor
$tr[T_1 T_2 \cdots T_i \cdots T_n]$,
which will be suppressed hereafter.

Maximally Helicity Violating (MHV)
amplitudes at the tree level are originally defined
\cite{Parke:1986gb}-\cite{Bern:1996je}
as those satisfying the condition
\textit{$\sum_i(2h_i-2)= -8$} with $h_i$
the helicities of external legs (defined as all incoming).
Since the chiral mass term is interpreted
as a VEV of a spacetime conformal
supergravity field $\overline{E}^{AB}$
[see \eqref{eqn:VEVEAB}]
we can think of the amplitudes with $n$ external legs
that contain the mass parameter at order $\mathcal{O}(M^k)$
as coming from diagrams with $(n+k)$ legs,
of which $k$ legs correspond to a background CSUGRA field
\textit{$\overline{E}^{AB}$}.
The helicity of this field is $0$, and
therefore mass-deformed SYM diagrams at order
$\mathcal{O}(M^k)$ that satisfy
\textit{$\sum_i(2h_i-2)= -8+2k$}
correpond to MHV diagrams in CSUGRA.
We can therefore generalize the term ``MHV'' to
{\it ``extended MHV''}
to describe those diagrams at order
$\mathcal{O}(M^k)$ that satisfy
\textit{$\sum_i(2h_i-2)= -8+2k$}.\footnote{At tree level, however, we only have
$\mathcal{O}(M^0)$ and $\mathcal{O}(M)$; the amplitudes at higher orders of $M$
all vanish. This can be understood by \eqref{eqn:twistor am}, in which $M$ gives
3 $\th_2$'s but the integrand needs to have exactly 4 $\th_2$ to yield nonzero result.}
The holomorphic structure of generalized
MHV amplitudes calculated in
this section will be discussed in \secref{sec:twmass}.

We will now present the results of the calculation
of various (extended) MHV amplitudes.
The Feynman diagrams and the detailed calculation are shown in \appref{app:amplitudes}
for interested readers.
We begin with \textit{$M^{AB}$}-independent
contribution to the MHV amplitudes.
These diagrams are the same as those of the undeformed
theory, and were calculated
in \cite{Parke:1986gb}\cite{Witten:2003nn}
with external gluons and in
\cite{Georgiou:2004gk}\cite{Wu:2004wz} with external gluinos.
We present them here for completeness, featuring
the use of the spinor notation.

\subsection{MHV amplitudes
(extended MHV at $\mathcal{O}(M^0)$)}\label{sec:MHV-M0}

\begin{itemize}
\item 4-gluon scattering amplitude:\footnote{
The notations $\langle i,j\rangle$
and $[i,j]$ are short for $\langle \lam_i,\lam_j\rangle$
and $[\tlam_i,\tlam_j]$ respectively.
We also set $\lam_{n+1}\equiv\lam_1$ for $n$ external legs.}
\be
A_{\mathcal{O}(M^0)}(+1,+1,-1,-1)
=\frac{ig^2}{2}\frac{{\langle3,4\rangle}^4}{\prod_{i=1}^4\langle
i,i+1 \rangle}. \label{eqn:AmpA(+1,+1,-1,-1)}
\ee
\item 2-gluon and 2-fermion scattering amplitude:
\be
A_{\mathcal{O}(M^0)}(+1/2,+1,-1,-1/2)=
{ig^2\varrho_4^A\wvrho_{1A}}\
\frac{\langle3,4\rangle^3\langle1,3\rangle}{\prod_{i=1}^4\langle i,i+1\rangle}.
\label{eqn:AmpA(+1/2,+1,-1,-1/2)}
\ee
\item 4-fermion scattering amplitude:
\bear
&&A_{\mathcal{O}(M^0)}(+1/2,+1/2,-1/2,-1/2)
\nn\\
&=& \frac{2ig^2{\langle3,4\rangle}^2}
{\prod_{i=1}^4 \langle i,i+1\rangle}
\Big\{
\varrho_4^A\wvrho_{1A}\wvrho_{2B}\varrho_3^B
\langle1,3\rangle\langle2,4\rangle +
\varrho_3^A\wvrho_{1A}\wvrho_{2B}\varrho_4^B\langle2,3\rangle\langle4,1\rangle
\Big\},
\label{eqn:AmpA(+1/2,+1/2,-1/2,-1/2)}
\eear
\item 2-fermion and 2-scalar scattering amplitude:
\bear
&&
A_{\mathcal{O}(M^0)}(+1/2,0,0,-1/2)
\nn\\
&=&\frac{2ig^2\langle3,4\rangle\langle2,4\rangle}{\prod_{i=1}^4 \langle i,i+1\rangle}
\left\{\frac{1}{2}
\varrho_4^A\wvrho_{1A}\varphi_2^{BC}\varphi_{3BC}\langle1,2\rangle\langle3,4\rangle
+\varrho_4^B\wvrho_{1A}\varphi_2^{CA}\varphi_{3BC}\langle2,3\rangle\langle4,1\rangle
\right\},
\nn\\ &&\qquad
\label{eqn:AmpA(+1/2,0,0,-1/2)}
\eear
where $\varphi_2$ and $\varphi_3$ are wavefunctions for the external scalars.

\end{itemize}

\subsection{Extended MHV amplitudes at $\mathcal{O}(M)$}
\label{sec:MHV-M1}

\begin{itemize}
\item 2-gluon and 2-fermion scattering amplitude:
\be
A_{\mathcal{O}(M)}(+1/2,+1,-1,+1/2)=
\frac{ig^2M^{AB}\wvrho_{1A}\wvrho_{4B}}{2}\
\frac{\langle3,1\rangle\langle3,4\rangle \langle4,1\rangle }
{\prod_{i=1}^4\langle i,i+1\rangle}.
\label{eqn:AmpA(+1/2,+1,-1,+1/2)}
\ee
\item 4-fermion scattering amplitude:
\bear
&&{
A_{\mathcal{O}(M)}(+1/2,+1/2,-1/2,+1/2)
}\nn\\
&=& \frac{2ig^2}{\prod_{1}^{4} \langle i, i+1\rangle}
\Big\{ \wvrho_{1A}M^{AB}\wvrho_{2B}\varrho_3^C\wvrho_{4C}
\langle1,2\rangle\langle2,3\rangle\langle3,1\rangle
\nn\\
&& \quad +  \wvrho_{1A}\varrho_3^A \wvrho_{2B} M^{BD}\wvrho_{4D} \langle2,4\rangle\langle2,3\rangle\langle3,4\rangle + \wvrho_{2B}\varrho_3^B \wvrho_{1A} M^{AD}
\wvrho_{4D}  \langle4,1\rangle\langle1,3\rangle\langle3,4\rangle \Big\}.
\nn\\
&&\qquad
\label{eqn:AmpA(+1/2,+1/2,-1/2,+1/2)}
\eear
\item 2-fermion and 2-scalar scattering amplitude:
\bear\label{eqn:AmpA(+1/2,0,0,+1/2)}
&&A_{\mathcal{O}(M)}(+1/2,0,0,+1/2)
\nn\\
&=&\frac{ig^2}{\prod_{i=1}^4 \langle i,i+1\rangle}
\Big\{
\langle2,3\rangle\langle3,4\rangle\langle4,2\rangle
\wvrho_{4C}M^{CB}\varphi_{3BD}\varphi_2^{DA}\wvrho_{1A}
\nn\\
&& \quad
+ \langle1,2\rangle\langle2,3\rangle\langle3,1\rangle
\wvrho_{4B}\varphi_{3}^{BD}\varphi_{2DA}M^{AD}\wvrho_{1D}
\nn\\
&& \quad
\frac{1}{2}
\left(\langle1,2\rangle\langle2,3\rangle\langle3,4\rangle
+ \langle1,2\rangle\langle1,4\rangle\langle3,4\rangle\right)
\wvrho_{4B}M^{BC}\wvrho_{1C}\varphi_{3}^{B'D'}\varphi_{2B'D'} \Big\}.
\eear

\end{itemize}

\section{Chiral B-model mass terms}\label{sec:chbmt}
We now compare the amplitudes calculated in
\secref{sec:scamp} with an integral over the moduli space
of holomorphic curves in twistor space.
Let us begin by reviewing
some facts about super twistor space \cite{Witten:2003nn}.
We denote the homogeneous
coordinates of the B-model target space
$\CP^{3|4}\setminus\CP^{1|4}$ by
$$
Z_1 = \lam^1,
\quad
Z_2=\lam^2,
\quad
Z_3=\mu_{\dot{1}},
\quad
Z_4=\mu_{\dot{2}},
\quad
\Theta^1,\dots,\Theta^4.
$$
It is convenient to define the two patches
\be\label{eqn:patches}
U\defineas\{Z^1\neq 0
\};
\qquad
U'\defineas\{
Z^2\neq 0
\}.
\ee
On the  patch $U$, the set
$$
Z\defineas \frac{Z_2}{Z_1},
\quad
X\defineas \frac{Z_3}{Z_1},
\quad
Y\defineas \frac{Z_4}{Z_1},
\quad
\Psi^A\defineas \frac{\Theta^A}{Z_1},
$$
is a good coordinate system.
On $U'$,
\be\label{eqn:XYZpr}
Z'\defineas \frac{Z_1}{Z_2}=\frac{1}{Z},
\quad
X'\defineas \frac{Z_3}{Z_2}=\frac{X}{Z},
\quad
Y'\defineas \frac{Z_4}{Z_2}=\frac{Y}{Z},
\quad
{\Psi'}^A\defineas \frac{\Theta^A}{Z_2} = \frac{1}{Z}\Psi^A,
\ee
is a good coordinate system.

Given a meromorphic function
\bear\label{eqn:expamsionA}
\cA(X,Y,Z,\Psi^1,\dots,\Psi^4) &=&
A+\Psi^A\chi_A
+\tfrac{1}{2}\Psi^A\Psi^B\phi_{AB}
\nn\\ &&
+\tfrac{1}{6}\epsilon_{ABCD}\Psi^A\Psi^B\Psi^C\wchi^D +\tfrac{1}{24}\epsilon_{ABCD}\Psi^A\Psi^B\Psi^C\Psi^D G,
\label{eqn:AXXPsiC}
\eear
where $A, \chi_A, \phi_{AB}, \wchi^D, G$
are holomorphic functions of $X,Y,Z$ with possible poles
at $Z=0$ and $Z=\infty$,
we can construct an on-shell wave-function of the $N=4$
fermion fields by (see appendix of \cite{Witten:2003nn}),
\be\label{eqn:psixC}
\begin{split}
\psi^{\a A}(x) &= \frac{1}{2\pi i}\oint_C \lam^\a
\wchi^A(x_{1\dot{1}}+x_{2\dot{1}}z,
        x_{1\dot{2}}+x_{2\dot{2}}z, z) dz,
\qquad
\lam^1\equiv 1,
\quad
\lam^2\equiv z,
\\
\bpsi^\dta_A(x) &= \frac{1}{2\pi i}\oint_C
\ppx{x_{1\dta}}\chi_A(x_{1\dot{1}}+x_{2\dot{1}}z,
       x_{1\dot{2}}+x_{2\dot{2}}z, z)dz.
\end{split}
\ee
The contour integrals are performed on, say, a  circle
around the origin.
There are also similar expressions for the bosons
$A, \phi_{AB}, G.$

We have a lot of freedom in choosing the holomorphic functions
$\wchi^A$ and $\chi_A$, and it is only their singular behavior
at $z=0$ and $z=\infty$ that is important, as we will now review.
We can deform the path $C$ of the
contour integrals \eqref{eqn:psixC}
to a small loop around the origin $z=0.$
This shows that these integrals are only sensitive
to the singular behavior of $\wchi^A$ and $\chi_A$
at $Z=0.$
Adding to $\wchi^A$ or $\chi_A$ a holomorphic function
of $X,Y,Z$ that is nonsingular for all $Z\neq\infty$
will not affect the physical wave-functions.
Similarly, we can perform the integrals \eqref{eqn:psixC}
in the coordinate system $X',Y',Z'.$
In these coordinates we set the superfield $\cA'$ to
$$
\cA'(X',Y',Z',{\Psi'}^1,\dots,{\Psi'}^4) =
\cA(X,Y,Z,\Psi^1,\dots,\Psi^4).
$$
The components of this field are
$$
\cA' =
A'+{\Psi'}^A\chi'_A
+\tfrac{1}{2}{\Psi'}^A{\Psi'}^B\phi'_{AB}
+\tfrac{1}{6}\epsilon_{ABCD}{\Psi'}^A{\Psi'}^B{\Psi'}^C\wchi'^D
+\tfrac{1}{24}\epsilon_{ABCD}
  {\Psi'}^A{\Psi'}^B{\Psi'}^C{\Psi'}^D G',
$$
Thus, the transformation rules for the
fermionic components are
$$
\wchi'^A(X', Y', Z') =
\frac{1}{{Z'}^3}\wchi^A(\frac{X'}{Z'}, \frac{Y'}{Z'}, \frac{1}{Z'}),
\qquad
\chi'_A(X', Y', Z') =
\frac{1}{Z'}\chi_A(\frac{X'}{Z'}, \frac{Y'}{Z'}, \frac{1}{Z'}).
$$
In these variables,
the contour integrals \eqref{eqn:psixC} can be written as
\be\label{eqn:psixCpr}
\begin{split}
\psi^{\a A}(x) &= -\frac{1}{2\pi i}\oint_C {\lam'}^\a
\frac{1}{z'^3}\wchi'^A(x_{1\dot{1}}z'+x_{2\dot{1}},
        x_{1\dot{2}}z'+x_{2\dot{2}}, z') dz',
\qquad
{\lam'}^1\equiv z',
\quad
{\lam'}^2\equiv 1,
\\
\bpsi^\dta_A(x) &=
-\frac{1}{2\pi i}\oint_C
\frac{1}{z'}\ppx{x_{1\dta}}\chi'_A(x_{1\dot{1}}z'+x_{2\dot{1}},
       x_{1\dot{2}}z'+x_{2\dot{2}}, z')dz'
\\
&=
-\frac{1}{2\pi i}\oint_C
\ppx{x_{2\dta}}\chi'_A(x_{1\dot{1}}z'+x_{2\dot{1}},
       x_{1\dot{2}}z'+x_{2\dot{2}}, z')dz'.
\end{split}
\ee
We require that
the fields $\wchi'^A(X', Y', Z')$ and $\chi'_A(X', Y', Z')$
be holomorphic in $X', Y', Z'$ for all finite $X', Y'$ and all
nonzero and finite $Z'.$
But we allow singularities at $Z'=0.$
In fact, similarly to the case
of \eqref{eqn:psixC},
the contour integrals are
only sensitive to the singular behavior of the fields at
$Z'=0.$
Thus, adding to $\cA'$ a holomorphic function
of $X',Y',Z', {\Psi'}^1,\dots, {\Psi'}^4$
that is nonsingular at $Z'=0$
will not affect the physical wavefunctions.

To summarize, there is a freedom in the choice of $\cA$,
\be\label{eqn:cAHom}
\cA(X,Y,Z,\Psi)\sim
\cA(X,Y,Z,\Psi)  + \cA_0(X,Y,Z,\Psi) + \cA_\infty(X,Y,Z,\Psi),
\ee
where $\cA_0$ is
an arbitrary meromorphic wavefunction that is
holomorphic at $Z\ne\infty$ (including $Z=0$),
and $\cA_\infty$ is
an arbitrary meromorphic wavefunction  that is
holomorphic at $Z\ne 0$ (including $Z=\infty$).
To check the holomorphicity requirement for $\cA_\infty$
one has to know what the good coordinates near
$Z=\infty$ are. In the undeformed case, these are given by
\eqref{eqn:XYZpr}.

In the next subsection we will reverse this logic
and find a deformation of the complex structure that
corresponds to a chiral mass term.
The idea is as follows.
First we find a solution to the Dirac equation
\eqref{eqn:DiracEq} in a form that augments \eqref{eqn:psixC}.
It will be given in terms of meromorphic functions
on twistor space that we will denote again by $\wchi^A$
and $\chi_A.$
Then, we define a superfield similarly to
\eqref{eqn:expamsionA}, and we look for an equivalence
in the form \eqref{eqn:cAHom}.
Since \eqref{eqn:psixC} will be augmented,
invariance of the physical wavefunctions
$\psi^{\a A}$ and $\bpsi^\dta_A$ under \eqref{eqn:cAHom}
will require a different definition of ``holomorphic at
$Z=\infty.$''
This will yield an augmentation of the transition functions
\eqref{eqn:XYZpr}, which will give us the
desired deformation of the complex structure.
Let's move on to the details!

\subsection{Super-complex structure deformation}\label{subsec:cpx}
As explained in \cite{Chiou:2005jn},
a chiral mass term can be incorporated into the
B-model twistor string theory as a certain
supercomplex structure deformation.
General deformations of the complex structure of
weighted projective superspaces
(and other holomorphic vector bundles) were
studied in \cite{Popov:2004rb}-\cite{Wolf:2004hp}.

A supercomplex structure deformation can be described
by changing the transition functions \eqref{eqn:XYZpr}.
The new transition functions that we need turned out to be
\be\label{eqn:XYZprDef}
Z'=\frac{1}{Z},
\quad
X'=\frac{X}{Z},
\quad
Y'=\frac{Y}{Z},
\quad
{\Psi'}^A = \frac{1}{Z}\Psi^A
+\frac{1}{6 Z^2}M^{AB}\epsilon_{BCDE}\Psi^C\Psi^D\Psi^E.
\ee
Let us recall how this deformation was derived
in \cite{Chiou:2005jn}.
We start with the free-field Dirac equations \eqref{eqn:DiracEq}.
The generic solution was given in \eqref{eqn:DiracSol}.
There, $\wvrho_A$ and $\varrho^A$ are both functions of
$\lam$ and $\tlam$, but we can Fourier transform them
with respect to $\tlam$ to obtain functions of twistor
space that we denote by $\chi_A$ and $\wchi^A.$
We can then write the solution to the Dirac equation
\eqref{eqn:DiracEq} as
\be\label{eqn:psixCM}
\begin{split}
\psi_\a^A(x) &= \frac{1}{2\pi i}\oint_C
\bigl\lbrack
\lam_\a
\wchi^A(z, x_{1\dot{1}}+x_{2\dot{1}}z,
           x_{1\dot{2}}+x_{2\dot{2}}z)
+M^{AB}\eta_\a\chi_B(z,
       x_{1\dot{1}}+x_{2\dot{1}}z,
       x_{1\dot{2}}+x_{2\dot{2}}z)
\bigr\rbrack dz,
\\
\bpsi^\dta_A(x) &= \frac{1}{2\pi i}\oint_C
\ppx{x_{1\dta}}\chi_A(z, x_{1\dot{1}}+x_{2\dot{1}}z,
       x_{1\dot{2}}+x_{2\dot{2}}z) dz,
\end{split}
\ee
where
$$
(\lam^1,\lam^2)\equiv (1, z),
\qquad
(\eta_1, \eta_2)\equiv (1, 0).
$$
[Equations \eqref{eqn:psixCM} can be compared to
\eqref{eqn:psixC} in the massless case.]
We can collect $\chi_A$ and $\wchi^A$ in a superfield
as in \eqref{eqn:expamsionA}.
Let us see what would be the analog of the equivalence
relation \eqref{eqn:cAHom}.
Obviously, $\bpsi^\dta_A$
and $\psi_{\a A}$ in \eqref{eqn:psixCM} do not change if
we add an arbitrary holomorphic function at $z\ne\infty$
to either $\wchi^A$ or $\chi_A$ or both.
Thus, the equivalence $\cA\sim\cA+\cA_0$ is the same as
in the massless case \eqref{eqn:cAHom}.

Things change, however, for $\cA_\infty$ in \eqref{eqn:cAHom}.
The coordinates $X', Y', Z'$ from \eqref{eqn:XYZpr}
are no longer good coordinates near $Z=\infty.$
If they were, we could define
\be\label{eqn:WrongDef}
\chi'_A(Z, X, Y) =
\frac{1}{Z}\chi_A(\frac{1}{Z}, \frac{X}{Z}, \frac{Y}{Z}),
\qquad
{\wchi}^{\prime\prime A}(Z, X, Y) =
\frac{1}{Z^3}\wchi^A(\frac{1}{Z}, \frac{X}{Z}, \frac{Y}{Z}),
\ee
(we use ${\wchi}^{\prime\prime A}$ because it is only a temporary
expression and we will modify it below)
to change, and we could write
the integrals in \eqref{eqn:psixCM} as follows.
We could substitute
\eqref{eqn:WrongDef} in the second equation of
\eqref{eqn:psixCM} to get
\be\label{eqn:WpsixCdnA}
\begin{split}
\bpsi^\dta_A(x) &= \frac{1}{2\pi i}\oint_C
\frac{1}{z}\ppx{x_{1\dta}}\chi_A(\frac{1}{z},
\frac{x_{1\dot{1}}}{z}+x_{2\dot{1}},
       \frac{x_{1\dot{2}}}{z}+x_{2\dot{2}}) dz
\\
&\qquad= -\frac{1}{2\pi i}\oint_C
\ppx{x_{2\dta}}\chi_A(z',
x_{1\dot{1}}z'+x_{2\dot{1}},
x_{1\dot{2}}z'+x_{2\dot{2}}) dz'
\end{split}
\ee
This last integrand is regular near $z'=0$
and therefore vanishes if $\chi_A$ is holomorphic near
$z'=0$, as in the massless case.
However, there would be a problem with the
integrand for $\bpsi^\dta_A.$
Substituting \eqref{eqn:WrongDef} in the first equation of
\eqref{eqn:psixCM}, we would get
\be\label{eqn:WpsixCMupA}
\begin{split}
\psi_\a^A(x) &= \frac{1}{2\pi i}\oint_C
\bigl\lbrack
\lam_\a
\frac{1}{z^3}
{\wchi}^{\prime\prime A}(\frac{1}{z}, \frac{x_{1\dot{1}}}{z}+x_{2\dot{1}},
           \frac{x_{1\dot{2}}}{z}+x_{2\dot{2}})
+\frac{1}{z}M^{AB}\eta_\a\chi'_B(\frac{1}{z},
       \frac{x_{1\dot{1}}}{z}+x_{2\dot{1}},
       \frac{x_{1\dot{2}}}{z}+x_{2\dot{2}})
\bigr\rbrack dz
\\
&= -\frac{1}{2\pi i}\oint_C
\bigl\lbrack
\lam_\a z'
{\wchi}^{\prime\prime A}(z', x_{1\dot{1}} z'+x_{2\dot{1}},
           x_{1\dot{2}}z'+x_{2\dot{2}})
+\frac{1}{z'} M^{AB}\eta_\a\chi'_B(z',
       x_{1\dot{1}}z'+x_{2\dot{1}},
       x_{1\dot{2}}z'+x_{2\dot{2}})
\bigr\rbrack dz'
\\
\end{split}
\ee
If ${\wchi}^{\prime \prime A}$ is regular at $z'=\infty$,
the first term in the integrand is  regular,
because for $\a=2$ we have $\lam_\a=1$ and for
$\a=1$ we have $\lam_\a=-z=-1/z'$.
However, the second term may have a pole for $\a=1$
because $\eta_1=1.$
The integrand of \eqref{eqn:WpsixCMupA} therefore
does not necessarily vanish.
This shows that \eqref{eqn:WrongDef} is incompatible with
\eqref{eqn:cAHom}.
We can fix this problem by a slight modification of
\eqref{eqn:WrongDef}. We define instead,
\be\label{eqn:NewDef}
\chi'_A(Z, X, Y) =
\frac{1}{Z}\chi_A(\frac{1}{Z}, \frac{X}{Z}, \frac{Y}{Z}),
\qquad
{\wchi}^{\prime A}(Z, X, Y) =
\frac{1}{Z^3}\wchi^A(\frac{1}{Z}, \frac{X}{Z}, \frac{Y}{Z})
-\frac{1}{Z^2}M^{AB}\chi_A(\frac{1}{Z}, \frac{X}{Z}, \frac{Y}{Z}).
\ee
This can be inverted to
\be\label{eqn:InvDef}
\chi_A(Z, X, Y) =
\frac{1}{Z}\chi'_A(\frac{1}{Z}, \frac{X}{Z}, \frac{Y}{Z}),
\qquad
\wchi^A(Z, X, Y) =
\frac{1}{Z^3}{\wchi}^{\prime A}(\frac{1}{Z}, \frac{X}{Z}, \frac{Y}{Z})
+\frac{1}{Z^2}M^{AB}\chi'_A(\frac{1}{Z}, \frac{X}{Z}, \frac{Y}{Z}).
\ee

Then
\be\label{eqn:psixCDef}
\begin{split}
\psi_\a^A(x) &= \frac{1}{2\pi i}\oint_C
\bigl\lbrack
\lam_\a
\frac{1}{z^3}
{\wchi}^{\prime A}(\frac{1}{z}, \frac{x_{1\dot{1}}}{z}+x_{2\dot{1}},
           \frac{x_{1\dot{2}}}{z}+x_{2\dot{2}})
\\
&\qquad
+\frac{1}{z}M^{AB}(\eta_\a+\frac{1}{z}\lam_\a)\chi'_B(\frac{1}{z},
       \frac{x_{1\dot{1}}}{z}+x_{2\dot{1}},
       \frac{x_{1\dot{2}}}{z}+x_{2\dot{2}})
\bigr\rbrack dz
\\
&= -\frac{1}{2\pi i}\oint_C
\bigl\lbrack
\lam_\a z'
{\wchi}^{\prime A}(z', x_{1\dot{1}} z'+x_{2\dot{1}},
           x_{1\dot{2}}z'+x_{2\dot{2}})
\\
&\qquad
+M^{AB}(\frac{1}{z'}\eta_\a+\lam_\a)
\chi_B(z',
       x_{1\dot{1}}z'+x_{2\dot{1}},
       x_{1\dot{2}}z'+x_{2\dot{2}})
\bigr\rbrack dz'
\\
\end{split}
\ee
Now the integrand is regular at $z'=\infty$ because
$$
\frac{1}{z'}\eta_1+\lam_1 
\stackrel{z'\rightarrow\infty}{\longrightarrow} 0,
\qquad
\frac{1}{z'}\eta_2+\lam_2 
\stackrel{z'\rightarrow\infty}{\longrightarrow} 1.
$$
Thus, the field redefinition \eqref{eqn:NewDef}
is compatible with the equivalence relation
\eqref{eqn:cAHom}.
These redefinitions \eqref{eqn:NewDef} are
the $\Psi$ and $\Psi\Psi\Psi$ components
of the superfield expression
$$
\cA'(X', Y', Z', {\Psi'}^A)
=\cA(X, Y, Z, \Psi^A),
$$
where the coordinates $X', Y', Z', {\Psi'}^A$ are defined in
\eqref{eqn:XYZprDef}.

\subsection{A note on the anti-chiral mass term}\label{subsec:ancpx}
One might wonder whether we could derive a similar
modification of the complex structure of
super twistor space for the anti-chiral mass term
deformation \textit{$M_{AB} \psi_\a^A\psi^{\a B}$}.
In this case, instead of the Dirac equations
\eqref{eqn:DiracEq} we get
$$
p_{\a\dta}\bpsi^\dta_A = M_{AB}\psi_\a^B,
\qquad
p_{\a\dta}\psi^{\a A} = 0.
$$
Instead of \eqref{eqn:psixCM},
the solution is now given by
\be\label{eqn:psixAnCM}
\begin{split}
\psi_\a^A(x) &= \frac{1}{2\pi i}\oint_C
\lam_\a
\wchi^A(z, x_{1\dot{1}}+x_{2\dot{1}}z,
           x_{1\dot{2}}+x_{2\dot{2}}z)
\bigr\rbrack dz,
\\
\bpsi^\dta_A(x) &= \frac{1}{2\pi i}\oint_C
\bigl\lbrack
\ppx{x_{1\dta}}\chi_A(z, x_{1\dot{1}}+x_{2\dot{1}}z,
       x_{1\dot{2}}+x_{2\dot{2}}z)
+M_{AB}\hat{\eta}_\dta\wchi^B(z,
       x_{1\dot{1}}+x_{2\dot{1}}z,
       x_{1\dot{2}}+x_{2\dot{2}}z)
\bigr\rbrack dz.
\end{split}
\ee
Here $\hat{\eta}_\dta$ has to satisfy
$$
\tlam^\dta\hat{\eta}_\dta = 1.
$$
In twistor space, however, we identify $\tlam^\dta$
with a differential operator
$$
\tlam^\dta \leftrightarrow i\ppx{\mu_\dta}.
$$
We can therefore set $\hat{\eta}_\dta$ to be the following
integral operator
\be\label{eqn:hateta}
\hat{\eta}_{\dot{1}}f(z,\mu_{\dot{1}}, \mu_{\dot{2}})
\defineas 0,
\qquad
\hat{\eta}_{\dot{2}}f(z,\mu_{\dot{1}}, \mu_{\dot{2}})
\defineas -i \int_0^{\mu_{\dot{2}}}
f(z,\mu_{\dot{1}}, s) ds,
\ee
where $f$ is an arbitrary holomorphic function.
Now the field redefinition \eqref{eqn:WrongDef} is good enough:
\be\label{eqn:WpsixAnCM}
\begin{split}
\bpsi^\dta_A(x) &=
\frac{1}{2\pi i}\oint_C
\bigl\lbrack
\ppx{x_{1\dta}}\chi_A(z, x_{1\dot{1}}+x_{2\dot{1}}z,
       x_{1\dot{2}}+x_{2\dot{2}}z)
+M_{AB}\hat{\eta}_\dta\wchi^B(z,
       x_{1\dot{1}}+x_{2\dot{1}}z,
       x_{1\dot{2}}+x_{2\dot{2}}z)
\bigr\rbrack dz
\nn\\
&=
\frac{1}{2\pi i}\oint_C
\bigl\lbrack
\frac{1}{z}\ppx{x_{1\dta}}
\chi'_A(\frac{1}{z}, \frac{1}{z}x_{1\dot{1}}+x_{2\dot{1}},
       \frac{1}{z}x_{1\dot{2}}+x_{2\dot{2}})
\nn\\ &\qquad\qquad
+\frac{1}{z^3}M_{AB}\hat{\eta}_\dta{\wchi}^{\prime \prime B}(\frac{1}{z},
       \frac{1}{z}x_{1\dot{1}}+x_{2\dot{1}},
       \frac{1}{z}x_{1\dot{2}}+x_{2\dot{2}})
\bigr\rbrack dz
\\
&=
-\frac{1}{2\pi i}\oint_C
\bigl\lbrack
z'\ppx{x_{1\dta}}
\chi'_A(z', x_{1\dot{1}}z'+x_{2\dot{1}},
            x_{1\dot{2}}z'+x_{2\dot{2}})
\nn\\ &\qquad\qquad
+{z'}^3M_{AB}\hat{\eta}_\dta{\wchi}^{\prime \prime B}(z',
       x_{1\dot{1}}z'+x_{2\dot{1}},
       x_{1\dot{2}}z'+x_{2\dot{2}})
\bigr\rbrack \frac{dz'}{{z'}^2}
\\
\end{split}
\ee
Using \eqref{eqn:hateta} we see that
for $\dta=\dot{1}$ the integrand is regular at $z'=0.$
For $\dta=\dot{2}$ we get
\be\label{eqn:WpsixAnCM2}
\begin{split}
\bpsi^{\dot{2}}_A(x) &=
-\frac{1}{2\pi i}\oint_C
\bigl\lbrack
{z'}^2\ppx{x_{2\dot{2}}}
\chi'_A(z', x_{1\dot{1}}z'+x_{2\dot{1}},
            x_{1\dot{2}}z'+x_{2\dot{2}})
\nn\\ &\qquad\qquad
+{z'}^3 M_{AB}
\int_0^{x_{1\dot{2}}z'+x_{2\dot{2}}}
{\wchi}^{\prime B}(z',
       x_{1\dot{1}}z'+x_{2\dot{1}},
       s')\frac{ds'}{z'}
\bigr\rbrack \frac{dz'}{{z'}^2},
\\
\end{split}
\ee
where we used the definition \eqref{eqn:hateta}
for $\dta=\dot{2}$, and we changed variables from $s$
to $s'=z' s$ in the
integral that defines $\hat{\eta}_{\dot{2}}.$
We now see that the integrand in
\eqref{eqn:WpsixAnCM2} is regular at $z'=\infty.$
Therefore, no deformation of complex structure is needed!
However, it was suggested  in \cite{Chiou:2005jn},
that the anti-chiral
mass parameter \textit{$M_{AB}$}
enters in a phase of D1-instanton terms.
The argument was based on a proposed indentification of
\textit{$M_{AB}$} with a VEV of one of the conformal
supergravity fields discussed in \cite{Berkovits:2004jj}.
We will not explore this further in the present paper.

\subsection{Deformed Holomorphic Curve}
\label{subsec:deg1}
In the undeformed twistor space,
a curve of degree $d=1$ in $\CP^{3|4}$ is given by
(see equation (4.46) of \cite{Witten:2003nn})
a set of linear equations:
\be\label{eqn:RC1}
X = -x_{1\dot{1}} -x_{2\dot{1}} Z,
\qquad
Y = -x_{1\dot{2}} -x_{2\dot{2}} Z,
\qquad
\Psi^A = -\theta_1^A -\theta_2^A Z,
\ee
where $x_{\a\dta}$ and $\theta_\a^A$ are moduli.
On the patch $U'$
[defined in \eqref{eqn:patches}]
we can write \eqref{eqn:RC1} as
\be\label{eqn:RC1p}
X' = -x_{1\dot{1}}Z' -x_{2\dot{1}},
\qquad
Y' = -x_{1\dot{2}}Z' -x_{2\dot{2}},
\qquad
{\Psi}^A = -\theta_1^A Z' -\theta_2^A.
\ee
After the deformation \eqref{eqn:XYZprDef},
equations \eqref{eqn:RC1p} no longer hold.
Instead, equations \eqref{eqn:RC1} imply
\bear
{\Psi'}^A &=& \frac{1}{Z}\Psi^A
+\frac{1}{6 Z^2}M^{AB}\epsilon_{BCDE}\Psi^C\Psi^D\Psi^E
\nn\\
&=&
-\theta_1^A Z' -\theta_2^A
-\frac{1}{6 Z'}M^{AB}\epsilon_{BCDE}
(\theta_1^C Z' +\theta_2^C)
(\theta_1^D Z' +\theta_2^D)
(\theta_1^E Z' +\theta_2^E)
\nn\\
&=&
-\frac{1}{6} M^{AB}\epsilon_{BCDE}\theta_1^C\theta_1^D\theta_1^E {Z'}^2
-(\theta_1^A
+\frac{1}{2}M^{AB}\epsilon_{BCDE}\theta_1^C\theta_1^D\theta_2^E) Z'
\nn\\ &&
-(\theta_2^A
+\frac{1}{2}M^{AB}\epsilon_{BCDE}\theta_1^C\theta_2^D\theta_2^E)
-\frac{1}{2 Z'}M^{AB}\epsilon_{BCDE}\theta_2^C\theta_2^D\theta_2^E.
\nn
\eear
Unless the last term is zero, this is not an acceptable
holomorphic curve.
We can cancel the last term by slightly modifying
equations \eqref{eqn:RC1} in the patch $U$ as follows.
We set
\be\label{eqn:RC1m}
\Psi^A = -\theta_1^A -\theta_2^A Z
+\frac{1}{6} M^{AB}\epsilon_{BCDE}\theta_2^C\theta_2^D\theta_2^E Z^2.
\ee
Then
\bear
{\Psi'}^A &=& \frac{1}{Z}\Psi^A
+\frac{1}{6 Z^2}M^{AB}\epsilon_{BCDE}\Psi^C\Psi^D\Psi^E
\nn\\
&=&
-\frac{1}{6} M^{AB}
  \epsilon_{BCDE}\theta_1^C\theta_1^D\theta_1^E {Z'}^2
-(\theta_1^A
+\frac{1}{2} M^{AB}
   \epsilon_{BCDE}\theta_1^C\theta_1^D\theta_2^E) Z'
\nn\\ &&
-(\theta_2^A
+\frac{1}{2} M^{AB}\epsilon_{BCDE}\theta_1^C\theta_2^D\theta_2^E
+\frac{1}{2} M^{AB}\epsilon_{BCDE}M^{EF}\epsilon_{FGHI}
\theta_1^C\theta_1^D\theta_2^G\theta_2^H\theta_2^I)
\nn
\eear
The poles in $Z'$ vanish --
the terms proportional to $1/Z'^2$ vanishes
because there are too many $\theta_2$'s, and
the terms proportional to $1/Z'$
vanish because the symmetric mass parameter
\textit{$M^{AB}$}
is coupled with the antisymmetric $\epsilon-$tensor.
In the next section we will express the scattering amplitudes
calculated in \secref{sec:scamp} as an integral over
supertwistor space with support on the deformed holomorphic
curve that we just found:
\be\label{eqn:RC1mm}
\begin{split}
X &= -x_{1\dot{1}} -x_{2\dot{1}} Z,
\qquad
Y = -x_{1\dot{2}} -x_{2\dot{2}} Z,
\\
\Psi^A &= -\theta_1^A -\theta_2^A Z
+\frac{1}{6}
  M^{AB}\epsilon_{BCDE}\theta_2^C\theta_2^D\theta_2^E Z^2.
\\
\end{split}
\ee

\section{Twistor amplitudes with mass terms}\label{sec:twmass}

As explained in \cite{Witten:2003nn},
a supersymmetric Yang-Mills amplitude $A(f_i)$
can be obtained from a twistor scattering amplitude
$\tilde{A}(\lambda_i^{\alpha}, \mu_i^{\dot{\alpha}}, \Psi_i^{A})$
by multiplying by the appropriate external wavefunctions
\textit{$f_i(\lambda_i^{\alpha}, \mu_i^{\dot{\alpha}}, \Psi_i^{A})$}
and integrating out all the supertwistor coordinates.
In particular,
the tree-level MHV amplitudes for 4 external particles
(without the mass deformation) can be written as
\bear\label{eqn:4epTw}
\tilde{A}(\lambda_i^{\alpha}, \mu_{i\dot{\alpha}}, \Psi_i^{A}) &=&
2ig^2  \int d^4 x \,d^8 \theta_a^A
\prod_{i=1}^{4} \delta^2 ( \mu_{i\dot{\alpha}} + x_{\alpha\dot{\alpha}} \lambda_i^{\alpha})
\delta^{4}(\Psi_i^A + \theta_{\alpha}^A \lambda_i^{\alpha}) \frac{1}{\prod_{i=1}^{4} \langle i, i+1\rangle}, \nn\\
A(f_i) &=& \prod_{i=1}^{4} \int d^2 \lambda_i^{\alpha} d^2 \mu_{i\dot{\alpha}}d^4
\Psi_i^{A} f_i(\lambda_i^{\alpha}, \mu_{i\dot{\alpha}}, \Psi_i^{A})
\tilde{A}(\lambda_i^{\alpha}, \mu_i^{\dot{\alpha}}, \Psi_i^{A}).
\eear
The $\delta$-functions in the integral imply that the twistor amplitude vanishes unless the external points
$(\lambda_i^{\alpha}, \mu_{i\dot{\alpha}}, \Psi_i^{A})$ all lie in the same holomorphic curve described by
\be
\mu_{i\dot{\alpha}} + x_{\alpha\dot{\alpha}} \lambda_i^{\alpha}=0,
\qquad
\Psi_i^A + \theta_{\alpha}^A \lambda_i^{\alpha}=0,
\ee
for some suitable $(x_{\a\dta},\th_\a^A)$.

In the presence of our mass term
the complex structure is deformed and,
according to \eqref{eqn:RC1mm},
the holomorphic curve is changed to
\be\label{eqn:curve}
\mu_{i\dot{\alpha}} + x_{\alpha\dot{\alpha}} \lambda_i^{\alpha}=0,
\qquad
\Psi_i^A + \theta_1^A + \theta_2^A Z_i
-\frac{1}{6}
  M^{AB}\epsilon_{BCDE}\theta_2^C\theta_2^D\theta_2^E Z_i^2=0.
\ee
We claim that \eqref{eqn:4epTw} is modified to
\bear\label{eqn:4epTw-with-m}
\tilde{A}(\lambda_i^{'\alpha}, \mu'_{i\dot{\alpha}}, \Psi_i^{A}) &=&
 2ig^2  \int d^4 x \,d^8 \theta_a^A  \frac{1}{\prod_{i=1}^{4} \langle \lam'_i, \tlam'_{i+1}\rangle} \nn\\
&&\ \times \prod_{i=1}^{4}
\delta^2 ( \mu'_{i\dot{\alpha}} + x_{\alpha\dot{\alpha}} \lambda_i^{'\alpha})
\delta^{4}(\Psi_i^A + \theta_1^A + \theta_2^A Z'_i -
\frac{1}{6}M^{AB}\epsilon_{BCDE}\theta_2^C\theta_2^D\theta_2^E Z_i^{'2}),\nn\\
A(f_i) &=& \prod_{i=1}^{4} \int d^2 \lambda_i^{'\alpha} d^2 \mu'_{i\dot{\alpha}}d^4
\Psi_i^{A} f_i(\lambda_i^{'\alpha}, \mu'_{i\dot{\alpha}}, \Psi_i^{A})
\tilde{A}(\lambda_i^{'\alpha}, \mu'_{i\dot{\alpha}}, \Psi_i^{A})
\eear
for all ``extended MHV'' amplitudes (defined in \secref{sec:scamp}) of 4 external particles.

In particular, if the external function is in a momentum state,
$f_i$ is of the form of a Fourier
transform of a $\delta$-function peaked at some fixed momentum; i.e.,
\be
f_i^{(p_i=\lam_i\tlam_i)}(\lambda_i^{'\alpha}, \mu'_{i\dot{\alpha}}, \Psi_i^{A})=
f_i(\Psi_i^A)\frac{1}{(2\pi)^4}
\int d^2\tlam_i^{'\dta}\
\delta^2(\lam_i^\a-\lam_i^{'\a})
\delta^2(\tlam_i^\dta-\tlam_i^{'\dta})
\exp({i\tlam_i^{'\dta}\mu'_{i\dta}}),
\ee
where $p_i^{\a\dta}=\lam_i^\a \tlam_i^\dta$ are the momenta for external particles.
The integral over bosonic coordinates in \eqref{eqn:4epTw-with-m} yields
\bear
&&\frac{1}{(2\pi)^4}\int d^2 \lambda_i^{'\alpha} \,d^2 \mu'_{i\dot{\alpha}} \int d^4x
\,g(\lam_i^{'\a})\,\delta^2 ( \mu'_{i\dot{\alpha}} + x_{\alpha\dot{\alpha}} \lambda_i^{'\alpha})\nn\\
&&\qquad\times
\int d^2\tlam_i^{'\dta}\, \delta^2(\lam_i^\a-\lam_i^{'\a})
\delta^2(\tlam_i^\dta-\tlam_i^{'\dta})\exp({i\tlam_i^{'\dta}\mu'_{i\dta}})\nn\\
&=&\frac{1}{(2\pi)^4}\int d^4x \exp(-i\sum_i x_{\a\dta}\lam_i^\a \tlam_i^\dta)\,g(\lam_i^\a)
=\delta^4(\sum_i p_i)\,g(\lam_i^\a),
\eear
where $g(\lam_i^{'\a})$ denotes the $\lam'$-dependence of
$\tilde{A}(\lambda_i^{'\alpha}, \mu'_{i\dot{\alpha}}, \Psi_i^{A})$ apart from
$\delta^2 ( \mu'_{i\dot{\alpha}} + x_{\alpha\dot{\alpha}} \lambda_i^{'\alpha})$
and the result gives rise
the momentum conservation factor. Consequently, \eqref{eqn:4epTw-with-m} reduces to
\bear\label{eqn:twistor am}
A(f_i)&=&2ig^2\delta^4(\sum_i p_i)
\int \prod_{i}^{4} d^4\Psi_{i}^{A} \int d^8\theta_{a}^{A}
\prod_{i=1}^{4}\delta^{4}(\Psi_i^A + \theta_1^A + \theta_2^A Z_i -
\frac{1}{6}M^{AB}\epsilon_{BCDE}\theta_2^C\theta_2^D\theta_2^E Z_i^2)
\nn \\
&& ~\times \frac{1}{\prod_{i=1}^{4} \langle i, i+1\rangle} f_1(\Psi_1^A)f_2(\Psi_2^A)f_3(\Psi_3^A)f_4(\Psi_4^A),
\eear
where the external function $f_i(\Psi_i^A)$ is given by $A,$ $\Psi^A\chi_A$, $\frac{1}{2}\Psi^A\Psi^B\phi_{AB}$,
$\frac{1}{6}\epsilon_{ABCD}\Psi^A\Psi^B\Psi^C\tilde{\chi}^D$ and
$\frac{1}{24}\epsilon_{ABCD}\Psi^A\Psi^B\Psi^C\Psi^D G$ for the external particle with
spin $+1$, $+1/2$, $0$, $-1/2$ and $-1$ respectively.

In the following, we compute various
amplitudes in momentum space by \eqref{eqn:twistor am}
(ignore the momentum conservation factor)
and compare them with the results we have calculated by
Feynman rules in \secref{sec:scamp}, with the
identification:
$A=G=\frac{1}{\sqrt{2}}$, $\chi=\wvrho$,
$\tilde{\chi}=\varrho$ and $\phi=\varphi$.
In all cases, both results agree with each other
and we thus verify the claim in
\eqref{eqn:4epTw-with-m}.
Finally, in \secref{sec:3phi},
we study a simple but interesting
case --- the 3-scalar interaction.

\subsection{MHV amplitudes (extended MHV at $\mathcal{O}(M^0)$)}

The Grassmanian integral over $d^8\th^A_a$ in
\eqref{eqn:twistor am} gives a nonzero result only if
the integrand has exactly 8 Grassman $\th$'s.
In the cases of MHV amplitudes, the external functions
$f_i(\Psi_i^A)$ altogether have 8 fermionic
coordinates $\Psi$'s, each of which gives 1 or 3 $\th$'s by the
$\delta$-function.
The mass term in \eqref{eqn:curve} is of the order
$\mathcal{O}(\th^3)$
and thus has too many $\th$'s to contribute in
\eqref{eqn:twistor am} for MHV amplitudes.
As a result, the mass deformation
does not affect the MHV amplitudes and the amplitudes
are the same as if no mass deformation.
In the following, we compute various MHV amplitudes by
\eqref{eqn:twistor am}
and compare them with the results in \secref{sec:MHV-M0}.

\subsubsection{$A_{\mathcal{O}(M^0)}(+1,+1,-1,-1)$}

The external functions for this case are $A_1$, $A_2$,
$\frac{1}{24} \epsilon_{ABCD} \Psi^{A}\Psi^{B}\Psi^{C}\Psi^{D} G_3$ and
$\frac{1}{24} \epsilon_{ABCD} \Psi^{A}\Psi^{B}\Psi^{C}\Psi^{D} G_4$.
The integral \eqref{eqn:twistor am} gives the amplitude
\bear
&& 2ig^2
\int \prod_{i}^{4} d^4\Psi_{i}^{A}
\int d^8\theta_{a}^{A}\prod_{i=1}^{4}\delta^{4}(
\Psi_i^A
+ \theta_1^A
+ \theta_2^A Z_i
-\frac{1}{6}M^{AB}\epsilon_{BCDE}
  \theta_2^C\theta_2^D\theta_2^E Z_i^2)
\nn \\
&& ~\times \frac{1}{\prod_{i=1}^{4} \langle i, i+1\rangle}
(A_1)(A_2)
\left(\frac{1}{24} \epsilon_{ABCD}
\Psi_{3}^{A}\Psi_{3}^{B}\Psi_{3}^{C} \Psi_{3}^{D} G_3\right)
\left(\frac{1}{24} \epsilon_{A'B'C'D'}
\Psi_{3}^{A'}\Psi_{3}^{B'}\Psi_{3}^{C'} \Psi_{3}^{D'} G_4\right)
\nn \\
&=&2ig^2\frac{A_1A_2G_3G_4}{\prod_{i=1}^{4} \langle i, i+1\rangle}(Z_3-Z_4)^4
=2ig^2\frac{A_1A_2G_3G_4}{\prod_{i=1}^{4} \langle i, i+1\rangle}\langle3,4\rangle^4,
\eear
which agrees with \eqref{eqn:AmpA(+1,+1,-1,-1)}.

\subsubsection{$A_{\mathcal{O}(M^0)}(+1/2,+1,-1,-1/2)$}
The external wavefunctions are $\Psi^A \chi_{A}$, $A_2$,
$\frac{1}{24} \epsilon_{ABCD} \Psi^{A}\Psi^{B}\Psi^{C}\Psi^{D} G_3$ and
$\frac{1}{6} \epsilon_{ABCD} \Psi^{A}\Psi^{B}\Psi^{C}
\tilde{\chi}^{D}$. The amplitude is
\bear\label{eqn:thPsInt}
&&2ig^2 \int \prod_{i}^{4} d^4\Psi_{i}^{A}
\int d^8\theta_{a}^{A} \prod_{i=1}^{4}\delta^{4}(
\Psi_i^A
+ \theta_1^A
+ \theta_2^A Z_i
-\frac{1}{6}M^{AB}\epsilon_{BCDE}
  \theta_2^C\theta_2^D\theta_2^E Z_i^2)
\nn\\
&& ~\times \frac{1}{\prod_{1}^{4} \langle i, i+1\rangle}
 (\Psi_1^A \chi_{1A})(A_2)
(\frac{1}{24} \epsilon_{A'B'C'D'}
 \Psi_{3}^{A'}\Psi_{3}^{B'}\Psi_{3}^{C'} \Psi_{3}^{D'} G_3)(
\frac{1}{6}  \epsilon_{A''B''C''D''}
\Psi_{4}^{A''}\Psi_{4}^{B''}\Psi_{4}^{C''}
\tilde{\chi}_{4}^{D''})
\nn\\
&=& - 2ig^2 \int d^8\theta_{a}^{A}
\frac{\chi_{1A} A_2 G_3 \tilde{\chi}_{4}^{D''}}{
\prod_{1}^{4} \langle i, i+1\rangle}7
 (\theta_1^A + \theta_2^A Z_1 + \cdots) \frac{1}{24}
\epsilon_{A'B'C'D'}(\theta_1^{A'} + \theta_2^{A'} Z_3
+ \cdots )
\nn\\
&& ~\times (\theta_1^{B'} + \theta_2^{B'} Z_3 + \cdots)
(\theta_1^{C'} + \theta_2^{C'} Z_3 + \cdots)
(\theta_1^{D'} + \theta_2^{D'} Z_3 + \cdots)
\nn\\
&& ~\times \frac{1}{6}
\epsilon_{{A''}{B''}{C''}{D''}}(\theta_1^{A''}
+ \theta_2^{A''} Z_4 + \cdots)(\theta_1^{B''}
+ \theta_2^{B''} Z_4 + \cdots )(\theta_1^{C''}
+ \theta_2^{C''} Z_4 + \cdots)
\nn\\
&=& 2ig^2\frac{(Z_3 -Z_1)(Z_4-Z_3)^3}{
\prod_{1}^{4} \langle i, i+1\rangle}
\tilde{\chi}_{4}^{A}\chi_{1A} A_2 G_3
= 2ig^2\frac{\langle1,3\rangle\langle3,4\rangle^3}{
\prod_{1}^{4} \langle i, i+1\rangle}
\tilde{\chi}_{4}^{A}\chi_{1A} A_2 G_3,
\eear
where $\cdots$ represents the mass-deformed part of the holomorphic curve.
The result agrees with
\eqref{eqn:AmpA(+1/2,+1,-1,-1/2)}

\subsubsection{$A_{\mathcal{O}(M^0)}(+1/2,+1/2,-1/2,-1/2)$}
The external wavefunctions for this amplitude are $\Psi^A \chi_{A}$, $\Psi^A \chi_{A}$,
$\frac{1}{24} \epsilon_{ABCD} \Psi^{A}\Psi^{B}\Psi^{C}\Psi^{D} G_3$ and
$\frac{1}{24} \epsilon_{ABCD} \Psi^{A}\Psi^{B}\Psi^{C}\Psi^{D} G_3$.
The amplitude is
\bear
&&2ig^2 \int \prod_{i}^{4} d^4\Psi_{i}^{A}
\int d^8\theta_{a}^{A}\prod_{i=1}^{4}\delta^{4}(
\Psi_i^A
+ \theta_1^A
+ \theta_2^A Z_i
- \frac{1}{6}M^{AB}\epsilon_{BCDE}
\theta_2^C\theta_2^D\theta_2^E Z_i^2)
\frac{1}{\prod_{1}^{4}\langle i, i+1\rangle}
\nn \\
&&\quad\times (\Psi_1^A \chi_{1A})(\Psi_2^{A'} \chi_{2{A'}})
(\frac{1}{6} \epsilon_{{A''}{B''}{C''}{D''}}
\Psi_{3}^{A''}\Psi_{3}^{B''}\Psi_{3}^{C''}
\tilde{\chi}_{3}^{D''})(
\frac{1}{24} \epsilon_{A^{'''}B^{'''}C^{'''}D^{'''}}
\Psi^{A^{'''}}\Psi^{B^{'''}}\Psi^{C^{'''}}\Psi^{D^{'''}} G_3)
\nn \\
&=& \frac{2ig^2\langle3,4\rangle^2}{\prod_{1}^{4}
\langle i, i+1\rangle}\Big\{
\tilde{\chi_4}^{A}\chi_{1A}\chi_{2A'}\tilde{\chi_3}^{A'}
\Big[(Z_3-Z_1)Z_4^3 +(Z_1Z_2-Z_2Z_3-2Z_3^2-2Z_1Z_3)Z_4^2
\nn \\
&& \qquad \qquad
+ (Z_3^3-Z_1Z_3^2-2Z_2Z_3^2-2Z_1Z_2Z_3)Z_4
+Z_1Z_2Z_3^2 -Z_2Z_3^3 \Big]
\nn \\
&&+\tilde{\chi_3}^{A} \chi_{1A}\chi_{2A'}\tilde{\chi_4}^{A'}
\Big[(Z_2-Z_3)Z_4^3 +(Z_1Z_3-Z_1Z_2-2Z_3^2-2Z_2Z_3)Z_4^2
\nn \\
&& \qquad \qquad
+ (-Z_3^3+Z_2Z_3^2-2Z_1Z_3^2+2Z_1Z_2Z_3)Z_4
-Z_1Z_2Z_3^3 +Z_1Z_3^3 \Big] \Big\}
\nn \\
&=& \frac{2ig^2\langle3,4\rangle^2}{\prod_{1}^{4}
\langle i, i+1\rangle}\Big\{
\tilde{\chi_4}^{A}\chi_{1A}\chi_{2A'}\tilde{\chi_3}^{A'}
\langle1,3\rangle\langle2,4\rangle
+\tilde{\chi_3}^{A} \chi_{1A}\chi_{2A'}\tilde{\chi_4}^{A'}
\langle2,3\rangle\langle4,1\rangle\Big\},
\eear
which agrees with \eqref{eqn:AmpA(+1/2,+1/2,-1/2,-1/2)}.

\subsubsection{$A_{\mathcal{O}(M^0)}(+1/2,0,0,-1/2)$}

The external wavefunctions are $\Psi_1^A \chi_{1A}$, $\frac{1}{2!} \Psi_{2}^{A}\Psi_{2}^{B}\phi_{2AB}$,
$\frac{1}{2!} \Psi_{3}^{A}\Psi_{3}^{B}\phi_{3AB}$ and
$\frac{1}{3!} \epsilon_{ABCD} \Psi_{4}^{A}\Psi_{4}^{B}\Psi_{4}^{C} \tilde{\chi}_4^{D}$.
The amplitude is
\bear
&& 2ig^2\int \prod_{i}^{4} d^4\Psi_{i}^{A}
\int d^8\theta_{a}^{A} \prod_{i=1}^{4}\delta^{4}(
\Psi_i^A + \theta_1^A + \theta_2^A Z_i
-\frac{1}{6}M^{AB}
 \epsilon_{BCDE}\theta_2^C\theta_2^D\theta_2^E Z_i^2)\,
\frac{1}{\prod_{i=1}^{4} \langle i, i+1\rangle}
\nn \\
&& ~\times  \left(\Psi_1^A \chi_{1A}\right)
\left(\frac{1}{2!} \Psi_{2}^{A'}\Psi_{2}^{B'}\phi_{2A'B'}\right)
\left(\frac{1}{2!} \Psi_{3}^{A''}\Psi_{3}^{B''}\phi_{3A''B''}\right)
\left(\frac{1}{3!} \epsilon_{A'''B'''C'''D'''} \Psi_{4}^{A'''}\Psi_{4}^{B'''}\Psi_{4}^{C'''} \tilde{\chi}_4^{D'''}\right)
\nn \\
&& = \frac{2ig^2}{\prod_{i=1}^{4}\langle i, i+1\rangle}
\Big\{ \tilde{\chi}_{4}^{A'}\phi_{3A'B'}\phi_2^{B'A}\chi_{1A}
(-Z_3Z_4^3 +Z_1Z_3Z_4^2+Z_3^2Z_4^2
+2Z_2Z_3Z_4^2+2Z_1Z_2Z_3Z_4
\nn \\
&& \qquad
+Z_1Z_2^2Z_4+Z_2^2Z_3Z_4-Z_1Z_2^2Z_3)
+ \tilde{\chi}_{4}^{A'}\phi_{2A'B'}\phi_3^{B'A}\chi_{1A}
(-Z_2Z_4^3 +Z_1Z_2Z_4^2+Z_2^2Z_4^2
\nn \\
&& \qquad
+2Z_2Z_3Z_4^2+2Z_1Z_2Z_3Z_4+Z_1Z_3^2Z_4+Z_2Z_3^2Z_4-Z_1Z_2Z_3^2) \nn \\
&& \qquad + \frac{1}{2}\tilde{\chi}_{4}^A\chi_{1A}\phi^{3A'B'}
\phi_{2A'B'}(-Z_1Z_4^3 -Z_2^2Z_3^2+2Z_1Z_3Z_4^2
+2Z_1Z_2Z_4^2+2Z_2Z_3^2Z_4 + 2Z_2^2Z_3Z_4) \Big\}
\nn \\
&& = \frac{2ig^2\langle3,4\rangle\langle2,4\rangle}{
\prod_{i=1}^{4}\langle i, i+1\rangle}
\Big\{
\frac{1}{2}\tilde{\chi}_{4}^{A}\chi_{1A}\phi_3^{A'B'}
\phi_{2A'B'}\langle1,2\rangle\langle3,4\rangle
+\tilde{\chi}_{4}^{A'}\phi_{3A'B'}\phi_2^{B'A}
\chi_{1A}\langle2,3\rangle\langle4,1\rangle
\Big\},
\eear
where the identities \eqref{eqn:phi4} and \eqref{eqn:phi6} are used.
The result agrees with \eqref{eqn:AmpA(+1/2,0,0,-1/2)}.

\subsection{Extended MHV amplitudes at $\mathcal{O}(M)$}

In the cases of extended MHV amplitudes at $\mathcal{O}(M)$,
the external functions $f_i(\Psi_i^A)$ altogether
have 6 fermionic coordinates $\Psi$'s.
In order to have 8 $\th$'s for the integrand in
\eqref{eqn:twistor am}, the mass term has to contribute exactly once.
Therefore, the resulting amplitudes are
of the order $\mathcal{O}(M)$.

\subsubsection{$A_{\mathcal{O}(M)}(+1/2,+1,-1,+1/2)$}
The external wavefunctions are $\Psi_1^A \chi_{1A}$, $A_{2}$,
$\frac{1}{24} \epsilon_{ABCD} \Psi_{3}^{A}\Psi_{3}^{A}\Psi_{3}^{A}\Psi_{3}^{D} G_3$, and $\Psi_4^A \chi_{4A}$.
The amplitude by \eqref{eqn:twistor am} is thus
\bear
&& 2ig^2\int \prod_{i}^{4} d^4\Psi_{i}^{A} \int d^8\theta_{a}^{A} \prod_{i=1}^{4}\delta^{4}(\Psi_i^A + \theta_1^A + \theta_2^A Z_i -
\frac{1}{6}M^{AB}\epsilon_{BCDE}\theta_2^C\theta_2^D\theta_2^E Z_i^2)
\nn \\
&& ~\times \frac{1}{\prod_{i=1}^{4} \langle i, i+1\rangle}
(\Psi_1^A \chi_{1A})(A_2)
(\frac{1}{24} \epsilon_{A''B''C''D''} \Psi_{3}^{A''}\Psi_{3}^{B''}\Psi_{3}^{C''} \Psi_{3}^{D''} G_3)(\Psi_4^{A'''} \chi_{4{A'''}})
\nn \\
&=& -2ig^2\int d^8\theta_{a}^{A} \frac{(Z_1 Z_4 + Z_3^2 - Z_3 Z_4 - Z_3 Z_1)(Z_1 - Z_4)}{\prod_{i=1}^{4} \langle i, i+1\rangle} \nn \\
&& ~\times M^{A{A'''}}\chi_{1A} A_2 G_3 \chi_{4{A'''}} \frac{1}{36}\epsilon_{A''FGH}\theta_{1}^{A'}\theta_{1}^{F}\theta_{1}^{G}\theta_{1}^{H}\epsilon_{IJKL}\theta_{2}^{I}\theta_{2}^{J}\theta_{2}^{K}\theta_{2}^{L}
\nn \\
&=& 2ig^2\frac{\langle3,1\rangle\langle3,4\rangle\langle4,1\rangle}{\prod_{i=1}^{4} \langle i, i+1\rangle}\chi_{4B}  M^{BA}\chi_{1A} G_3  A_2.
\eear
The result concurs with \eqref{eqn:AmpA(+1/2,+1,-1,+1/2)}.

\subsubsection{$A_{\mathcal{O}(M)}(+1/2,+1/2,-1/2,+1/2)$}

The external wavefunctions are $\Psi_1^A \chi_{1A}$, $\Psi_2^A \chi_{2A}$,
$\frac{1}{6} \epsilon_{ABCD} \Psi_{3}^{A}\Psi_{3}^{A}\Psi_{3}^{A} \tilde{\chi}_{3}^{D}$ and
$\Psi_4^A \chi_{4A}$.
The integral \eqref{eqn:twistor am} gives the amplitude
\bear
&&2ig^2 \int \prod_{i}^{4} d^4\Psi_{i}^{A} \int d^8\theta_{a}^{A}\prod_{i=1}^{4}\delta^{4}(\Psi_i^A + \theta_1^A + \theta_2^A Z_i - \frac{1}{6}M^{AB}\epsilon_{BCDE}\theta_2^C\theta_2^D\theta_2^E Z_i^2)
\nn \\
&& \quad\times\frac{1}{\prod_{1}^{4} \langle i, i+1\rangle} (\Psi_1^A \chi_{1A})(\Psi_2^{A'} \chi_{2{A'}})
(\frac{1}{6} \epsilon_{{A''}{B''}{C''}{D''}} \Psi_{3}^{A''}\Psi_{3}^{B''}\Psi_{3}^{C''} \tilde{\chi}_{3}^{D''})(\Psi_4^{A'''} \chi_{4{A'''}})
\nn \\
&=&  \frac{2ig^2}{\prod_{1}^{4} \langle i, i+1\rangle}
\Big\{ \chi_{1A}M^{AA'} \chi_{2A'} \tilde{\chi}^{3A''}\chi_{4A''} \Big( Z_1Z_2(Z_1-Z_2) + Z_3(Z_2^2-Z_1^2)+Z_3^2(Z_1-Z_2) \Big) \nn \\
&& + \chi_{1A}\tilde{\chi}^{3A} \chi_{2A'}M^{A'A'''}\chi_{4A'''} \Big( Z_1Z_4(Z_4-Z_1) + Z_3(Z_1^2-Z_4^2)+Z_3^2(Z_4-Z_1) \Big)  \nn \\
&& +  \chi_{2A'} \tilde{\chi}^{3A'}\chi_{1A} M^{AA'''}\chi_{4A'''} \Big( Z_2Z_4(Z_4-Z_2) + Z_3(Z_2^2-Z_4^2)+Z_3^2(Z_4-Z_2) \Big) \Big\}
\nn \\
&=&  \frac{2ig^2}{\prod_{1}^{4} \langle i, i+1\rangle}
\Big\{ \chi_{1A}M^{AA'} \chi_{2A'} \tilde{\chi}^{3A''} \chi_{4A''} \langle1,2\rangle\langle2,3\rangle\langle3,1\rangle
+\chi_{1A}\tilde{\chi}^{3A} \chi_{2A'}M^{A'A'''}  \chi_{4A'''} \langle2,4\rangle\langle2,3\rangle\langle3,4\rangle\nn\\
&& \qquad \qquad+  \chi_{2A'} \tilde{\chi}^{3A'}\chi_{1A} M^{AA'''}  \chi_{4A'''}\langle4,1\rangle\langle1,3\rangle\langle3,4\rangle \Big\},
\eear
which agrees with \eqref{eqn:AmpA(+1/2,+1/2,-1/2,+1/2)}.

\subsubsection{$A_{\mathcal{O}(M)}(+1/2,0,0,+1/2)$}

The external wavefunctions are given
by $\Psi_1^A \chi_{1A}$,
$\frac{1}{2!} \Psi_{2}^{A}\Psi_{2}^{B}\phi_{2AB}$,
$\frac{1}{2!} \Psi_{3}^{A}\Psi_{3}^{B}\phi_{3AB}$
and $\Psi_1^{A} \chi_{1A}$.
The amplitude is
\bear
&& 2ig^2\int \prod_{i}^{4} d^4\Psi_{i}^{A} \int d^8\theta_{a}^{A} \prod_{i=1}^{4}\delta^{4}(\Psi_i^A + \theta_1^A + \theta_2^A Z_i -
\frac{1}{6}M^{AB}\epsilon_{BCDE}\theta_2^C\theta_2^D\theta_2^E Z_i^2)
\nn \\
&& ~\times \frac{1}{\prod_{i=1}^{4} \langle i, i+1\rangle} (\Psi_1^A \chi_{1A})
(\frac{1}{2!} \Psi_{2}^{A'}\Psi_{2}^{B'}\phi_{2A'B'})
(\frac{1}{2!} \Psi_{3}^{A''}\Psi_{3}^{B''}\phi_{3A''B''})
(\Psi_1^{A'''} \chi_{1{A'''}})
\nn \\
&=& \frac{ig^2}{2\prod_{i=1}^{4}\langle i, i+1\rangle}
\left \{
\chi_{4A'''}M^{A'''A}\chi_{1A}\phi_3^{BC}\phi_{2BC}
\right.\nn\\
&&\qquad\qquad\qquad\qquad\times\Big(Z_1^2Z_4 - Z_1Z_4^2 + Z_2Z_4^2 - Z_2^2 Z_4 Z_1 Z_3^2 - Z_1^2 Z_3\Big)\nn\\
&&\qquad\qquad\qquad \left.
+ 2\chi_{4A'''}M^{A'''A'}\phi_{3A'B'}\phi_2^{B'A}\chi_{1A} \Big(Z_2 Z_4^2 - Z_2^2 Z_4 + Z_3^2 Z_4 - Z_3 Z_4^2\Big)
\right.\nn\\
&&\qquad \qquad\qquad \left.
+ 2\chi_{4A'''}\phi_3^{A'''B'}\phi_{2B'A'}M^{A'A}\chi_{1A} \Big(Z_1^2 Z_2 - Z_1 Z_2^2 + Z_1 Z_3^2 - Z_1^2 Z_3\Big)
\right.\nn\\
&&\qquad \qquad\qquad \left.
+ 2 \epsilon^{AB'B''A'''} M^{A''A'}\chi_{4{A'''}}\chi_{1A}\phi_3^{A''B''}\phi_{2A'B'} \Big(Z_2Z_3^2 - Z_2^2 Z_3\Big)
\right \} \nn \\
&=& \frac{ig^2}{2\prod_{i=1}^{4}\langle i, i+1\rangle}
\left \{
\chi_{4{A'''}}M^{A'''A}\chi_{1A}\phi_3^{BC}\phi_{2BC}\Big(\langle1,2\rangle\langle2,3\rangle\langle3,4\rangle + \langle1,2\rangle\langle3,4\rangle\langle1,4\rangle\Big)
\right.\nn\\
&&\qquad\qquad\qquad \left.
+ 2\chi_{4A'''}M^{A'''A'}\phi_{3A'B'}\phi_2^{B'A}\chi_{1A} \langle2,3\rangle\langle3,4\rangle\langle4,2\rangle \right. \nn\\
&&\qquad \qquad\qquad \left.
+ 2\chi_{4A'''}\phi_3^{A'''B'}\phi_{2B'A'}M^{A'A}\chi_{1A} \langle1,2\rangle\langle2,3\rangle\langle3,1\rangle \right \},
\eear
where the identities \eqref{eqn:phi4}, \eqref{eqn:phi8} and $M^{A'A''}\phi_{2A'A''}=0$ (due to the symmetry of $M$ and
antisymmetry of $\phi$) are used. The result concurs with \eqref{eqn:AmpA(+1/2,0,0,+1/2)}.

\subsubsection{$A_{\mathcal{O}(M)}(0,0,0)$}\label{sec:3phi}
Finally, we study the simple but instructive case: 3-scalar amplitude, i.e., $A_{\mathcal{O}(M)}(0,0,0)$.
For the amplitudes of 3 massless particles, the momentum conservation implies that $p_i$ are collinear.
Angular momentum conservation further forces the amplitude to vanish for 3-gluon scattering. For
3-scalar scattering, this is not the case and in fact this is
the only nonvanishing 3-particle
extended MHV of $\mathcal{O}(M)$.
The amplitude can be obtained as well from the integral in super-twistor space [by a formula similar to
\eqref{eqn:twistor am} but with only 3 external functions and the prefactor $2g^2$ replaced by $g/2$].
It is
\bear
&& \frac{ig}{2}\int \prod_{i}^{3} d^4\Psi_{i}^{A} \int d^8\theta_{a}^{A} \prod_{i=1}^{3}\delta^{4}(\Psi_i^A + \theta_1^A + \theta_2^A Z_i -
\frac{1}{6}M^{AB}\epsilon_{BCDE}\theta_2^C\theta_2^D\theta_2^E Z_i^2)\,
\nn \\
&& ~\times
\frac{1}{\prod_{i=1}^{3} \langle i, i+1\rangle}
\left(\frac{1}{2!} \Psi_{1}^{A}\Psi_{1}^{B}\phi_{1AB}\right)
\left(\frac{1}{2!} \Psi_{2}^{A'}\Psi_{2}^{B'}\phi_{2A'B'}\right)
\left(\frac{1}{2!} \Psi_{3}^{A''}\Psi_{3}^{B''}\phi_{3A''B''}\right)
\nn \\
&=&  \frac{ig}{2\prod_{i=1}^{3}\langle i, i+1\rangle}
\Big\{
Z_1^2Z_2(M^{AA'}\phi_{1AB}\phi_{2A'B'}\phi_3^{BB'})+Z_1^2Z_3(M^{AA'}\phi_{1AB}\phi_{3A'B'}\phi_2^{BB'})\nn\\
&&\qquad\qquad
+Z_2^2Z_3(M^{AA'}\phi_{2AB}\phi_{3A'B'}\phi_1^{BB'})+Z_2^2Z_1(M^{AA'}\phi_{2AB}\phi_{1A'B'}\phi_3^{BB'})\nn\\
&&\qquad\qquad
+Z_3^2Z_1(M^{AA'}\phi_{3AB}\phi_{1A'B'}\phi_2^{BB'})+Z_3^2Z_2(M^{AA'}\phi_{3AB}\phi_{2A'B'}\phi_1^{BB'})
\Big\}\nn\\
&=&\frac{ig (M^{AA'}\phi_{1AB}\phi_{2A'B'}\phi_3^{BB'})}{2\prod_{i=1}^{3}\langle i, i+1\rangle}
\left\{
Z_1^2Z_2-Z_1^2Z_3+Z_2^2Z_3-Z_2^2Z_1+Z_3^2Z_1-Z_3^2Z_2
\right\}\nn\\
&=&-\frac{ig}{2}\left(M^{AA'}\phi_{1AB}\phi_{2A'B'}\phi_3^{BB'}\right),
\eear
and we used the identities
\eqref{eqn:phi6} and \textit{$M^{AB}\phi_{iAB}=0$}.

This result agrees with the tree-level Feynman diagram calculation.
At tree level, we have only one diagram as in
\figref{fig:scalar-vertex}, where the Feynman rule for the vertex
together with $\varphi_\mathcal{I}$, $\varphi_\mathcal{J}$ and
$\varphi_\mathcal{K}$ for the external legs trivially yields the
same result. This confirms again that the mass-deformed
$3\phi$-interaction discussed in \secref{subsec:3phi} and
\appref{app:3phi} is required to have the correct holomorphic
structure in super-twistor space in the presence of the chiral mass
deformation.

\section{$N=1$ Supersymmetry}
\label{sec:susy}
The chiral-mass deformation that we studied in
\secref{sec:cfmt}-\secref{sec:twmass}
depends on 10 complex parameters
\textit{$M^{AB}$} ($A,B=1,\dots,4$) and in general breaks
$N=4$ supersymmetry completely.
In this section we will study a subset of these deformations
-- those that preserve $N=1$ supersymmetry.

The unperturbed $N=4$ super Yang-Mills theory has 8
supersymmetry generators $Q_{\a A}$ and $\bQ_{\dta}^A.$
The $N=1$ deformation that we will study in this section
will preserve
the generators with $A=4$ and break those with $A=1,2,3.$
We will use indices $i,j,k,\dots =1,2,3$ instead of $A$,
when the summation excludes $A=4.$

In $N=1$ superspace notation the theory contains
a vector multiplet $V$ with associated chiral field strength
multiplet $W_\a$ and its complex conjugate $\bW^\dta.$
In addition there are 3 chiral multiplets $\Phi^i$
and their complex conjugates $\bPhi_i$, where $i=1,2,3$
is an $SU(3)$ flavor index.
The Lagrangian is given by
\be\label{eqn:susyLag}
\begin{split}
2\gYM^2 \Lag &= \int d^2\bth d^2\th\,
\tr\bigl\{\bPhi_i e^V\Phi^i\bigr\}
+\int d^2\th
\tr\bigl\{W_\a W^\a + \epsilon_{ijk}\Phi^i[\Phi^j,\Phi^k]
\bigr\}
\\
&
+\int d^2\bth
\tr\bigl\{\bW^\dta\bW_\dta
+ \epsilon^{ijk}\bPhi_i[\bPhi_j,\bPhi_k]
+M^{ij}\bPhi_i\bPhi_j
\bigr\}
\\
\end{split}
\ee
Note that the chiral mass term only deforms the
$d^2\bth$ integral, and the chiral and anti-chiral
superpotentials are not the complex conjugates of each other!
(The situation is reminiscent of the deformations
used in \cite{Dijkgraaf:2002xd}.)

Integrating out the auxiliary fields in the superfields,
we find that the mass deformation adds the following extra
terms to the potential
\be\label{eqn:mass deform}
\Delta U =
\frac{1}{2}\tr\bigl\{M^{ij}\bpsi_{\dta i}\bpsi^\dta_j
+\gYM M^{ij}\epsilon_{jkl}\phi_i^* [\phi^k, \phi^l]
\bigr\},
\ee
where $\phi_i$ is the $\th=\bth=0$ component of
$\Phi_i/\gYM.$

Now let us turn to twistor space \textit{$\CP^{3|4}$}.
Let us first identify the action of the $N=1$ supersymmetry
generators on the undeformed twistor space.
It is
$$
\delta\lam_\a = 0,
\qquad
\delta\mu_\dta = \bspr_\dta\Theta^4,
\qquad
\delta\Theta^i = 0\quad (i=1,2,3),
\qquad
\delta\Theta^4 = \spr_\a\lam^\a,
$$
where $\spr$ and $\bspr$
are the anti-commuting SUSY parameters.
Using the holomorphic coordinates on patch $U$ from
\secref{sec:chbmt},
we can rewrite the SUSY transformations as
\be\label{eqn:XYZsusy}
\delta X = \bspr_{\dot{1}}\Psi^4,
\qquad
\delta Y = \bspr_{\dot{2}}\Psi^4,
\qquad
\delta Z = 0,
\qquad
\delta \Psi^i = 0,\quad (i=1,2,3),
\qquad
\delta \Psi^4 = \spr_1 +\spr_2 Z.
\ee
The mass-deformed space, given by
the transition functions \eqref{eqn:XYZprDef},
which can be written in our case as
\be\label{eqn:XYZprij}
Z'=\frac{1}{Z},
\quad
X'=\frac{X}{Z},
\quad
Y'=\frac{Y}{Z},
\quad
{\Psi'}^i = \frac{1}{Z}\Psi^i
+\frac{1}{2 Z^2}M^{ij}\epsilon_{jkl}\Psi^k\Psi^l\Psi^4,
\quad
{\Psi'}^4 = \frac{1}{Z}\Psi^4,
\ee
is not invariant under the same SUSY transformations
\eqref{eqn:XYZsusy}, because they would imply
$$
\delta {\Psi'}^i =
\frac{1}{2 Z^2}M^{ij}\epsilon_{jkl}\Psi^k\Psi^l
  (\spr_1 +\spr_2 Z)
=
\tfrac{1}{2}M^{ij}\epsilon_{jkl}{\Psi'}^k{\Psi'}^l
 (\spr_1 +\frac{1}{Z'}\spr_2),
$$
which is ill-defined near $Z'=0.$
We can fix this by modifying the SUSY transformation
\eqref{eqn:XYZsusy} to
\be\label{eqn:XYZsusyM}
\delta X = \bspr_{\dot{1}}\Psi^4,
\qquad
\delta Y = \bspr_{\dot{2}}\Psi^4,
\qquad
\delta Z = 0,
\qquad
\delta \Psi^i =
-\tfrac{1}{2}\spr_2 M^{ij}\epsilon_{jkl}\Psi^k\Psi^l,
\qquad
\delta \Psi^4 = \spr_1 +\spr_2 Z.
\ee
This modified transformation law still satisfies the
correct commutation relations, as can be seen after some algebra
and using \textit{$M^{ij}=M^{ji}$}.
Thus, the deformed twistor space associated with
\eqref{eqn:susyLag} is indeed supersymmetric.

\section{Summary and discussion}\label{sec:disc}

In this paper we studied
a new deformation of twistor string theory;
we tested the proposal that
the deformation of twistor space
to a space whose complex structure is
defined by the transition functions \eqref{eqn:XYZprij}
is associated with the
deformation of $N=4$ super Yang-Mills theory given
by the following Lagrangian:
\be
\begin{split}
g^2\Lag &=
\frac{1}{4}\tr\Bigl(
F_{\mu\nu}F^{\mu\nu}+2D_\mu\phi_\mathcal{I} D^\mu\phi_\mathcal{I}
-[\phi_\mathcal{I},\phi_\mathcal{J}]^2\Bigr)
+\frac{i}{2}\tr\Bigl(\bpsi\gamma^\mu D_\mu\psi
+i\bpsi\Gamma^\mathcal{I}[\phi_\mathcal{I},\psi]\Bigr)
\nn\\ &
\qquad+\frac{i}{2}\tr\Bigl(M^{AB}\bpsi^\dta_A\bpsi_{\dta B}
+\frac{1}{4}M^{\mathcal{I}\mathcal{J}\mathcal{K}}
\phi_\mathcal{I} [\phi_\mathcal{J}, \phi_\mathcal{K}]
\Bigr).
\nn
\end{split}
\ee
Here \textit{$M^{AB}=M^{BA}$} is the mass parameter in the
representation $\rep{10}$ of the R-symmetry group $SU(4)$,
and \textit{$M^{\mathcal{I}\mathcal{J}\mathcal{K}}$}
is linearly related to \textit{$M^{AB}$} and is given
in \eqref{eqn:MIJK}.

We calculated tree-level 4-point scattering amplitudes
up to order $O(M)$
and we checked that these amplitudes can be reproduced
from an integral over a moduli space of holomorphic
curves in the deformed twistor space, just like the
undeformed case.

Among other things, twistor string theory is interesting
in that it opens a window into the nonperturbative aspects
of topological string theory on supermanifolds.
There has been a lot of progress recently in understanding
the nonperturbative aspects of topological string
theories on ordinary manifolds
(see for instance \cite{Nekrasov:2004js}-\cite{Nekrasov:2005bb}).

The perturbative open topological string theory
with target space \textit{$\CP^{3|4}$}
reproduces a self-dual truncation of
$N=4$ SYM theory \cite{Witten:2003nn}.
Extensions to other weighted projective target
spaces were demonstrated in
\cite{Popov:2004rb}-\cite{Wolf:2004hp}.
It was also suggested in \cite{Witten:2003nn} that
D1-instantons in the topological string theory
complete the self-dual truncation to a full $N=4$ SYM theory.
In fact, the integral \eqref{eqn:4epTw} (copied from
\cite{Witten:2003nn}) is the one-instanton contribution
to the amplitude.
Our results suggest a possible extension of these ideas
to a 10-parameter family of deformations of
the target space \textit{$\CP^{3|4}$}.

Other deformations of twistor string theory have been
studied in \cite{Kulaxizi:2004pa}\cite{Ahn:2004yu},
and orbifolds of twistor string theory were
studied in \cite{Park:2004bw}\cite{Giombi:2004xv}.
For example, Kulaxizi and  Zoubos \cite{Kulaxizi:2004pa}
translated the so-called $\beta$-deformations
of $N=4$ SYM \cite{Leigh:1995ep}-\cite{Lunin:2005jy}
into a non-anticommutativity among the fermionic
coordinates of super twistor space.
It would be interesting to add a chiral mass term
to these deformations and to the orbifold constructions.

Another possible direction for further study
is the reduction to $D=3$ and lower dimensions.
The relevant target space for $D=3$ is the weighted projective
space \textit{$W\CP^{2|1,1,1,1}$}.
This reduction was studied in
\cite{Chiou:2005jn}\cite{Popov:2005uv}\cite{Saemann:2005ji}
and involves minitwistor space
\cite{Hitchin:1982gh}\cite{AtiyahHitchin}.
Other reductions of twistor string theory have been
recently proposed in \cite{Lechtenfeld:2005xi}.
It would be interesting to further study the corresponding
reduction of the complex structure deformation that 
was described in the present paper.

\acknowledgments
It is a pleasure to thank Mina Aganagic,
Eric Gimon and P. Ho\v{r}ava for helpful discussions.
This work was supported in part by the Director,
Office of Science,
Office of High Energy and Nuclear Physics,
of the U.S. Department of
Energy under Contract DE-AC03-76SF00098,
and in part by the NSF under grant PHY-0098840.


\appendix
\section{Notation and useful formulae}\label{app:notation}
\subsection{Spinors}\label{app:spinors}
Our metric is in Minkowski signature $(+,-,-,-)$.
Spinor indices of type $(1/2,0)$ and $(0,1/2)$ are raised and lowered with antisymmetric tensors
$\epsilon_{\a\b}$, $\epsilon_{\dta\dtb}$
and their inverses $\epsilon^{\a\b}$, $\epsilon_{\dta\dtb}$:
\be
\lam_\a=\epsilon_{\a\b}\lam^\b,
\qquad
\lam^\a=\epsilon^{\a\b}\lam_\b,
\qquad
\tlam_\dta=\epsilon_{\dta\dtb}\tlam^\dtb,
\qquad
\tlam^\dta=\epsilon^{\dta\dtb}\tlam_\dtb,
\ee
with
\be
\epsilon_{\a\b}=\epsilon_{\dta\dtb}
=-\epsilon^{\a\b}=-\epsilon^{\dta\dtb},
\qquad
\epsilon_{12}=1.
\ee
The Lorentz invariants $\langle\lam_1,\lam_2\rangle$
and $[\tlam_1,\tlam_2]$ are defined as
\be
\langle\lam_1,\lam_2\rangle
=-\langle\lam_2,\lam_1\rangle
=\epsilon_{\a\b}\lam_1^\a\lam_2^{\b}
=\lam_1^\a\lam_{2\a}=-\lam_{1\a}\lam_2^\a,
\ee
and
\be
[\tlam_1,\tlam_2]=-[\tlam_2,\tlam_1]=\epsilon^{\dta\dtb}\tlam_{1\dta}\tlam_{2\dtb}
=\tlam_{1\dta}\tlam_2^\dta=-\tlam_1^\dta\tlam_{2\dta}.
\ee
The vector representation of $SO(3,1)$ can be
represented as the tensor
product of two spinor representation of opposite chirality
via:
\be
p_{\a\dta}=\sigma_{\a\dta}^\mu p_\mu,
\qquad
p^{\dta\a}=\bsig^{\mu\dta\a} p_\mu,
\ee
where $\sigma^\mu=(\textbf{1},\vec{\sigma})$,
$\bsig^\mu=(\textbf{1},-\vec{\sigma})$
and $\vec{\sigma}$ are Pauli matrices.

Some useful formulae for $\sigma$-matrices are listed below:
\be\label{eqn:sigma1}
\sigma^\mu_{\a\dta}\sigma_{\mu\b\dtb}
=2\epsilon_{\a\b}\epsilon_{\dta\dtb}, \qquad
\bsig_\mu^{\dta\a}\sigma^\mu_{\b\dtb}
=2\delta^\a_\b\delta^\dta_\dtb,
\ee
\be\label{eqn:sigma2}
{(\sigma^\mu\bsig^\nu+\sigma^\nu\bsig^\mu)_\a}^\b=2\eta^{\mu\nu}{\delta_\a}^\b,
\qquad
{(\bsig^\mu\sigma^\nu+\bsig^\nu\sigma^\mu)_\dta}^\dtb=2\eta^{\mu\nu}{\delta^\dta}_\dtb.
\ee and
\be\label{eqn:sigma3}
\tr \sigma^\mu \bsig^\nu = 2\eta^{\mu\nu}.
\ee

The inner product of two vectors gives
\be\label{eqn:sigma6}
W_\mu V^\mu=\eta^{\mu\nu}W_\mu V_\nu=
\frac{1}{2}\tr[\sigma^\mu\bsig^\nu]W_\mu V_\nu=
\frac{1}{2}W_{\a\dta}V^{\dta\a}=\frac{1}{2}W^{\dta\a}V_{\a\dta}.
\ee

If $p^\mu$ is lightlike, we can decompose $p^\mu$ as
\be\label{eqn:sigma4}
p_{\a\dta}=\lam_\a\tlam_\dta.
\ee
Furthermore, if $q^\mu$ is also lightlike (written as $q_{\a\dta}=\mu_\a\tilde{\mu}_\dta$)
we have
\be\label{eqn:sigma5}
p\cdot q=-\frac{1}{2}\langle\lam,\mu\rangle[\tlam,\tilde{\mu}].
\ee

\subsection{$SU(4)$ $R$-symmetry indices}\label{app:susy-indices}
In $N=4$ super Yang-Mills theory,
the scalar field $\phi_\mathcal{I}$ is real and in the
representation $\textbf{6}$ of $SU(4)$. Since $\textbf{6}$ is the antisymmetric part of $\textbf{4}\times\textbf{4}$
or $\bar{\textbf{4}}\times\bar{\textbf{4}}$,
we can exchange $\phi_\mathcal{I}$
($\mathcal{I}$ is an index of $\textbf{6}$) for the complex antisymmetric field
$\phi_{AB}=-\phi_{BA}$ ($A,B$: indices of $\textbf{4}$) or
$\phi^{AB}=-\phi^{BA}$ ($A,B$: indices of $\bar{\textbf{4}}$) by
\be\label{eqn:phi1}
\phi_{AB}=\Gamma^\mathcal{I}_{AB}\phi_\mathcal{I}, \qquad
\phi^{AB}=\Gamma^{\mathcal{I}AB}\phi_\mathcal{I},
\ee
where $\Gamma^\mathcal{I}$'s satisfy
\be\label{eqn:phi2}
\delta_{\mathcal{I}\mathcal{J}}\Gamma^{\mathcal{I}AB}
\Gamma^\mathcal{J}_{CD}=\frac{1}{2}(\delta^A_C\delta^B_D-\delta^A_D\delta^B_C),
\qquad \delta_{\mathcal{I}\mathcal{J}}\Gamma^{\mathcal{I}}_{AB}
\Gamma^\mathcal{J}_{CD}=\frac{1}{2}\epsilon_{ABCD},
\ee
and
\be\label{eqn:phi3}
\Gamma^{\mathcal{I}AB}=\frac{1}{2}\ \epsilon^{ABCD}\Gamma^{\mathcal{I}}_{CD}, \qquad
\Gamma^{\mathcal{I}}_{AB}=\frac{1}{2}\ \epsilon_{ABCD}\Gamma^{\mathcal{I}CD}.
\ee
The reality condition on $\phi$ now reads as
\be\label{eqn:phi4}
\phi^{AB}=\frac{1}{2}\ \epsilon^{ABCD}\phi_{CD}, \qquad \phi_{AB}=\frac{1}{2}\ \epsilon_{ABCD}\phi^{CD},
\qquad \text{or} \quad (\phi_{AB})^*=\phi^{AB}.
\ee
It follows that
\be\label{eqn:phi5}
\phi_1^\mathcal{I}\phi_{2\mathcal{I}}=\phi_2^\mathcal{I}\phi_{1\mathcal{I}}=
\d^{\mathcal IJ}\phi_{1\mathcal{I}}\phi_{2\mathcal{J}}
=\phi_1^{AB}\phi_{2AB}.
\ee

Some useful formulae regarding antisymmetry of
$\phi_{AB}$ are listed below:
\be\label{eqn:phi6}
-\frac{1}{2}\,\delta^A_B\,\phi_{2CD}\phi_3^{CD}=
\phi_2^{AC}\phi_{3CB}+\phi_3^{AC}\phi_{2CB}, \ee \be\label{eqn:phi7}
-\frac{1}{2}\,M^{AB}\phi_2^{CD}\phi_{3CD}=
M^{BC}\left(\phi_{2CD}\phi_3^{DA}+\phi_{3CD}\phi_2^{DA}\right)
=M^{AC}\left(\phi_{2CD}\phi_3^{DB}+\phi_{3CD}\phi_2^{DB}\right),
\ee
and
\be\label{eqn:phi8}
\epsilon^{AB'B''A'''}\phi_3^{A''B''}\phi_{2A'B'}
=-\delta_{A''}^{A}\phi_3^{B'A'''}\phi_{2A'B'}
+\delta_{A''}^{A'''}\phi_3^{B'A}\phi_{2A'B'}
-\phi_3^{A'''A}\phi_{2A'A''}.
\ee

The mass parameter \textit{$M^{AB}=M^{BA}$} is in the irreducible representation
$\rep{10}$ of the R-symmetry group $SU(4).$
Using the double cover $SU(4)\rightarrow SO(6)$,
the representation $\rep{10}$ of $SU(4)$ is induced from an
irreducible representation of $SO(6)$ which can be realized as
self-dual anti-symmetric 3-tensors.
Explicitly, define
\be\label{eqn:MIJK}
M^{\mathcal{I}\mathcal{J}\mathcal{K}}\defineas
\Gamma^{[\mathcal{I}}_{AB}
\Gamma^{\mathcal{J}}_{CD}
\Gamma^{\mathcal{K}]}_{EF}
\epsilon^{ABCE}M^{DF}
\Longrightarrow
M^{\mathcal{I}\mathcal{J}\mathcal{K}}
=
\tfrac{1}{3!}
\epsilon_{\mathcal{I}\mathcal{J}\mathcal{K}\mathcal{P}\mathcal{Q}\mathcal{R}}
M^{\mathcal{P}\mathcal{Q}\mathcal{R}}.
\ee
Then, the 3-$\phi$ coupling from \eqref{eqn:mass deform1}
can be written as
\be
M^{AB}\epsilon^{CDEF}
\tr\bigl\{\phi_{AC} [\phi_{BD}, \phi_{EF}]\bigr\}
=
M^{\mathcal{I}\mathcal{J}\mathcal{K}}
\tr\bigl\{
\phi_{\mathcal{I}}\phi_{\mathcal{J}}\phi_{\mathcal{K}}
\bigr\}.
\ee
We define the $su(4)$-invariant symbol
\be
\Gamma^{\mathcal{I}\mathcal{J}\mathcal{K}}_{AB}
\defineas
\Gamma^{[\mathcal{I}}_{DE}
\Gamma^{\mathcal{J}}_{CA}
\Gamma^{\mathcal{K}]}_{FB}
\epsilon^{CDEF}.
\ee
It is symmetric in $AB$ and anti-symmetric
in $\mathcal{I}\mathcal{J}\mathcal{K}$ and satisfies
the self-duality relation
\be
\Gamma^{\mathcal{I}\mathcal{J}\mathcal{K}}_{AB}
=\frac{1}{3!}
{\epsilon^{\mathcal{I}\mathcal{J}\mathcal{K}}}_{
\mathcal{P}\mathcal{Q}\mathcal{R}}
\Gamma^{\mathcal{P}\mathcal{Q}\mathcal{R}}_{AB}.
\ee
We can then write
\be
M^{\mathcal{I}\mathcal{J}\mathcal{K}} \equiv
\Gamma^{\mathcal{I}\mathcal{J}\mathcal{K}}_{AB}M^{AB}.
\ee

\section{Feynman rules with chiral mass terms}
\label{app:Feynman rules}
The Lagrangian of $D=4$, $N=4$ Yang-Mills theory is given by
\be\label{eqn:susy4Lag}
\Lag =
\frac{1}{4g^2}\tr\left(F_{\mu\nu}F^{\mu\nu}+2D_\mu\phi_\mathcal{I} D^\mu\phi_\mathcal{I}
-[\phi_\mathcal{I},\phi_\mathcal{J}]^2\right)
+\frac{i}{2g^2}\tr\left(\bpsi\gamma^\mu D_\mu\psi
+i\bpsi\Gamma^\mathcal{I}[\phi_\mathcal{I},\psi]\right),
\ee
where $\bpsi^a=(\psi^\a,\bpsi_\dta)$
is a Dirac spinor, and we treat $\psi^\a$ and $\bpsi_\dta$
independently.\footnote{In $N=4$ SYM
$\psi_a$ is actually a Majorana spinor.
But in order
to get the Feynman rules
by analogy with ordinary QED, we treat
the two chiralities independently
and in the end identify external
fermions as anti-fermions to take into account that $\psi$
is Majorana. See also \appref{app:anti-fermion}.}

In terms of $N=1$ superfields the Lagrangian is
\bear\label{eqn:susy1Lag}
g^2 \Lag &=& \int d^2\bth d^2\th\,
\tr\bigl\{\bPhi_i e^V\Phi^i\bigr\}
+\int d^2\th
\tr\bigl\{W_\a W^\a + \epsilon_{ijk}\Phi^i[\Phi^j,\Phi^k]
\bigr\}
\nn\\
&&+\int d^2\bth
\tr\bigl\{\bW^\dta\bW_\dta + \epsilon^{ijk}\bPhi_i[\bPhi_j,\bPhi_k]
\bigr\},
\nn\\
&&\text{chiral superfields:}\qquad
\Phi_i=\phi_i+\sqrt{2}\th\psi_i+\th\th F_i,\quad i=1,2,3
\nn\\
&&\text{vector superfield:}\qquad
V=-\th\sigma^\mu\bar{\th}A_\mu+i\th\th\bar{\th}\bar{\lam}
-i{\bar \th\th}\th\lam+\frac{1}{2}\th\th{\bar \th\th}D,
\eear
where we identify the component fields of \eqref{eqn:susy1Lag}
with those of \eqref{eqn:susy4Lag} according to
\be\label{eqn:field relation}
\phi_{A=i,B=j}=\epsilon_{ijk}\phi^k,
\qquad
\phi_{A=4,B=i}=\phi_i^*,
\ee
and
\be
(\psi_{A=i},\overline{\psi}_{A=i})=(\psi_{i},\overline{\psi}_{i}),
\qquad (\psi_{A=4},\overline{\psi}_{A=4})=(\lam,\bar{\lam}),
\ee
where $\phi^{AB}=-\phi^{BA}=\Gamma^{\mathcal{I}AB}\phi_\mathcal{I}=\epsilon^{ABCD}\phi_{CD}/2$
(as defined in \appref{app:susy-indices}).

Now, if the chiral mass term $M^{ij}\bPhi_i\bPhi_j$ is added to \eqref{eqn:susy1Lag}
as discussed in \eqref{eqn:susyLag}, $N=4$ supersymmetry is broken to $N=1$ and the extra term \eqref{eqn:mass deform}
leads to
\bear\label{eqn:mass deform1}
\Delta U &=&
\tr\bigl\{M^{AB}\bpsi_{\dta A}\bpsi^\dta_B
+\frac{1}{4}\,g M^{AB}\epsilon^{CDEF}\phi_{AC}
[\phi_{BD}, \phi_{EF}]
\bigr\},
\eear
which is added to \eqref{eqn:susy4Lag}
(with $M^{A=4,B}=M^{B,A=4}=0$).\footnote{
The equality of the chiral
spinor mass terms in \eqref{eqn:mass deform}
and in \eqref{eqn:mass deform1} is obvious,
while the equality of the mass-deformed
$3\phi$-interaction terms
is less transparent and will be discussed in \appref{app:3phi}.}

In this paper (unless otherwise mentioned),
we considered the general chiral mass term
(i.e., $M^{4A}=M^{A4}$ could be nonzero)
and the mass deformation had the form
\eqref{eqn:mass deform1}
(thus breaking $N=1$ supersymmetry in general).
In the following,
we first present the Feynman rules involving the chiral spinor mass term in
\ref{app:FermionProp}--\ref{app:anti-fermion}
and later in \ref{app:3phi} we present the Feynman rule for
the $3\phi$-interaction.

\subsection{Fermion propagators}\label{app:FermionProp}
When the chiral spinor mass term $M^{AB}\bpsi_{\dta A}\bpsi^\dta_B$ is added to \eqref{eqn:susy4Lag},
the Dirac part of the modified Lagrangian reads (the color group factor is ignored)
\be \label{eqn:DiracLag}
\Lag_{Dirac}=\bpsi(i\gamma^\mu\partial_\mu)\psi-M^{AB}\bpsi_{\dta A}\bpsi^\dta_B
=(\psi^\a_A,\bpsi_{\dta A})
\begin{pmatrix} 0 & {\delta_A}^B p_\mu\sigma^\mu_{\a\dtb} \\
 {\delta^A}_B \,p_\mu\bsig^{\mu\dta\b}&  -M^{AB}{\delta^\dta}_\dtb \\
\end{pmatrix}
\begin{pmatrix} \psi_{\b B} \\
 \bpsi^\dtb_B \\
\end{pmatrix}
\ee

The spinor propagator is $i\times$(inverse of the middle operator) on the right of
\eqref{eqn:DiracLag}. With the identities in \eqref{eqn:sigma2},
the propagator is given by
\be \label{eqn:propagator}
\frac{i}{p^2}
\begin{pmatrix} M^{AB}{\delta_\a}^\b & {\delta^A}_B p_\mu\sigma^\mu_{\a\dtb} \\
 {\delta_A}^B p_\mu\bsig^{\mu\dta\b}&  0 \\
\end{pmatrix}
=\frac{i}{p^2}
\begin{pmatrix} M^{AB}{\delta_\a}^\b & {\delta^A}_B p_{\a\dtb} \\
 {\delta_A}^B p^{\dta\b}&  0 \\
\end{pmatrix}.
\ee
The corresponding Feynman rules are listed in \figref{fig:FermionProp}.

\FIGURE{
\begin{picture}(470,100)

\put(30,70){\begin{picture}(200,60)
  \thicklines
  \put(0,0){\vector(1,0){45}}
  \put(45,0){\line(1,0){35}}
  \put(0,5){$\a,A$}
  \put(75,5){$\dtb,B$}
  \put(40,7){$p$}
  \put(115,0){$=i\frac{p^{\dtb\a}}{p^2}{\delta^A}_B$}
\end{picture}}

\put(260,70){\begin{picture}(200,60)
  \thicklines
  \put(0,0){\vector(1,0){45}}
  \put(45,0){\line(1,0){35}}
  \put(0,5){$\dta,A$}
  \put(75,5){$\b,B$}
  \put(40,7){$p$}
  \put(115,0){$=i\frac{p_{\b\dta}}{p^2}{\delta_A}^B$}
\end{picture}}

\put(30,5){\begin{picture}(200,60)
  \thicklines
  \put(0,0){\vector(1,0){45}}
  \put(45,0){\line(1,0){35}}
  \put(0,5){$\a,A$}
  \put(75,5){$\b,B$}
  \put(40,7){$p$}
  \put(115,0){$=i\frac{{\delta_\b}^\a}{p^2}M^{AB}$}
\end{picture}}

\put(260,5){\begin{picture}(200,60)
  \thicklines
  \put(0,0){\vector(1,0){45}}
  \put(45,0){\line(1,0){35}}
  \put(0,5){$\dta,A$}
  \put(75,5){$\dtb,B$}
  \put(40,7){$p$}
  \put(115,0){$=0$}
\end{picture}}

\end{picture}
\caption{
The fermion propagators in the presence of
a chiral mass term.
}\label{fig:FermionProp}
}

\subsection{Solutions of Dirac equation}
With the anti-chiral mass term, the Dirac equation
can be read off from \eqref{eqn:DiracLag} as
\be
\begin{pmatrix} 0 & {\delta_A}^B p_\mu\sigma^\mu_{\a\dtb} \\
 {\delta^A}_B p_\mu\bsig^{\mu\dta\b}&  -M^{AB}{\delta^\dta}_\dtb \\
\end{pmatrix}
\begin{pmatrix} \psi_{\b B} \\
 \bpsi^\dtb_B \\
\end{pmatrix}
=0.
\ee
The solutions were described in \secref{subsec:eom}.
Consider the positive-frequency solutions,
i.e., $\psi_\a(x)=\psi_\a(p)e^{-ip\cdot x}$
and $\bpsi^\dta(x)=\bpsi^\dta(p)e^{-ip\cdot x}$.
These $\psi_\a(p)$ and $\bpsi^\dta(p)$ obey the equation of motion \eqref{eqn:DiracEq}.
It is easy to see that $p^2=p_{\a\dta}p^{\dta\a}=0$, and thus the momentum is lightlike and can be decomposed as \eqref{eqn:lamtlam}.
A basis for the solutions is given by \eqref{eqn:DiracSol},
which is invariant under
\be\label{eqn:sol gauge}
\eta'_\a\rightarrow \eta_\a+\zeta\lam_\a,
\qquad
\varrho'^A\rightarrow \varrho^A-\zeta M^{AB}\wvrho_B,
\qquad
\wvrho'_B\rightarrow\wvrho_B
\ee
for any arbitrary number $\zeta$.

\subsection{Helicities and incoming functions}\label{app:helicity}
In the presence of an anti-chiral mass term, the helicity and chirality
no longer coincide. However, since the 4-momentum
$p$ is still lightlike helicity is Lorentz invariant
and can be used to specify the polarization of incoming
and outgoing fermions.

When the lightlike $p$ is written as
\eqref{eqn:lamtlam} and if we treat
$\wvrho_A(\lam,\tlam)$, $\varrho^A(\lam,\tlam)$ and $\eta_\a(\lam,\tlam)$
in the solution \eqref{eqn:DiracSol} as continuous functions of $\lam$ and $\tlam$,
the helicity operator is given by
\be\label{eqn:helicity}
\hat{h}=\lam_\a\frac{\partial}{\partial\lam_\a}
-\tlam_{\dta}\frac{\partial}{\partial\tlam_\dta},
\ee
which gives eigenvalues $-2h$ when acting on the function $\psi(p)$
if $\psi(p)e^{i p\cdot x}$ is a momentum eigenstate \cite{Witten:2003nn}.
To find the solutions of positive and negative helicities, we first study
some properties of helicities:
\begin{lem}\label{lem:1}
$\hat{h}f(E)=0$ if $f(\lam,\tlam)$ is a function of the energy $E$ only.
\begin{proof}
$2E=(p_0+p_3)+(p_0-p_3)=p_{1{\dot 1}}+p_{2{\dot 2}}
=\lam_1\tlam_{\dot 1}+\lam_2\tlam_{\dot 2}$. It is easy to show $\hat{h}E=0$
and therefore $\hat{h}f(E)=0.$
\end{proof}
\end{lem}
\begin{lem}\label{lem:2}
$\hat{h}\eta_\a=-\eta_\a$, if $\eta_\a$ is given by\footnote{
The exact reason for the choice \eqref{eqn:eta} will be
clear when we study the normalization condition \eqref{eqn:ortho}.
The choice \eqref{eqn:eta} satisfies \eqref{eqn:etalam}.
Also note that in Minkowski signature we have $\lam_\a^*=\pm\tlam_\dta$ and we choose the
$+$ sign here for positive-frequency solutions ($-$ sign is for negative-frequency solutions).}
\be\label{eqn:eta}
\eta_1=\frac{\lam_2^*}{2E}=\frac{\tlam_{\dot 2}}{2E},
\qquad
\eta_2=-\frac{\lam_1^*}{2E}=-\frac{\tlam_{\dot 1}}{2E}.
\ee
\begin{proof}
Follows immediately from
$E=\lam_1\tlam_{\dot 1}+\lam_2\tlam_{\dot 2}$.
\end{proof}
\end{lem}

Using \eqref{eqn:DiracSol}, let's write the solution as a Dirac spinor\footnote{
The Dirac spinor $\psi_a^A=\bpsi^\dta_A+\psi_{\a}^A$
is a shorthand for
$\psi_a^A=\begin{pmatrix}0 \\ \bpsi^\dta_A\end{pmatrix}
+\begin{pmatrix}\psi_{\a}^A \\ 0\end{pmatrix}
=\begin{pmatrix}\psi_{\a}^A \\ \bpsi^\dta_A\end{pmatrix}$.
}
\be
\psi_a^A=\bpsi^\dta_A+\psi_{\a}^A
= \tlam^\dta\wvrho_A
 +\lam_\a\varrho^A +M^{AB}\eta_\a\wvrho_B.
\ee
The eigenvalue problem $\hat{h}\psi_a=-2h\psi_a$ now becomes
\bear
\hat{h}\psi_a^A&=&\hat{h}\left(\tlam^\dta\wvrho_A
+\lam_\a\varrho^A +M^{AB}\eta_\a\wvrho_B\right) \nn \\
&=&-\tlam^\dta\wvrho_A+\tlam^\dta(\hat{h}\wvrho_A)+\lam_\a\varrho^A
+\lam_\a(\hat{h}\varrho^A)+M^{AB}\eta_\a(\hat{h}\wvrho_B)
-M^{AB}\eta_\a\wvrho_B
\eear
by Lemma \ref{lem:2}. Furthermore, by Lemma \ref{lem:1}, if we choose $\varrho=0$ and
$\wvrho=\wvrho(E)$, we get the positive-helicity state with $h=+1/2$; if we choose
$\wvrho=0$ and $\varrho=\varrho(E)$, we get the negative-helicity state with $h=-1/2$.
Therefore,we have a basis of helicity states:
\be
u_{+a}^A=\wvrho_A(E)\tlam^\dta+M^{AB}\wvrho_B(E)\eta_\a, \qquad
u_{-a}^A=\varrho^A(E)\lam_\a.
\ee

The normalization condition will fix $\varrho$, $\wvrho$ and $\eta$.
Firstly, we consider
the orthogonality of $u^+$ and $u^-$:
\be\label{eqn:ortho}
u_-^{\dag A} u^B_+ =(\varrho^{A*}\lam_1^*,\varrho^{A*}\lam_2^*,0,0)
\begin{pmatrix} M^{BC}\wvrho_C\eta_1 \\ M^{BC}\wvrho_C\eta_2 \\ \wvrho_B\tlam^{\dot 1} \\ \wvrho_B\tlam^{\dot 2} \\
\end{pmatrix}
=\varrho^{A*}M^{BC}\wvrho_C\left(\lam_1^*\eta_1+\lam_2^*\eta_2\right)=0.
\ee
This together with \eqref{eqn:etalam} enforces the solution in \eqref{eqn:eta}.

Secondly, consider
\be
u_-^{\dag A} u_-^B =\varrho^{A*}\varrho^B(\lam_1^*,\lam_2^*,0,0)
\begin{pmatrix} \lam_1 \\ \lam_2 \\ 0 \\ 0 \\
\end{pmatrix}
=\varrho^{A*}\varrho^B\left(\lam_1\tlam_{\dot 1}+\lam_2\tlam_{\dot 2}\right)=2E\varrho^{A*}\varrho^B.
\ee
To have the correct normalization condition, namely $u_-^{\dag A} u_-^B =2E\delta^{AB}$,
$\varrho^A$'s have to satisfy
\be\label{eqn:varrho}
\varrho^{A*}\varrho^B=\delta^{AB}.
\ee

Finally, we compute
\bear
u_+^{\dag A} u_+^B &=&\left((M^{AC}\wvrho_C\,\eta_1)^*,(M^{AC}\wvrho_C\,\eta_2)^*,
(\wvrho_A\tlam^{\dot 1})^*,(\wvrho_A\tlam^{\dot 2})^*\right)
\begin{pmatrix} M^{BD}\wvrho_D\eta_1 \\ M^{BD}\wvrho_D\eta_2 \\ \wvrho_B\tlam^{\dot 1} \\ \wvrho_B\tlam^{\dot 2} \\
\end{pmatrix} \nn \\
&=&(M^{AC}\wvrho_C)^*\,M^{BD}\wvrho_D\left(|\eta_1|^2+|\eta_2|^2\right)+
\wvrho_A^*\wvrho_B\left(\tlam^{\dot 1}\lam^1+\tlam^{\dot 2}\lam^2\right) \nn \\
&=&2E\left(\frac{(M^{AC}\wvrho_C)^*\,M^{BD}\wvrho_D}{4E^2}+\wvrho_A^*\wvrho_B\right)
\eear
by \eqref{eqn:eta}. To have $u_+^{\dag A}u_+^B = 2E\delta^{AB}$, we have to set
\be\label{eqn:wvrho}
\frac{(M^{AC}\wvrho_C)^*\,M^{BD}\wvrho_D}{4E^2}+\wvrho_A^*\wvrho_B=\delta^{AB}.
\ee
This in general is not possible. The failure to orthogonalize
the $+$ helicity part is due to the
fact that the Hamiltonian is not Hermitian (CPT is violated).
Nevertheless, for an arbitrarily given
$\wvrho_A(E)$ and $\varrho_A(E)$,
$u_{+a}^A=\wvrho_A\tlam^\dta+M^{AB}\wvrho_B\eta_\a$
can always be normalized and used for incoming states.

To summarize, the basis of normalized helicity states is:
\be\label{eqn:positive sol}
u_{+a}^A(p)=\wvrho_A\tlam^\dta+M^{AB}\wvrho_B\eta_\a, \qquad
u_{-a}^A(p)=\varrho^A\lam_\a, \ee Following the recipe of field
theory, in momentum space, we use $u_\pm(p)$ for the incoming
fermion state functions with $\pm 1/2$ helicities.\footnote{When
studying the holomorphic structure of scattering amplitudes
connected to twistor string theory, we relax the conditions
\eqref{eqn:eta}, \eqref{eqn:varrho} and \eqref{eqn:wvrho}. Instead,
we use the freedom \eqref{eqn:sol gauge} to set $\eta_\a$ to
$\eta_\a=(1,0)$ while scaling
$\lam^\a$ to $(1,Z=\lam^2/\lam^1).$}

\subsection{Outgoing functions}
To find the outgoing states, we cannot just take the Dirac conjugate (i.e., $\bar{u}_\pm^{aA}$) of
\eqref{eqn:positive sol} as in ordinary field theory,
because CPT is no longer invariant.
Instead, we should restore
CPT symmetry by adding the anti-chiral mass term
\textit{$M_{AB}=({M^{AB}})^*$}
and get the new solution $u_\pm^{aA}$ and its Dirac conjugate
$\bar{u}_\pm^{aA}$. In the end, we take $M_{AB}\rightarrow 0$
(formally keeping $M^{AB}$ fixed)
and $\bar{u}_\pm^{aA}$ in this limit will be the
outgoing states for our theory with only a chiral mass term.

With both chiral and anti-chiral masses, the momentum is no longer
lightlike. However, we can take the relativistic limit ($p\gg M$)
and still decompose $p=\lam\tlam$.
At the order $\mathcal{O}(M)$,
the helicity is still well-defined,
and repeating the calculation above leads to
\bear
u_{+a}^A(p)&\approx&\wvrho_A\tlam^\dta+M^{AB}\wvrho_B\eta_\a
=\begin{pmatrix}M^{AB}\wvrho_B\eta_\a
\\
\wvrho_A\tlam^\dta\end{pmatrix},
\nn\\
u_{-a}^A(p)&\approx&\varrho^A\lam_\a
+M_{AB}\varrho^B\tilde{\eta}^\dta
=\begin{pmatrix}\varrho^A\lam_\a
\\
M_{AB}\varrho^B\tilde{\eta}^\dta\end{pmatrix},
\eear
where $\lam^\a\eta_\a=\tlam_\dta\tilde{\eta}^\dta=1$.
This gives the Dirac conjugate:
\bear
\bar{u}_+^{aA}(p)&\approx&\varrho^A\lam^\a-M_{AB}\varrho^B\tilde{\eta}_\dta
=\left(\varrho^A\lam^\a,\ -M_{AB}\varrho^B\tilde{\eta}_\dta\right), \nn\\
\bar{u}_-^{aA}(p)&\approx&-M^{AB}\wvrho_B\eta^\a+\wvrho_A\tlam_\dta
=\left(-M^{AB}\wvrho_B\eta^\a,\ \wvrho_A\tlam_\dta\right),
\eear
by the identities
${M^{AB}}^*=M_{AB}$, $({\wvrho_A})^*=\varrho^A$
and $(\eta_\a)^*=-\tilde{\eta}_{\dta}$.\footnote{
In Minkowski signature,
we have $(\lam^{\a})^*=\pm\tlam^\dta$ and we choose
the $+$ sign here
for positive frequency solutions.
In order to satisfy $\lam^\a\eta_\a=\tlam_\dta\tilde{\eta}^\dta=1$,
we have to choose $(\eta_\a)^*=-\tilde{\eta}_{\dta}$
correspondingly with an extra minus sign.}
Setting $M_{AB}=0$, we get the outgoing state functions:
\be\label{eqn:negtive sol}
\bar{u}_+^{aA}(p)=\varrho^A\lam^\a,
\qquad
\bar{u}_-^{aA}(p)=-M^{AB}\wvrho_B\eta^\a+\wvrho_A\tlam_\dta.
\ee
Equivalently, \eqref{eqn:positive sol} and \eqref{eqn:negtive sol} give the Feynman rules for external fermions
depicted in \figref{fig:extFermion}.

\FIGURE{
\begin{picture}(470,270)

\put(30,200){\begin{picture}(200,60)
  \thicklines
  \put(0,15){\vector(1,0){45}}
  \put(45,15){\line(1,0){35}}
  \put(80,15){\circle*{5}}
  \put(0,20){$+$}
  \put(65,20){$\a$}
  \put(40,22){$p$}
  \put(15,0){\text{incoming}}
  \put(115,10){$=\,M^{AB}\wvrho_B \eta_\a$}
\end{picture}}

\put(260,200){\begin{picture}(200,60)
  \thicklines
  \put(0,15){\vector(1,0){45}}
  \put(45,15){\line(1,0){35}}
  \put(80,15){\circle*{5}}
  \put(0,20){$+$}
  \put(65,20){$\dta$}
  \put(40,22){$p$}
  \put(15,0){\text{incoming}}
  \put(115,10){$=\,\wvrho_A\tlam^\dta$}
\end{picture}}

\put(30,135){\begin{picture}(200,60)
  \thicklines
  \put(0,15){\vector(1,0){45}}
  \put(45,15){\line(1,0){35}}
  \put(80,15){\circle*{5}}
  \put(0,20){$-$}
  \put(65,20){$\a$}
  \put(40,22){$p$}
  \put(15,0){\text{incoming}}
  \put(115,10){$=\,\varrho^A\lam_\a$}
\end{picture}}

\put(260,135){\begin{picture}(200,60)
  \thicklines
  \put(0,15){\vector(1,0){45}}
  \put(45,15){\line(1,0){35}}
  \put(80,15){\circle*{5}}
  \put(0,20){$-$}
  \put(65,20){$\dta$}
  \put(40,22){$p$}
  \put(15,0){\text{incoming}}
  \put(115,10){$=\,0$}
\end{picture}}

\put(30,70){\begin{picture}(200,60)
  \thicklines
  \put(0,15){\line(1,0){40}}
  \put(80,15){\vector(-1,0){40}}
  \put(80,15){\circle*{5}}
  \put(0,20){$+$}
  \put(65,20){$\a$}
  \put(40,22){$p$}
  \put(15,0){\text{outgoing}}
  \put(115,10){$=\,\varrho^A\lam^\a$}
\end{picture}}

\put(260,70){\begin{picture}(200,60)
  \thicklines
  \put(0,15){\line(1,0){40}}
  \put(80,15){\vector(-1,0){40}}
  \put(80,15){\circle*{5}}
  \put(0,20){$+$}
  \put(65,20){$\dta$}
  \put(40,22){$p$}
  \put(15,0){\text{outgoing}}
  \put(115,10){$=\,0$}
\end{picture}}

\put(30,5){\begin{picture}(200,60)
  \thicklines
  \put(0,15){\line(1,0){40}}
  \put(80,15){\vector(-1,0){40}}
  \put(80,15){\circle*{5}}
  \put(0,20){$-$}
  \put(65,20){$\a$}
  \put(40,22){$p$}
  \put(15,0){\text{outgoing}}
  \put(115,10){$=\,-M^{AB}\wvrho_B \eta^\a$}
\end{picture}}

\put(260,5){\begin{picture}(200,60)
  \thicklines
  \put(0,15){\line(1,0){40}}
  \put(80,15){\vector(-1,0){40}}
  \put(80,15){\circle*{5}}
  \put(0,20){$-$}
  \put(65,20){$\dta$}
  \put(40,22){$p$}
  \put(15,0){\text{outgoing}}
  \put(115,10){$=\,\wvrho_A\tlam_\dta$}
\end{picture}}

\end{picture}
\caption{
The Feynman rules for external fermions.
}\label{fig:extFermion}
}

\subsection{Negative-frequency solutions}\label{app:anti-fermion}
Similarly, we can solve for the negative-frequency solutions:
$\psi(x)=v(p)e^{ip\cdot x}$.
Repeating the calculation above, we find a basis
of normalized helicity states for anti-fermions:\footnote{In fact,
since CPT is violated, ``anti-fermion'' is not an appropriate
term to describe the negative-frequency solution.
Nevertheless, we use this name anyway for convenience.}
\be
v_+^A(p)=\wvrho_A\tlam^\dta-M^{AB}\wvrho_B\eta_\a, \qquad
v_-^A(p)=\varrho^A\lam_\a.
\ee
To compute $\bar{v}_\pm(p)$,
we follow the method discussed after \eqref{eqn:negtive sol}.
For negative-frequency solutions, however,
we have $(\lam^{\a})^*=-\tlam^\dta$ and accordingly
we should choose $(\wvrho_A)^*=\varrho^A$
and $(\eta_\a)^*=\tilde{\eta}_{\dta}$ (contrary to the
positive-frequency case). This leads to
\be
\bar{v}_+^A(p)=-\varrho^A\lam^\a, \qquad
\bar{v}_-^A(p)=-\wvrho_A\tlam_\dta-M^{AB}\wvrho_B\eta^\a.
\ee
In momentum space $\bar{v}_\pm(p)$ is used for the incoming
anti-fermions with $\pm 1/2$ helicities and $v_\pm(p)$ for the outgoing anti-fermions
with $\pm 1/2$ helicities. The corresponding Feynman rules for external anti-fermions
are depicted in \figref{fig:extantiFermion}.

Notice that since $\bpsi^a=(\psi^\a,\bpsi_\dta)$ is Majorana,
the anti-fermions with adjoint color $T_i$ and helicity $\pm$
are identical to the fermions with $\bar{T}_i$ and helicity $\mp$. We can treat any
external fermionic legs as either ``fermions'' or ``anti-fermions.'' The Feynman rules
in \figref{fig:extFermion} and \figref{fig:extantiFermion} turn out to give the same resulting
amplitude regardless of which way we choose, as long as all the
directions of the arrows are consistent
with the vertex rules depicted in \figref{fig:vertex}.

\FIGURE{
\begin{picture}(470,270)

\put(30,200){\begin{picture}(200,60)
  \thicklines
  \put(0,15){\line(1,0){40}}
  \put(80,15){\vector(-1,0){40}}
  \put(80,15){\circle*{5}}
  \put(0,20){$+$}
  \put(65,20){$\a$}
  \put(40,30){$p$}
  \put(35,23){\vector(1,0){15}}
  \put(15,0){\text{incoming}}
  \put(115,10){$=\,-\varrho^A\lam^\a$}
\end{picture}}

\put(260,200){\begin{picture}(200,60)
  \thicklines
  \put(0,15){\line(1,0){40}}
  \put(80,15){\vector(-1,0){40}}
  \put(80,15){\circle*{5}}
  \put(0,20){$+$}
  \put(65,20){$\dta$}
  \put(40,30){$p$}
  \put(35,23){\vector(1,0){15}}
  \put(15,0){\text{incoming}}
  \put(115,10){$=\,0$}
\end{picture}}

\put(30,135){\begin{picture}(200,60)
  \thicklines
  \put(0,15){\line(1,0){40}}
  \put(80,15){\vector(-1,0){40}}
  \put(80,15){\circle*{5}}
  \put(0,20){$-$}
  \put(65,20){$\a$}
  \put(40,30){$p$}
  \put(35,23){\vector(1,0){15}}
  \put(15,0){\text{incoming}}
  \put(115,10){$=\,-M^{AB}\wvrho_B \eta^\a$}
\end{picture}}

\put(260,135){\begin{picture}(200,60)
  \thicklines
  \put(0,15){\line(1,0){40}}
  \put(80,15){\vector(-1,0){40}}
  \put(80,15){\circle*{5}}
  \put(0,20){$-$}
  \put(65,20){$\dta$}
  \put(40,30){$p$}
  \put(35,23){\vector(1,0){15}}
  \put(15,0){\text{incoming}}
  \put(115,10){$=\,-\wvrho_A\tlam_\dta$}
\end{picture}}

\put(30,70){\begin{picture}(200,60)
  \thicklines
  \put(0,15){\vector(1,0){45}}
  \put(45,15){\line(1,0){35}}
  \put(80,15){\circle*{5}}
  \put(0,20){$+$}
  \put(65,20){$\a$}
  \put(40,30){$p$}
  \put(50,23){\vector(-1,0){15}}
  \put(15,0){\text{outgoing}}
  \put(115,10){$=\,-M^{AB}\wvrho_B \eta_\a$}
\end{picture}}

\put(260,70){\begin{picture}(200,60)
  \thicklines
  \put(0,15){\vector(1,0){45}}
  \put(45,15){\line(1,0){35}}
  \put(80,15){\circle*{5}}
  \put(0,20){$+$}
  \put(65,20){$\dta$}
  \put(40,30){$p$}
  \put(50,23){\vector(-1,0){15}}
  \put(15,0){\text{outgoing}}
  \put(115,10){$=\,\wvrho_A\tlam^\dta$}
\end{picture}}

\put(30,5){\begin{picture}(200,60)
  \thicklines
  \put(0,15){\vector(1,0){45}}
  \put(45,15){\line(1,0){35}}
  \put(80,15){\circle*{5}}
  \put(0,20){$-$}
  \put(65,20){$\a$}
  \put(40,30){$p$}
  \put(50,23){\vector(-1,0){15}}
  \put(15,0){\text{outgoing}}
  \put(115,10){$=\,\varrho^A\lam_\a$}
\end{picture}}

\put(260,5){\begin{picture}(200,60)
  \thicklines
  \put(0,15){\vector(1,0){45}}
  \put(45,15){\line(1,0){35}}
  \put(80,15){\circle*{5}}
  \put(0,20){$-$}
  \put(65,20){$\dta$}
  \put(40,30){$p$}
  \put(50,23){\vector(-1,0){15}}
  \put(15,0){\text{outgoing}}
  \put(115,10){$=\,0$}
\end{picture}}

\end{picture}
\caption{
The Feynman rules for external anti-fermions.
}\label{fig:extantiFermion}
}

\subsection{Mass-deformed $3\phi$-interaction}\label{app:3phi}
We first study the case that
$M^{AB}$ is restricted to $M^{ij}$ (i.e. $M^{A=4,B}=M^{B,A=4}=0$).
With the field identification \eqref{eqn:field relation},
the mass-deformed $3\phi$-interaction term in
\eqref{eqn:mass deform} is given by
\be\label{eqn:3int1}
M^{ij}\epsilon_{jkl}\,\phi_i^*[\phi^k,\phi^l]
=\frac{1}{4}\,M^{ij}\epsilon_{jkl}\,\phi_{4i}\,\epsilon^{kmn}\epsilon^{lpq}\,[\phi_{mn},\phi_{pq}]
=\frac{1}{2}\,M^{ij}\epsilon^{klm}\,\phi_{4i}[\phi_{jk},\phi_{lm}].
\ee
On the other hand, with $M^{AB}\rightarrow M^{ij}$,
the $3\phi$-interaction term in \eqref{eqn:mass deform1} reduces to
\bear\label{eqn:3int2}
&&\tr\left\{ M^{AB}\epsilon^{CDEF}\,\phi_{AC}[\phi_{BD},\phi_{EF}]\right\}\nn\\
&=&\tr\big\{M^{ij}\epsilon^{4klm}\,\phi_{i4}[\phi_{jk},\phi_{lm}]
+M^{ij}\epsilon^{k4lm}\,\phi_{ik}[\phi_{j4},\phi_{lm}]
+\cdots\big\}\nn\\
&=&2\tr \left\{M^{ij}\epsilon^{klm}\,\phi_{4i}[\phi_{jk},\phi_{lm}]\right\}.
\eear
Comparing \eqref{eqn:3int1} with \eqref{eqn:3int2}, we conclude that the $3\phi$-interaction term in
\eqref{eqn:mass deform} equals that in \eqref{eqn:mass deform1} when $M^{AB}\rightarrow M^{ij}$.

For general $M^{AB}$,
we then have the $3\phi$-interaction in the Lagrangian:
\bear
&&\frac{g}{4}\,\tr\left\{ M^{AB}\epsilon^{CDEF}\,\phi_{AC}[\phi_{BD},\phi_{EF}]\right\}=
\frac{g}{4}\,\tr\left\{ M^{AB}\epsilon^{CDEF}\,\phi_{BD}[\phi_{EF},\phi_{AC}]\right\}=\cdots
\nn\\
&=&\frac{g}{2}\,M^{AB}\Gamma^\mathcal{I}_{AC}\Gamma^\mathcal{J}_{BD}\Gamma^{\mathcal{K}CD}
\tr\left\{\phi_\mathcal{I}\,[\phi_\mathcal{J},\phi_\mathcal{K}]\right\}
=
\frac{g}{2}\,M^{AB}\Gamma^\mathcal{K}_{AC}\Gamma^\mathcal{I}_{BD}\Gamma^{\mathcal{J}CD}
\tr\left\{\phi_\mathcal{I}\,[\phi_\mathcal{J},\phi_\mathcal{K}]\right\}=\cdots\nn\\
&=&\frac{g}{4}\,M^{AB}\Gamma^{[\mathcal{I}}_{AC}\Gamma^\mathcal{J}_{BD}\Gamma^{\mathcal{K}]}_{EF}\epsilon^{CDEF}
\tr\left\{\phi_\mathcal{I}\,[\phi_\mathcal{J},\phi_\mathcal{K}]\right\}
=\frac{g}{4}\,M^{\mathcal IJK}\tr\left\{\phi_\mathcal{I}\,[\phi_\mathcal{J},\phi_\mathcal{K}]\right\}
\eear
where ``$\cdots$'' means cyclic permutation of indices.
The planar part of the corresponding Feynman rule (3-scalar vertex) is depicted in \figref{fig:scalar-vertex}.

\FIGURE{
\begin{picture}(420,125)

\put(40,0){\begin{picture}(200,60)

  \thinlines
  \put(8,16){\line(1,2){6}}
  \put(16,32){\line(1,2){6}}
  \put(24,48){\line(1,2){6}}
  \put(30,60){\line(-1,2){6}}
  \put(22,76){\line(-1,2){6}}
  \put(14,92){\line(-1,2){6}}
  \put(20,15){$\mathcal{I}$}
  \put(20,100){$\mathcal{K}$}
  \put(100,55){$=$}
  \put(130,55){$-i\frac{1}{4}\,g\,M^{\mathcal IJK}
               \quad=\quad-i\frac{1}{2}\,g\,M^{AB}\,\Gamma^\mathcal{I}_{AC}\Gamma^\mathcal{J}_{BD}\Gamma^{\mathcal{K}CD}$}

  %
  \thinlines
  \multiput(30,60)(17,0){3}{\line(1,0){13}}
  \put(70,45){$\mathcal{J}$}

\end{picture}}

\end{picture}
\caption{
Feynman rules for 3-scalar vertices
(planar part only and the algebraic factor for the color group ignored).
Here, $\mathcal{I}$, $\mathcal{J}$ and $\mathcal{K}$
are $SU(4)$ R-symmetry indices in the
representation $\textbf{6}$.
$(\mathcal{I}, \mathcal{J}, \mathcal{K})$
are in the counterclockwise order on the page, since for the planar diagram
we use the convention that the color group factor
is the trace of adjoint matrices in counterclockwise order.
}\label{fig:scalar-vertex}
}

\subsection{Other Feynman rules}\label{app:others}
The Feynman rules involving the gluons do not change with the (anti-)chiral mass
term. For our purpose, instead of arbitrary $\epsilon_\mu$, we use helicity
to describe the polarization.
To get a positive (negative) helicity polarization
vector, we set
\bear\label{eqn:gluon}
\epsilon_{\mu,+}\rightarrow\tilde{\epsilon}_{\a\dta}=\frac{\xi_\a\tlam_\dta}{\langle\xi,\lam\rangle},
\qquad
\epsilon_{\mu,-}\rightarrow\epsilon_{\a\dta}=\frac{\lam_\a\tilde{\xi}_\dta}{[\tlam,\tilde{\xi}]},\nn\\
\epsilon_{\mu,+}^*\rightarrow\epsilon_{\a\dta}=\frac{\lam_\a\tilde{\xi}_\dta}{[\tlam,\tilde{\xi}]},
\qquad
\epsilon_{\mu,-}^*\rightarrow\tilde{\epsilon}_{\a\dta}=\frac{\xi_\a\tlam_\dta}{\langle\xi,\lam\rangle},
\eear
where $\xi$($\tilde{\xi}$) is arbitrary but not a multiple of $\lam$($\tlam$) (See \cite{Witten:2003nn}).
Feynman rules for external gluons are shown in \figref{fig:gluon}.

\FIGURE{
\begin{picture}(470,120)

\put(3,70){\begin{picture}(200,60)
  \thinlines
  \multiput(0,15)(12,0){7}{\qbezier(0,0)(3,3)(6,0)}
   \multiput(6,15)(12,0){6}{\qbezier(0,0)(3,-3)(6,0)}
  \put(80,15){\circle*{5}}
  \put(0,20){$+$}
  \put(40,30){$p$}
  \put(35,23){\vector(1,0){15}}
  \put(15,0){\text{incoming}}
  \put(95,10){$=\,\epsilon_{\mu,+}\rightarrow\tilde{\epsilon}_{\a\dta}=
         \frac{\xi_\a\tlam_\dta}{\langle\xi,\lam\rangle}$}
\end{picture}}

\put(230,70){\begin{picture}(200,60)
  \thinlines
  \multiput(0,15)(12,0){7}{\qbezier(0,0)(3,3)(6,0)}
   \multiput(6,15)(12,0){6}{\qbezier(0,0)(3,-3)(6,0)}
  \put(80,15){\circle*{5}}
  \put(0,20){$-$}
  \put(40,30){$p$}
  \put(35,23){\vector(1,0){15}}
  \put(15,0){\text{incoming}}
  \put(95,10){$=\,\epsilon_{\mu,-}\rightarrow\epsilon_{\a\dta}=
         \frac{\lam_\a\tilde{\xi}_\dta}{[\tlam,\tilde{\xi}]}$}
\end{picture}}

\put(3,5){\begin{picture}(200,60)
  \thinlines
  \multiput(0,15)(12,0){7}{\qbezier(0,0)(3,3)(6,0)}
   \multiput(6,15)(12,0){6}{\qbezier(0,0)(3,-3)(6,0)}
  \put(80,15){\circle*{5}}
  \put(0,20){$+$}
  \put(40,30){$p$}
  \put(50,23){\vector(-1,0){15}}
  \put(15,0){\text{outgoing}}
  \put(95,10){$=\,\epsilon_{\mu,+}^*\rightarrow\epsilon_{\a\dta}=
         \frac{\lam_\a\tilde{\xi}_\dta}{[\tlam,\tilde{\xi}]}$}
\end{picture}}

\put(230,5){\begin{picture}(200,60)
  \thinlines
  \multiput(0,15)(12,0){7}{\qbezier(0,0)(3,3)(6,0)}
   \multiput(6,15)(12,0){6}{\qbezier(0,0)(3,-3)(6,0)}
  \put(80,15){\circle*{5}}
  \put(0,20){$-$}
  \put(40,30){$p$}
  \put(50,23){\vector(-1,0){15}}
  \put(15,0){\text{outgoing}}
  \put(95,10){$=\,\epsilon_{\mu,-}^*\rightarrow\tilde{\epsilon}_{\a\dta}=
         \frac{\xi_\a\tlam_\dta}{\langle\xi,\lam\rangle}$}
\end{picture}}

\end{picture}
\caption{
The Feynman rules for external gluons.
}\label{fig:gluon}
}

All other Feynman rules are exactly the same as those in
the massless theory.
In particular, we list the fermion-gluon vertices in
\figref{fig:vertex}, fermion-scalar vertices in
\figref{fig:scalarvertex} and scalar-gluon vertex in \figref{fig:scalar-gluon-vertex}.

\FIGURE{
\begin{picture}(420,125)

\put(20,0){\begin{picture}(200,60)

  \thicklines
  \put(10,20){\vector(1,2){10}}
  \put(20,40){\line(1,2){10}}
  \put(30,60){\vector(-1,2){10}}
  \put(20,80){\line(-1,2){10}}
  \put(20,15){$\a,A$}
  \put(20,100){$\dtb,B$}
  \put(75,55){$\mu$}
  \put(110,55){$=ig\, \bsig^{\mu\dtb\a}{\delta_A}^B$}

  \thinlines
  \multiput(30,60)(12,0){4}{\qbezier(0,0)(3,3)(6,0)}
   \multiput(36,60)(12,0){3}{\qbezier(0,0)(3,-3)(6,0)}

\end{picture}}

\put(245,0){\begin{picture}(200,60)

  \thicklines
  \put(10,20){\vector(1,2){10}}
  \put(20,40){\line(1,2){10}}
  \put(30,60){\vector(-1,2){10}}
  \put(20,80){\line(-1,2){10}}
  \put(20,15){$\dta,A$}
  \put(20,100){$\b,B$}
  \put(75,55){$\mu$}
  \put(110,55){$=ig\, \sigma^\mu_{\b\dta}{\delta^A}_B$}

  \thinlines
  \multiput(30,60)(12,0){4}{\qbezier(0,0)(3,3)(6,0)}
   \multiput(36,60)(12,0){3}{\qbezier(0,0)(3,-3)(6,0)}

\end{picture}}

\end{picture}
\caption{
Feynman rules for fermion-gluon vertices. Here, $A$ and $B$ are $SU(4)$ R-symmetry
indices in the $\textbf{4}$ or $\bar{\textbf{4}}$ representation.
The algebraic factor for the color group is ignored.
}\label{fig:vertex}
}

\FIGURE{
\begin{picture}(420,125)

\put(20,0){\begin{picture}(200,60)

  \thicklines
  \put(10,20){\vector(1,2){10}}
  \put(20,40){\line(1,2){10}}
  \put(30,60){\vector(-1,2){10}}
  \put(20,80){\line(-1,2){10}}
  \put(20,15){$\a,A$}
  \put(20,100){$\b,B$}
  \put(75,55){$\mathcal{I}$}
  \put(110,55){$=2ig\, \Gamma^\mathcal{I}_{BA}{\delta^\a}_\b$}

  \thinlines
  \put(30,60){\line(1,0){5}}
  \put(40,60){\line(1,0){10}}
  \put(55,60){\line(1,0){10}}

\end{picture}}

\put(245,0){\begin{picture}(200,60)

  \thicklines
  \put(10,20){\vector(1,2){10}}
  \put(20,40){\line(1,2){10}}
  \put(30,60){\vector(-1,2){10}}
  \put(20,80){\line(-1,2){10}}
  \put(20,15){$\dta,A$}
  \put(20,100){$\dtb,B$}
  \put(75,55){$\mathcal{I}$}
  \put(110,55){$=2ig\, \Gamma^{\mathcal{I}BA}{\delta_\dta}^\dtb$}

  \thinlines
  \put(30,60){\line(1,0){5}}
  \put(40,60){\line(1,0){10}}
  \put(55,60){\line(1,0){10}}

\end{picture}}

\end{picture}
\caption{
Feynman rules for fermion-scalar vertices. Here, $A$ and $B$ are $SU(4)$ R-symmetry
indices in the $\textbf{4}$ or $\bar{\textbf{4}}$
representation and $\mathcal{I}$
is the index in $\textbf{6}$ representation.
The algebraic factor for the color group is ignored.
}\label{fig:scalarvertex}
}

\FIGURE{
\begin{picture}(420,125)

\put(150,0){\begin{picture}(200,60)

  \thinlines
  \put(8,16){\line(1,2){6}}
  \put(16,32){\vector(1,2){6}}
  \put(24,48){\line(1,2){6}}
  \put(30,60){\line(-1,2){6}}
  \put(22,76){\vector(-1,2){6}}
  \put(14,92){\line(-1,2){6}}
  \put(20,15){$\mathcal{I}$}
  \put(20,100){$\mathcal{J}$}
  \put(5,38){$p_1$}
  \put(5,75){$p_2$}
  \put(90,55){$=$}
  \put(110,55){$i\frac{1}{2}\,g\, (p_1+p_2)^\mu{\delta}^{\mathcal IJ}$}

  \thinlines
  \multiput(30,60)(12,0){4}{\qbezier(0,0)(3,3)(6,0)}
   \multiput(36,60)(12,0){3}{\qbezier(0,0)(3,-3)(6,0)}
  \put(70,50){$\mu$}

\end{picture}}

\end{picture}
\caption{
Feynman rule for the scalar-gluon vertex.
Here, $\mathcal{I}$ and $\mathcal{J}$ are $SU(4)$ R-symmetry
indices in the $\textbf{6}$ representation; $p_1$ and $p_2$
represent the physical momenta if the corresponding dashed line
happens to be an external leg.
The algebraic factor for the color group is ignored.
}\label{fig:scalar-gluon-vertex}
}

\section{Detailed computation for Feynman diagrams}\label{app:amplitudes}
In this section, we present the calculation
of the tree-level planar Feynman diagrams in detail for the
scattering amplitudes presented in \secref{sec:scamp}. Some techniques used here can be found in \cite{Dixon:1996wi}.

\subsection{MHV amplitudes (extended MHV at $\mathcal{O}(M^0)$)}\label{app:MHV-M0}
In this subsection, we calculate MHV diagrams without mass contribution.

\subsubsection{$A_{\mathcal{O}(M^0)}(+1,+1,-1,-1)$}
This is a 4-gluon scattering amplitude.
The Feynman diagrams are shown in \figref{fig:ggggppmm}.
Accordingly, the amplitude of 4 gluons are the same as that without
the mass term, which is given by
(See, e.g., \cite{Cachazo:2005ga})
\be
A_{\mathcal{O}(M^0)}(+1,+1,-1,-1)
=\frac{ig^2}{2}\frac{{\langle3,4\rangle}^4}{\prod_{i=1}^4\langle
i,i+1 \rangle}. \label{eqn:AmpA(+1,+1,-1,-1) app} \ee

\FIGURE{
\begin{picture}(420,125)


\put(0,120){\text{(a)}}
\put(60,0){\begin{picture}(150,60)

  \thinlines
  \put(80,10){$p_2,+$}
  \multiput(30,60)(12,-12){4}{\qbezier(0,0)(1,-5)(6,-6)}
   \multiput(36,54)(12,-12){3}{\qbezier(0,0)(5,-1)(6,-6)}

  \multiput(30,60)(12,12){4}{\qbezier(0,0)(1,5)(6,6)}
   \multiput(36,66)(12,12){3}{\qbezier(0,0)(5,1)(6,6)}
  \put(80,110){$p_3,+$}

  \put(-40,10){$p_1,+$}
  \multiput(30,60)(-12,-12){4}{\qbezier(0,0)(-1,-5)(-6,-6)}
   \multiput(24,54)(-12,-12){3}{\qbezier(0,0)(-5,-1)(-6,-6)}

  \multiput(30,60)(-12,12){4}{\qbezier(0,0)(-1,5)(-6,6)}
   \multiput(24,66)(-12,12){3}{\qbezier(0,0)(-5,1)(-6,6)}
  \put(-40,110){$p_4,+$}

\end{picture}}

\put(175,120){\text{(b)}}
\put(205,0){\begin{picture}(150,60)

  \thinlines

  \multiput(30,30)(12,-12){3}{\qbezier(0,0)(1,-5)(6,-6)}
   \multiput(36,24)(12,-12){2}{\qbezier(0,0)(5,-1)(6,-6)}

  \multiput(30,90)(12,12){3}{\qbezier(0,0)(1,5)(6,6)}
   \multiput(36,96)(12,12){2}{\qbezier(0,0)(5,1)(6,6)}

  \multiput(30,30)(-12,-12){3}{\qbezier(0,0)(-1,-5)(-6,-6)}
   \multiput(24,24)(-12,-12){2}{\qbezier(0,0)(-5,-1)(-6,-6)}

  \multiput(30,90)(-12,12){3}{\qbezier(0,0)(-1,5)(-6,6)}
   \multiput(24,96)(-12,12){2}{\qbezier(0,0)(-5,1)(-6,6)}

  \multiput(30,30)(0,20){3}{\qbezier(0,0)(-5,5)(0,10)}
   \multiput(30,40)(0,20){3}{\qbezier(0,0)(5,5)(0,10)}
\end{picture}}

\put(300,120){\text{(c)}}
\put(325,0){\begin{picture}(150,60)

  \thinlines
  \multiput(30,60)(12,0){3}{\qbezier(0,0)(3,3)(6,0)}
   \multiput(36,60)(12,0){2}{\qbezier(0,0)(3,-3)(6,0)}

  \multiput(60,60)(12,-24){3}{\qbezier(0,0)(1,-10)(6,-12)}
   \multiput(66,48)(12,-24){2}{\qbezier(0,0)(5,-2)(6,-12)}

  \multiput(60,60)(12,24){3}{\qbezier(0,0)(1,10)(6,12)}
   \multiput(66,72)(12,24){2}{\qbezier(0,0)(5,2)(6,12)}

  \multiput(30,60)(-12,-24){3}{\qbezier(0,0)(-1,-10)(-6,-12)}
   \multiput(24,48)(-12,-24){2}{\qbezier(0,0)(-5,-2)(-6,-12)}

  \multiput(30,60)(-12,24){3}{\qbezier(0,0)(-1,10)(-6,12)}
   \multiput(24,72)(-12,24){2}{\qbezier(0,0)(-5,2)(-6,12)}

\end{picture}}

\end{picture}
\caption{
Planar Feynman diagrams that contribute to the MHV amplitude
$A_{\mathcal{O}(M^0)}(+1,+1,-1,-1)$.
In order to directly apply the Feynman rules
as in \appref{app:Feynman rules},
in the figures we are not using the convention that
all external legs are incoming (instead,
all depicted momenta and helicities are physical).
}\label{fig:ggggppmm}
}

\subsubsection{$A_{\mathcal{O}(M^0)}(+1/2,+1,-1,-1/2)$}

\FIGURE{
\begin{picture}(420,125)


\put(40,120){\text{(a)}}
\put(110,0){\begin{picture}(150,60)

  \thicklines
  \put(-40,0){$+,A,p_1$}
  \put(0,0){\vector(1,1){15}} 
   \put(15,15){\line(1,1){15}}
  \put(30,30){\vector(0,1){30}} 
   \put(30,60){\line(0,1){30}}
  \put(30,90){\vector(-1,1){15}} 
   \put(15,105){\line(-1,1){15}}
  \put(-40,125){$+,B,p_4$}

  \put(10,20){$\dta$}
  \put(20,40){$\a$}
  \put(20,75){$\dtb$}
  \put(10,90){$\b$}

  \thinlines
  \put(65,0){$+,p_2$}
  \multiput(30,30)(12,-12){3}{\qbezier(0,0)(1,-5)(6,-6)}
   \multiput(36,24)(12,-12){2}{\qbezier(0,0)(5,-1)(6,-6)}

  \multiput(30,90)(12,12){3}{\qbezier(0,0)(1,5)(6,6)}
   \multiput(36,96)(12,12){2}{\qbezier(0,0)(5,1)(6,6)}
  \put(65,125){$+,p_3$}

\end{picture}}

\put(250,120){\text{(b)}}
\put(280,0){\begin{picture}(150,60)

  \thicklines
  \put(-5,7){$+$}
  \put(0,0){\vector(1,2){15}} 
   \put(15,30){\line(1,2){15}}
  \put(30,60){\vector(-1,2){15}} 
   \put(15,90){\line(-1,2){15}}
  \put(-5,107){$+$}
  \put(10,40){$\dta$}
  \put(10,70){$\a$}

  \thinlines
  \multiput(30,60)(12,0){3}{\qbezier(0,0)(3,3)(6,0)}
   \multiput(36,60)(12,0){2}{\qbezier(0,0)(3,-3)(6,0)}

  \put(90,7){$+$}
  \multiput(60,60)(12,-24){3}{\qbezier(0,0)(1,-10)(6,-12)}
   \multiput(66,48)(12,-24){2}{\qbezier(0,0)(5,-2)(6,-12)}

  \multiput(60,60)(12,24){3}{\qbezier(0,0)(1,10)(6,12)}
   \multiput(66,72)(12,24){2}{\qbezier(0,0)(5,2)(6,12)}
  \put(90,107){$+$}

\end{picture}}

\end{picture}
\caption{
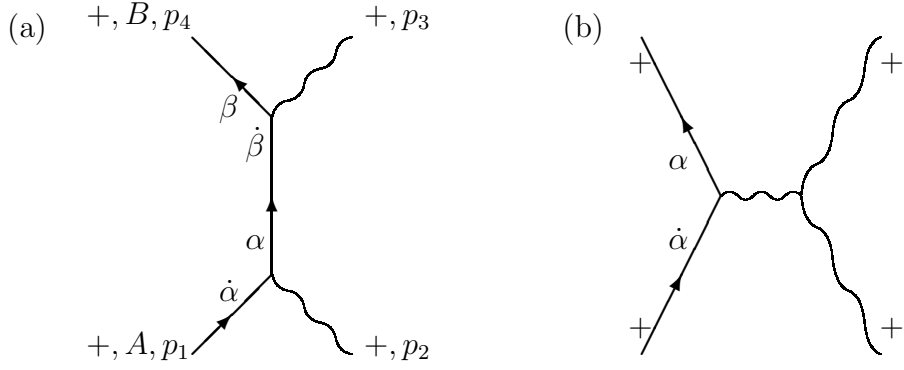
Planar Feynman diagrams for the MHV amplitude $A_{\mathcal{O}(M^0)}(+1/2,+1,-1,-1/2)$.
Wavy lines are gluons and solid lines are fermions.
Time is in the vertical upward direction.
}\label{fig:fggfppmm}
}

This is a 2-gluon and 2-fermion scattering amplitude.
The Feynman rules give two contributions listed in \figref{fig:fggfppmm}:
\be
A_a=
\epsilon^*_{3\,\nu+}\varrho_4^B\lam_4^\b
\left(ig\sigma^{\nu}_{\b\dtb}\right)
\left[\frac{i(p_1+p_2)^{\dtb\a}{\delta^A}_B}{(p_1+p_2)^2}\right]
\left(ig\sigma^\mu_{\a\dta}\right)
\wvrho_{1A}\tlam^\dta_1\,\epsilon_{2\,\mu+}.
\ee
Here, \textit{$\epsilon_{2\,\nu+}$} and \textit{$\epsilon^*_{3\,\nu+}$}
are gluon polarization vectors for the particles with momenta
$p_2$ and $p_3$,
respectively. \textit{$\sigma^{\nu}_{\b\dtb}$} are Pauli matrices.
By the rules in \eqref{eqn:gluon} and in \figref{fig:gluon},
we have
\be
\epsilon_{\mu+}(p_2)\sigma^{\mu}_{\a\dta}
=\frac{\xi_{2\a}\tlam_{2 \dta}}{\langle\xi_2,2\rangle},
\qquad
\epsilon^*_{\nu+}(p_3)\sigma^{\nu}_{\b\dtb}
=\frac{\lam_{3\b}\tilde{\xi}_{3\dtb}}{[3,\tilde{\xi}_3]},
\ee
where $\xi_2$ and $\xi_3$ are arbitrary spinors.
\bear
A_a&=&
\frac{ig^2\varrho_4^A\wvrho_{1A}}
{\langle1,2\rangle[1,2][3,\tilde{\xi}_3]\langle\xi_2,\lam_2\rangle}\
\lam^{\b}_{4}\lam_{3\b}\tilde{\xi}_{3\dtb}\left(\tlam_1^\dtb\lam_1^\a+\tlam_2^\dtb\lam_2^\a\right)
\xi_{2\a}\tlam_{2\dta}\tlam^{\dta}_1\nn \\
&=&
\frac{ig^2\varrho_4^A\wvrho_{1A}}
{\langle1,2\rangle[3,\tilde{\xi}_3]\langle\xi_2,\lam_2\rangle}\
\langle3,4\rangle\left([\tilde{\xi}_3,1]\langle1,\xi_2\rangle+[\tilde{\xi}_3,2]\langle2,\xi_2\rangle\right).
\eear
Gauge-fixing the external gluon polarizations by taking $\xi_2=\lam_3$
and $\tilde{\xi}_3=\tlam_2$, we get\footnote{Henceforth, an arrow $\rightarrow$ represents
a particular gauge choice.}
\be
A_a\rightarrow
{ig^2\varrho_4^A\wvrho_{1A}}\
\frac{\langle3,4\rangle\langle1,3\rangle[2,1]}{\langle1,2\rangle\langle2,3\rangle[2,3]}
=
-{ig^2\varrho_4^A\wvrho_{1A}}\
\frac{\langle3,4\rangle^3\langle1,3\rangle}{\prod_{i=1}^4\langle i,i+1\rangle},
\ee
where we have used the identity
\be
\frac{[2,1]}{[2,3]}=\frac{\langle4,3\rangle}{\langle4,1\rangle},
\ee
which follows from momentum conservation, $\sum_{i=1}^2 \lam_i^\a\tlam_i^\dta=\sum_{i=3}^4 \lam_i^\a\tlam_i^\dta$.

The diagram of \figref{fig:fggfppmm}b gives
\bear
A_b&=&
\epsilon^*_{3\,\nu+}(ig)
\left[\eta^{\mu\nu}(p_2+p_3)^\varrho
+\eta^{\nu\varrho}(p_2-2p_3)^\mu+\eta^{\rho\mu}(p_3-2p_2)^\nu\right]\epsilon_{2\,\mu+}\nn \\
& &\mbox{}\times
\left[i\frac{\eta_{\sigma\rho}}{(p_2-p_3)^2}\right]
\varrho_4^B\lam_4^\a
\left(ig\sigma^\sigma_{\a\dta}{\delta^A}_B\right)
\wvrho_{1A}\tlam_1^\dta,
\eear
where $p_2^\mu\epsilon_{2\mu+}=\epsilon^{*}_{3\nu+}p_3^\nu=0$ and
$\eta^{\mu\nu}\epsilon^*_{3\nu+}\epsilon_{2\mu+}$
can be expressed again in the form of \figref{fig:gluon} with the help of the identity \eqref{eqn:sigma6}.

It follows that
\bear \label{eqn:result-1b}
A_b&=&
\frac{-ig^2\varrho_4^A\wvrho_{1A}}
{2\langle2,3\rangle[2,3][3,\tilde{\xi}_3]\langle\xi_2,2\rangle}\
\lam_4^\a\tlam_1^\dta\nn\\
&& \mbox{}\times
\left\{
\tilde{\xi}^\dtb_3\lam^\b_3\tlam_{2\dtb}\,\xi_{2\b}(p_2+p_3)_{\a\dta}
-2\tilde{\xi}_{3\dta}\lam_{3\a}\tlam_{2\dtb}\,\xi_{2\b}p_3^{\dtb\b}
-2\tilde{\xi}^\dtb_3\lam^\b_3\,\xi_{2\a}\tlam_{2\dta} p_{2\dtb\b}
\right\}\nn\\
&=&
\frac{-ig^2\varrho_4^A\wvrho_{1A}}
{2\langle2,3\rangle[2,3][3,\tilde{\xi}_3]\langle\xi_2,2\rangle}
\left\{
\langle4,2\rangle[2,1]\langle3,\xi_2\rangle[2,\tilde{\xi}_3]+
\langle4,3\rangle[3,1]\langle3,\xi_2\rangle[2,\tilde{\xi}_3]\right.\nn\\
&&\qquad\qquad\left.-2\langle4,3\rangle[\tilde{\xi}_3,1][2,3]\langle3,\xi_2\rangle
-2\langle4,\xi_2\rangle\langle3,2\rangle[2,1][2,\tilde{\xi}_3]
\right\}\nn\\
&\rightarrow&
0
\eear
in the gauge $\xi_2=\lam_3$ and $\tilde{\xi}_3=\tlam_2$.

Therefore,
\be
A_{\mathcal{O}(M^0)}(+1/2,+1,-1,-1/2)=A_a+A_b=
{ig^2\varrho_4^A\wvrho_{1A}}\
\frac{\langle3,4\rangle^3\langle1,3\rangle}{\prod_{i=1}^4\langle i,i+1\rangle}.
\label{eqn:AmpA(+1/2,+1,-1,-1/2) app}
\ee

\subsubsection{$A_{\mathcal{O}(M^0)}(+1/2,+1/2,-1/2,-1/2)$}

\FIGURE{
\begin{picture}(420,125)


\put(40,120){\text{(a)}}
\put(80,0){\begin{picture}(150,60)

  \thicklines
  \put(-25,7){$+,A$}
  \put(0,0){\vector(1,2){15}} 
   \put(15,30){\line(1,2){15}}
  \put(30,60){\vector(-1,2){15}} 
   \put(15,90){\line(-1,2){15}}
  \put(-25,107){$+,D$}
  \put(10,40){$\dta$}
  \put(10,70){$\a$}

  \thicklines
  \put(95,7){$+,B$}
  \put(90,0){\vector(-1,2){15}} 
   \put(75,30){\line(-1,2){15}}
  \put(60,60){\vector(1,2){15}} 
   \put(75,90){\line(1,2){15}}
  \put(95,107){$+,C$}
  \put(80,40){$\dtb$}
  \put(80,70){$\b$}

  \thinlines
  \multiput(30,60)(12,0){3}{\qbezier(0,0)(3,3)(6,0)}
   \multiput(36,60)(12,0){2}{\qbezier(0,0)(3,-3)(6,0)}

\end{picture}}

\put(250,120){\text{(b)}}
\put(300,0){\begin{picture}(150,60)

  \thicklines
  \put(-20,0){$-$}
  \put(30,30){\vector(-1,-1){18}} 
   \put(0,0){\line(1,1){12}}
   \put(15,5){\vector(1,1){10}}
  \put(60,0){\vector(-1,1){15}} 
   \put(45,15){\line(-1,1){15}}
  \put(0,120){\vector(1,-1){18}} 
   \put(18,102){\line(1,-1){12}}
   \put(25,105){\vector(-1,1){10}}
  \put(30,90){\vector(1,1){15}} 
   \put(45,105){\line(1,1){15}}
  \put(-20,115){$-$}
  \put(65,0){$+$}
  \put(65,115){$+$}

  \put(10,20){$\dta$}
  \put(45,20){$\dtb$}
  \put(45,90){$\gamma$}
  \put(10,90){$\delta$}

  \thinlines
  \put(30,30){\line(0,1){5}}
  \put(30,40){\line(0,1){10}}
  \put(30,55){\line(0,1){10}}
  \put(30,70){\line(0,1){10}}
  \put(30,85){\line(0,1){5}}

\end{picture}}

\end{picture}
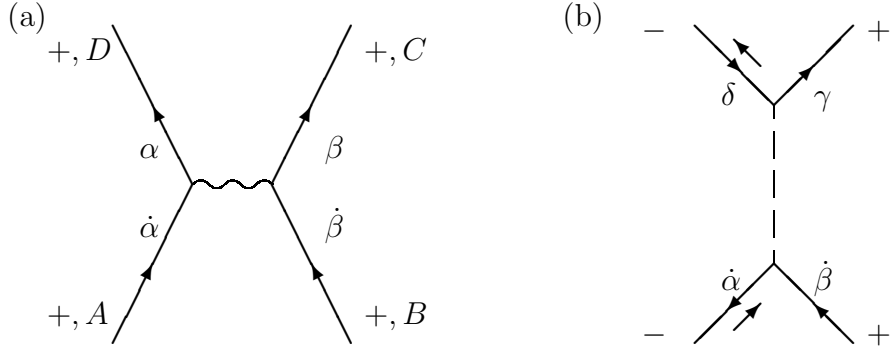
\caption{ Planar Feynman diagrams for
$A_{\mathcal{O}(M^0)}(+1/2,+1/2,-1/2,-1/2)$. In (b), the helicities for the 1st
and 4th particles are flipped since we treat them as anti-fermions.
}\label{fig:ffffppmm}
}


This is a 4-fermion scattering amplitude.
The Feynman diagram in \figref{fig:ffffppmm}(a) gives
\bear
A_a
&=&
(\varrho_4^D\lam_4^\a)(\wvrho_{1A}\tlam_1^\dta)
\left(ig\sigma^\mu_{\a\dta}{\delta^A}_D\right)
\left[\frac{-i\eta_{\mu\nu}}{(p_1-p_4)^2}\right]
\left(ig\sigma^\nu_{\b\dtb}{\delta^B}_C\right)
(\varrho_3^C\lam_3^\b)(\wvrho_{2B}\tlam^\dtb_2)\nn\\
&=& 2ig^2\varrho_4^A\wvrho_{1A}\varrho_3^B\wvrho_{2B}
\frac{\langle3,4\rangle[1,2]}{\langle1,4\rangle[1,4]}
=2ig^2\varrho_4^A\wvrho_{1A}\varrho_3^B\wvrho_{2B}
\frac{{\langle3,4\rangle}^3\langle1,2\rangle}{\prod_{i=1}^4 \langle
i,i+1\rangle}
\eear
by the identity \eqref{eqn:sigma1}.

Since the anti-fermions with adjoint color $T_i$ and helicity $\pm$
are identical to the fermions with $\bar{T}_i$ and helicity $\mp$,
we should consider the s-channel as shown on \figref{fig:ffffppmm}(b),
which gives
\bear
A_b
&=&
(\varrho_4^D\lam_{4\delta})(\varrho_3^C\lam_3^\gamma)
\left(2ig\Gamma^\mathcal{J}_{CD}{\delta^\delta}_\gamma\right)
\left[\frac{-i\delta_{\mathcal{I}\mathcal{J}}}{(p_1+p_2)^2}\right]
\left(2ig\Gamma^{\mathcal{I}AB}{\delta_\dtb}^\dta\right)
(-\wvrho_{1A}\tlam_{1\dta})(\wvrho_{2B}\tlam^\dtb_2)\nn\\
&=&
-2ig^2(\varrho_4^A\wvrho_{1A}\varrho_3^B\wvrho_{2B}
-\varrho_3^A\wvrho_{1A}\varrho_4^B\wvrho_{2B})
\frac{\langle3,4\rangle}{\langle1,2\rangle}\nn\\
&=&-2ig^2(\varrho_4^A\wvrho_{1A}\varrho_3^B\wvrho_{2B}
-\varrho_3^A\wvrho_{1A}\varrho_4^B\wvrho_{2B})
\frac{{\langle3,4\rangle}^2\langle2,3\rangle\langle4,1\rangle}
{\prod_{i=1}^4 \langle i,i+1\rangle}
\eear
by the identity \eqref{eqn:phi2}.

Thus,
\bear
&&A_{\mathcal{O}(M^0)}(+1/2,+1/2,-1/2,-1/2)=A_a+A_b\nn\\
&=&
-\frac{2ig^2{\langle3,4\rangle}^2}
{\prod_{i=1}^4 \langle i,i+1\rangle}
\left\{
\varrho_4^A\wvrho_{1A}\varrho_3^B\wvrho_{2B}
\left(\langle2,3\rangle\langle4,1\rangle-\langle3,4\rangle\langle1,2\rangle\right)
\right.-
\left.
\varrho_3^A\wvrho_{1A}\varrho_4^B\wvrho_{2B}\langle2,3\rangle\langle4,1\rangle
\right\} \nn \\
&\longrightarrow& \frac{2ig^2{\langle3,4\rangle}^2}
{\prod_{i=1}^4 \langle i,i+1\rangle}
\left\{
\varrho_4^A\wvrho_{1A}\wvrho_{2B}\varrho_3^B
\langle1,3\rangle\langle2,4\rangle +
\varrho_3^A\wvrho_{1A}\wvrho_{2B}\varrho_4^B\langle2,3\rangle\langle4,1\rangle
\right\},
\label{eqn:AmpA(+1/2,+1/2,-1/2,-1/2) app}
\eear
where in the last line we scale $(\lam_i^1,\lam_i^2)=(1,Z_i)$ and thus
$\langle i,j \rangle=Z_j-Z_i$.

\subsubsection{$A_{\mathcal{O}(M^0)}(+1/2,0,0,-1/2)$}

\FIGURE{
\begin{picture}(420,125)


\put(40,120){\text{(a)}}
\put(80,0){\begin{picture}(150,60)

  \thicklines
  \put(-25,7){$+,A$}
  \put(0,0){\vector(1,2){15}} 
   \put(15,30){\line(1,2){15}}
  \put(30,60){\vector(-1,2){15}} 
   \put(15,90){\line(-1,2){15}}
  \put(-25,107){$+,B$}
  \put(10,40){$\dta$}
  \put(10,70){$\a$}

  \thinlines
  \put(95,7){$\mathcal{I}$}
  \multiput(60,60)(8,-16){4}{\line(1,-2){6}}
  \multiput(60,60)(8,16){4}{\line(1,2){6}}
  \put(95,107){$\mathcal{J}$}

  \thinlines
  \multiput(30,60)(12,0){3}{\qbezier(0,0)(3,3)(6,0)}
   \multiput(36,60)(12,0){2}{\qbezier(0,0)(3,-3)(6,0)}

\end{picture}}

\put(250,120){\text{(b)}}
\put(300,0){\begin{picture}(150,60)

  \thicklines
  \put(-20,0){$+$}
  \put(-6,-6){\vector(1,1){18}} 
   \put(12,12){\line(1,1){18}}
  \put(30,30){\vector(0,1){30}} 
   \put(30,60){\line(0,1){30}}
  \put(30,90){\vector(-1,1){18}} 
   \put(12,108){\line(-1,1){18}}
  \put(-20,125){$+$}

  \put(10,20){$\dta$}
  \put(20,40){$\dtb$}
  \put(20,75){$\b$}
  \put(10,90){$\g$}

  \thinlines
  \multiput(30,30)(13,-13){3}{\line(1,-1){10}}
  \multiput(30,90)(13,13){3}{\line(1,1){10}}

\end{picture}}

\end{picture}
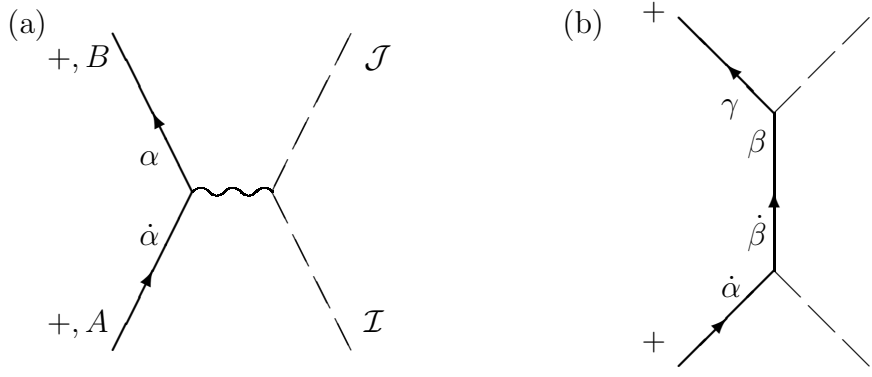
\caption{ Planar Feynman diagram for
$A_{\mathcal{O}(M^0)}(+1/2,0,0,-1/2)$. The dashed lines are
scalars.}\label{fig:fssfp00m}
}


The Feynman diagram in \figref{fig:fssfp00m}(a) gives:
\bear
A_a
&=&
(\varrho_4^B\lam_4^\a)(\wvrho_{1A}\tlam_1^\dta)
\left(ig\,\sigma^\mu_{\a\dta}{\delta^A}_B\right)
\left[\frac{-i\eta_{\mu\nu}}{(p_1-p_4)^2}\right]
\left(\frac{ig}{2}\,(p_2+p_3)^\nu\delta^{\mathcal{I}\mathcal{J}}\right)
\varphi_{2\mathcal{I}}\varphi_{3\mathcal{J}}\nn\\
&=& -ig^2\varrho_4^A\wvrho_{1A}\varphi_2^\mathcal{I}\varphi_{3\mathcal{I}}
\frac{\langle4,2\rangle[1,2]+\langle4,3\rangle[1,3]}{2\langle1,4\rangle[1,4]}\nn\\
&=& ig^2\varrho_4^A\wvrho_{1A}\varphi_2^{BC}\varphi_{3BC}
\frac{\langle1,2\rangle\langle2,4\rangle{\langle3,4\rangle}^2}{\prod_{i=1}^4 \langle
i,i+1\rangle},
\eear
where $\varphi_{3\mathcal{J}}$ and $\varphi_{2\mathcal{I}}$ are used for the external scalar particles
and \eqref{eqn:phi5} is used.

\noindent
\figref{fig:fssfp00m}(b) gives:
\bear
A_b&=&
\varphi_{3\mathcal{J}}\varrho_4^B\lam_4^\g
\left(2ig\Gamma^{\mathcal{J}}_{BD}{\d^\b}_\g\right)
\left[\frac{i(p_1+p_2)_{\b\dtb}{\delta_C}^D}{(p_1+p_2)^2}\right]
\left(2ig\Gamma^{\mathcal{I}CA}{\d_\dta}^\dtb\right)
\wvrho_{1A}\tlam^\dta_1\,\varphi_{2\mathcal{I}}\nn\\
&=&2ig^2\varrho_4^B\wvrho_{1A}\varphi_2^{CA}\varphi_{3BC}
\frac{\langle4,1\rangle[1,1]+\langle4,2\rangle[2,1]}{\langle1,2\rangle[1,2]}\nn\\
&=&2ig^2\varrho_4^B\wvrho_{1A}\varphi_2^{CA}\varphi_{3BC}
\frac{\langle2,3\rangle\langle3,4\rangle\langle4,1\rangle\langle2,4\rangle}{\prod_{i=1}^4 \langle
i,i+1\rangle}.
\eear
Altogether,
\bear
&&A_{\mathcal{O}(M^0)}(+1/2,0,0,-1/2)=A_a+A_b
\nn\\
&=&\frac{2ig^2\langle3,4\rangle\langle2,4\rangle}{\prod_{i=1}^4 \langle i,i+1\rangle}
\Big\{\frac{1}{2}
\varrho_4^A\wvrho_{1A}\varphi_2^{BC}\varphi_{3BC}\langle1,2\rangle\langle3,4\rangle
+\varrho_4^B\wvrho_{1A}\varphi_2^{CA}\varphi_{3BC}\langle2,3\rangle\langle4,1\rangle
\Big\}.
\nn\\ &&
\label{eqn:AmpA(+1/2,0,0,-1/2) app}
\eear

\subsection{Extended MHV amplitudes at $\mathcal{O}(M)$}\label{app:MHV-M1}

In this subsection we will calculate the extended MHV diagrams
with the contribution
of the mass \textit{$M^{AB}$} up to the first order.

\subsubsection{$A_{\mathcal{O}(M)}(+1/2,+1,-1,+1/2)$}

\FIGURE{
\begin{picture}(420,270)


\put(0,265){\text{(a)}}
\put(60,145){\begin{picture}(150,60)

  \thicklines
  \put(-40,0){$1,+,A$}
  \put(0,0){\vector(1,1){15}} 
   \put(15,15){\line(1,1){15}}
  \put(30,30){\vector(0,1){30}} 
   \put(30,60){\line(0,1){30}}
  \put(30,90){\vector(-1,1){15}} 
   \put(15,105){\line(-1,1){15}}
  \put(-40,125){$4,-,B$}

  \put(10,20){$\a$}
  \put(20,40){$\dta$}
  \put(20,75){$\b$}
  \put(10,90){$\dtb$}

  \thinlines
  \put(65,0){$2,+$}
  \multiput(30,30)(12,-12){3}{\qbezier(0,0)(1,-5)(6,-6)}
   \multiput(36,24)(12,-12){2}{\qbezier(0,0)(5,-1)(6,-6)}

  \multiput(30,90)(12,12){3}{\qbezier(0,0)(1,5)(6,6)}
   \multiput(36,96)(12,12){2}{\qbezier(0,0)(5,1)(6,6)}
  \put(65,125){$3,+$}

\end{picture}}

\put(170,265){\text{(b)}}
\put(210,145){\begin{picture}(150,60)

  \thicklines
  \put(-10,0){$+$}
  \put(0,0){\vector(1,1){15}} 
   \put(15,15){\line(1,1){15}}
  \put(30,30){\vector(0,1){30}} 
   \put(30,60){\line(0,1){30}}
  \put(30,90){\vector(-1,1){15}} 
   \put(15,105){\line(-1,1){15}}
  \put(-10,115){$-$}

  \put(10,20){$\dta$}
  \put(20,40){$\a$}
  \put(20,75){$\dtb$}
  \put(10,90){$\b$}

  \thinlines
  \put(60,2){$+$}
  \multiput(30,30)(12,-12){3}{\qbezier(0,0)(1,-5)(6,-6)}
   \multiput(36,24)(12,-12){2}{\qbezier(0,0)(5,-1)(6,-6)}

  \multiput(30,90)(12,12){3}{\qbezier(0,0)(1,5)(6,6)}
   \multiput(36,96)(12,12){2}{\qbezier(0,0)(5,1)(6,6)}
  \put(60,115){$+$}

\end{picture}}

\put(310,265){\text{(c)}}
\put(340,145){\begin{picture}(150,60)

  \thicklines
  \put(-10,0){$+$}
  \put(0,0){\vector(1,1){15}} 
   \put(15,15){\line(1,1){15}}
  \put(30,30){\vector(0,1){30}} 
   \put(30,60){\line(0,1){30}}
  \put(30,90){\vector(-1,1){15}} 
   \put(15,105){\line(-1,1){15}}
  \put(-10,115){$-$}

  \put(10,20){$\dta$}
  \put(20,40){$\a$}
  \put(20,75){$\b$}
  \put(10,90){$\dtb$}

  \thinlines
  \put(60,2){$+$}
  \multiput(30,30)(12,-12){3}{\qbezier(0,0)(1,-5)(6,-6)}
   \multiput(36,24)(12,-12){2}{\qbezier(0,0)(5,-1)(6,-6)}

  \multiput(30,90)(12,12){3}{\qbezier(0,0)(1,5)(6,6)}
   \multiput(36,96)(12,12){2}{\qbezier(0,0)(5,1)(6,6)}
  \put(60,115){$+$}

\end{picture}}

\put(50,120){\text{(d)}}
\put(80,0){\begin{picture}(150,60)

  \thicklines
  \put(-5,7){$+$}
  \put(0,0){\vector(1,2){15}} 
   \put(15,30){\line(1,2){15}}
  \put(30,60){\vector(-1,2){15}} 
   \put(15,90){\line(-1,2){15}}
  \put(-5,107){$-$}
  \put(10,40){$\a$}
  \put(10,70){$\dta$}

  \thinlines
  \multiput(30,60)(12,0){3}{\qbezier(0,0)(3,3)(6,0)}
   \multiput(36,60)(12,0){2}{\qbezier(0,0)(3,-3)(6,0)}

  \put(90,7){$+$}
  \multiput(60,60)(12,-24){3}{\qbezier(0,0)(1,-10)(6,-12)}
   \multiput(66,48)(12,-24){2}{\qbezier(0,0)(5,-2)(6,-12)}

  \multiput(60,60)(12,24){3}{\qbezier(0,0)(1,10)(6,12)}
   \multiput(66,72)(12,24){2}{\qbezier(0,0)(5,2)(6,12)}
  \put(90,107){$+$}

\end{picture}}

\put(230,120){\text{(e)}}
\put(260,0){\begin{picture}(150,60)

  \thicklines
  \put(-5,7){$+$}
  \put(0,0){\vector(1,2){15}} 
   \put(15,30){\line(1,2){15}}
  \put(30,60){\vector(-1,2){15}} 
   \put(15,90){\line(-1,2){15}}
  \put(-5,107){$-$}
  \put(10,40){$\dta$}
  \put(10,70){$\a$}

  \thinlines
  \multiput(30,60)(12,0){3}{\qbezier(0,0)(3,3)(6,0)}
   \multiput(36,60)(12,0){2}{\qbezier(0,0)(3,-3)(6,0)}

  \put(90,7){$+$}
  \multiput(60,60)(12,-24){3}{\qbezier(0,0)(1,-10)(6,-12)}
   \multiput(66,48)(12,-24){2}{\qbezier(0,0)(5,-2)(6,-12)}

  \multiput(60,60)(12,24){3}{\qbezier(0,0)(1,10)(6,12)}
   \multiput(66,72)(12,24){2}{\qbezier(0,0)(5,2)(6,12)}
  \put(90,107){$+$}

\end{picture}}

\end{picture}
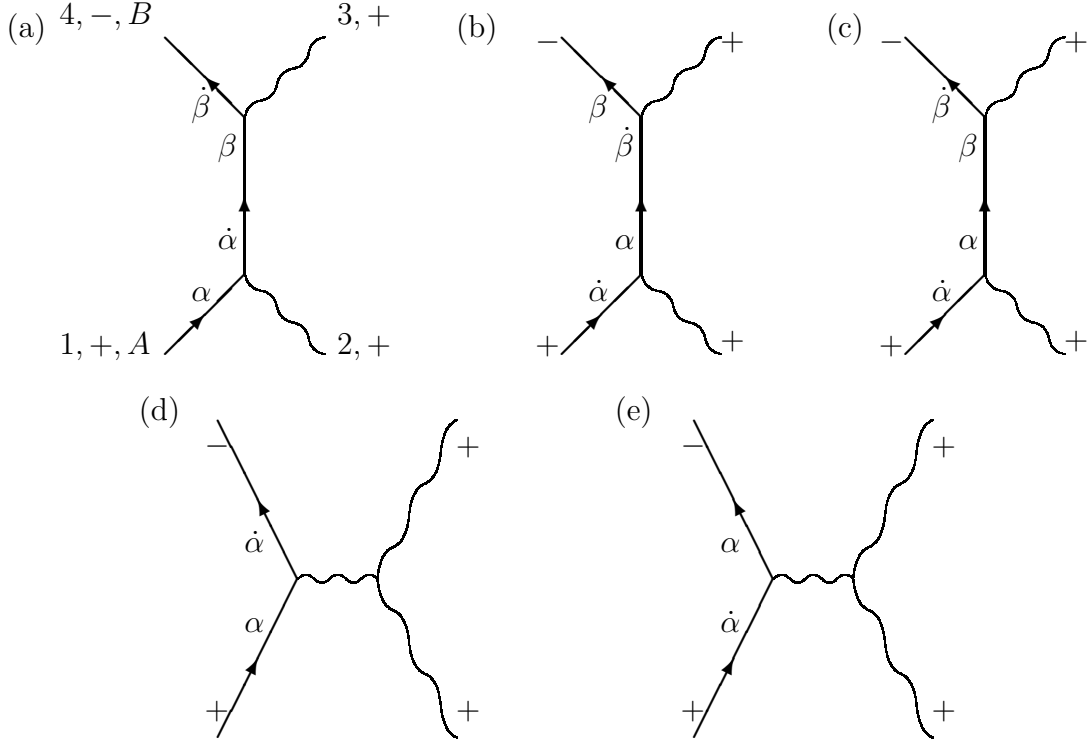
\caption{ Planar Feynman diagrams with two external fermions and two
external gluons corresponding to the extended MHV amplitude
$A_{\mathcal{O}(M)}(+1/2,+1,-1,+1/2)$. }\label{fig:fggfppmp}
}


The Feynman diagrams are listed in \figref{fig:fggfppmp}:
\bear
A_a&=&
\epsilon^*_{3\nu+}\wvrho_{4B}\tlam_{4\dtb}
\left(ig\bsig^{\nu\dtb\b}\right)
\left[\frac{i(p_1+p_2)_{\b\dta}\,{\delta_A}^B}{(p_1+p_2)^2}\right]
\left(ig\bsig^{\mu\dta\a}\right)
M^{AC}\wvrho_{1C}\eta_{1\a}\epsilon_{2\mu+}\nn\\
&=&
\frac{ig^2M^{AB}\wvrho_{1A}\wvrho_{4B}}
{\langle1,2\rangle[1,2][3,\tilde{\xi}_3]\langle\xi_2,2\rangle}\
\tlam_{4\dtb}\,\tilde{\xi}^\dtb_3\lam^\b_3
\left[\lam_{1\b}\tlam_{1\dta}+\lam_{2\b}\tlam_{2\dta}\right]\,
\tlam^\dta_2\xi^\a_2\eta_{1\a}\nn\\
&=&
{ig^2M^{AB}\wvrho_{1A}\wvrho_{4B}}
\frac{[4,\tilde{\xi}_3]\langle3,1\rangle\langle\xi_2,\eta_1\rangle}
{\langle1,2\rangle[3,\tilde{\xi}_3]\langle\xi_2,2\rangle}
\rightarrow
{ig^2M^{AB}\wvrho_{1A}\wvrho_{4B}}
\frac{[4,2]\langle3,1\rangle\langle3,\eta_1\rangle}
{\langle1,2\rangle[3,2]\langle3,2\rangle}
\nn \\
&=&
{ig^2M^{AB}\wvrho_{1A}\wvrho_{4B}}
\frac{\langle3,1\rangle\langle3,4\rangle
\left\{\langle3,1\rangle\langle3,\eta_1\rangle\right\}}
{\prod_{i=1}^4\langle i,i+1\rangle},
\eear
in the gauge
$\xi_2=\lam_2$ and $\tilde{\xi}_3=\tlam_3$.

Similarly, we have
\bear
A_b&=&
\epsilon^*_{3\nu+}(-M^{BC}\wvrho_{4C}\eta_4^\b)
\left(ig\sigma^\nu_{\b\dtb}\right)
\left[\frac{i(p_1+p_2)^{\dtb\a}\,{\delta^A}_B}{(p_1+p_2)^2}\right]
\left(ig\sigma^\mu_{\a\dta}\right)
\tilde{\rho_{1A}}\tlam_1^\dta\epsilon_{2\mu+}\nn\\
&=&
-\frac{ig^2M^{AB}\wvrho_{1A}\wvrho_{4B}}
{\langle1,2\rangle[1,2][3,\tilde{\xi}_3]\langle\xi_2,2\rangle}\
\eta_4^\b\lam_{3\b}\,\tilde{\xi}_{3\dtb}
\left[\tlam_1^\dtb\lam_1^\a+\tlam_2^\dtb\lam_2^\a\right]\,
\xi_{2\alpha}\tlam_{2\dta}\tlam_1^\dta\nn\\
&=&
-{ig^2M^{AB}\wvrho_{1A}\wvrho_{4B}}
\frac{\langle3,\eta_4\rangle[\tilde{\xi}_3,1]\langle1,\xi_2\rangle}
{\langle1,2\rangle[3,\tilde{\xi}_3]\langle\xi_2,2\rangle}\nn\\
&\rightarrow&
-{ig^2M^{AB}\wvrho_{1A}\wvrho_{4B}}
\frac{\langle3,\eta_4\rangle[1,2]\langle1,3\rangle}
{\langle1,2\rangle[3,2]\langle2,3\rangle} \nn \\
&=&
-{ig^2M^{AB}\wvrho_{1A}\wvrho_{4B}}
\frac{\langle3,1\rangle\langle3,4\rangle
\left\{\langle3,4\rangle\langle3,\eta_4\rangle\right\}}
{\prod_{i=1}^4\langle i,i+1\rangle}
\eear
in the same gauge.

Meanwhile,
\bear
A_c&=&
\epsilon^*_{3\,\nu+}\wvrho_{4B}\tlam_{4\dtb}
\left(ig\bsig^{\nu\dtb\b}\right)
\left[i\frac{M^{AB}{\delta_\b}^\a}{(p_1+p_2)^2}\right]
\left(ig\sigma_{\mu\a\dta}\right)
\wvrho_{1A}\tlam^\dta_1\,\epsilon_{2\,\mu+}\nn\\
&=&
\frac{ig^2M^{AB}\wvrho_{1A}\wvrho_{4B}}
{\langle1,2\rangle]1,2][3,\tilde{\xi}_3]\langle\xi_2,2\rangle}\
\tlam_{4\dtb}\,\tilde{\xi}^\dtb_3\lam^\a_3\,
\xi_{2\a}\tlam_{2\dta}\tlam^{\dta}_1
\rightarrow 0,
\eear
for $\lam^\a_3\xi_{2\a}\rightarrow\langle3,3\rangle=0$.

Furthermore, since the 3-gluon vertices in diagrams (d) and (e) have exactly the same structure as that
in \figref{fig:fggfppmm}(b),
we have the same vanishing result as \eqref{eqn:result-1b}:
\be
A_d\rightarrow 0, \qquad A_e\rightarrow 0,
\ee
as we are taking the same gauge, $\xi_2=\lam_2$ and $\tilde{\xi}_3=\tlam_3$.

As a result,
\bear
&&A_{\mathcal{O}(M)}(+1/2,+1,-1,+1/2)=A_a+A_b+A_c+A_d+A_e\nn\\
&=&\frac{ig^2M^{AB}\wvrho_{1A}\wvrho_{4B}}{2}\
\frac{\langle3,1\rangle\langle3,4\rangle
\left\{\langle3,1\rangle\langle3,\eta_1\rangle - \langle3,4\rangle\langle3,\eta_4\rangle\right\}}
{\prod_{i=1}^4\langle i,i+1\rangle} \nn \\
&\longrightarrow& \frac{ig^2M^{AB}\wvrho_{1A}\wvrho_{4B}}{2}\
\frac{\langle3,1\rangle\langle3,4\rangle \langle4,1\rangle }
{\prod_{i=1}^4\langle i,i+1\rangle},
\label{eqn:AmpA(+1/2,+1,-1,+1/2) app}
\eear
where in the last line we scale $(\lam_i^1,\lam_i^2)=(1,Z_i)$
and $(\eta_i^1,\eta_i^2)=(0,1)$; accordingly $\langle i,j\rangle=-\langle j,i\rangle=Z_j-Z_i$
and $\langle i, \eta_j\rangle=-\langle \eta_j, i\rangle=1$.\footnote{Henceforth, a long arrow $\longrightarrow$
represents the scaling $(\lam_i^1,\lam_i^2)=(1,Z_i)$ and $(\eta_i^1,\eta_i^2)=(0,1)$.}

\subsubsection{$A_{\mathcal{O}(M)}(+1/2,+1/2,-1/2,+1/2)$}

The Feynman diagrams are listed in
\figref{fig:ffffppmp}. Figures (a)-(d) are diagrams exchanging a gluon propagator
while (e) and (f) exchange a scalar propagator.

The Feynman rules give us
\bear A_a &=&
(-M^{DE}\wvrho_{4E}\eta_4^\d)(\wvrho_{1A}\tlam_1^\dta)
\left(ig\sigma^\mu_{\d\dta}{\delta^A}_D\right)
\left[\frac{-i\eta_{\mu\nu}}{(p_1-p_4)^2}\right]
\left(ig\sigma^\nu_{\g\dtb}{\delta^B}_C\right)
(\varrho_3^C\lam_3^\g)(\wvrho_{2B}\tlam^\dtb_2)\nn\\
&=&
-2ig^2(M^{AD}\wvrho_{1A}\wvrho_{4D})(\varrho_3^C\wvrho_{2C})
\frac{\langle3,\eta_4\rangle[1,2]}{\langle1,4\rangle[1,4]}\nn\\
&=&-2ig^2(M^{AD}\wvrho_{1A}\wvrho_{4D})(\varrho_3^C\wvrho_{2C})
\frac{\langle1,2\rangle\langle3,4\rangle\left\{\langle3,4\rangle\langle3,\eta_4\rangle\right\}}
{\prod_{i=1}^4 \langle i,i+1\rangle}
\eear
by the identity \eqref{eqn:sigma1}.
\bear
A_b
&=&
(\wvrho_{4D}\tlam_{4\dtd})(M^{AE}\wvrho_{1E}\eta_{1\a})
\left(ig\bsig^{\mu\dtd\a}{\delta_A}^D\right)
\left[\frac{-i\eta_{\mu\nu}}{(p_1-p_4)^2}\right]
\left(ig\sigma^\nu_{\g\dtb}{\delta^B}_C\right)
(\varrho_3^C\lam_3^\g)(\wvrho_{2B}\tlam^\dtb_2)\nn\\
&=&
-2ig^2(M^{AD}\wvrho_{1A}\wvrho_{4D})(\varrho_3^C\wvrho_{2C})
\frac{\langle3,\eta_1\rangle[2,4]}{\langle1,4\rangle[1,4]}\nn\\
&=&2ig^2(M^{AD}\wvrho_{1A}\wvrho_{4D})(\varrho_3^C\wvrho_{2C})
\frac{\langle1,2\rangle\langle3,4\rangle\left\{\langle3,1\rangle\langle3,\eta_1\rangle\right\}}
{\prod_{i=1}^4 \langle i,i+1\rangle}
\eear
by the identity \eqref{eqn:sigma1}.
\bear
A_c &=& (\wvrho_{4D}\tlam_{4}^{\dtd})(\varrho_3^C\lam_{3}^{\g})
\left(ig\sigma^{\mu}_{\g\dtd}{\delta^D}_C\right)
\left[\frac{-i\eta_{\mu\nu}}{(p_1+p_2)^2}\right]
\left(ig\bsig^{\nu\dta\b}{\delta_B}^A\right)
(-\wvrho_{1A}\lam_{1\dta})(M^{BF}\wvrho_{2F}\eta_{2\b})\nn\\
&=&
2ig^2(M^{AB}\wvrho_{1A}\wvrho_{2B})(\varrho_3^C\wvrho_{4C})
\frac{\langle3,\eta_2\rangle[1,4]}{\langle1,2\rangle[1,2]}\nn\\
&=&2ig^2(M^{AB}\wvrho_{1A}\wvrho_{2B})(\varrho_3^C\wvrho_{4C})
\frac{\langle2,3\rangle^2\langle\eta_2,3\rangle\langle4,1\rangle}{\prod_{i=1}^4 \langle i,i+1\rangle},
\eear


\FIGURE{
\begin{picture}(420,270)


\put(-10,265){\text{(a)}}
\put(40,150){\begin{picture}(150,60)

  \thicklines
  \put(-35,7){$1,+,A$}
  \put(0,0){\vector(1,2){15}} 
   \put(15,30){\line(1,2){15}}
  \put(30,60){\vector(-1,2){15}} 
   \put(15,90){\line(-1,2){15}}
  \put(-35,107){$4,-,D$}
  \put(10,40){$\dot{\alpha}$}
  \put(10,70){$\delta$}

  \thicklines
  \put(90,7){$2,+,B$}
  \put(90,0){\vector(-1,2){15}} 
   \put(75,30){\line(-1,2){15}}
  \put(60,60){\vector(1,2){15}} 
   \put(75,90){\line(1,2){15}}
  \put(90,107){$3,+,C$}
  \put(80,40){$\dtb$}
  \put(80,70){$\gamma$}

  \thinlines
  \multiput(30,60)(12,0){3}{\qbezier(0,0)(3,3)(6,0)}
   \multiput(36,60)(12,0){2}{\qbezier(0,0)(3,-3)(6,0)}

\end{picture}}

\put(175,265){\text{(b)}}
\put(200,150){\begin{picture}(150,60)

  \thicklines
  \put(-5,7){$+$}
  \put(0,0){\vector(1,2){15}} 
   \put(15,30){\line(1,2){15}}
  \put(30,60){\vector(-1,2){15}} 
   \put(15,90){\line(-1,2){15}}
  \put(-5,107){$-$}
  \put(10,70){$\dot{\delta}$}
  \put(10,40){$\alpha$}

  \thicklines
  \put(95,7){$+$}
  \put(90,0){\vector(-1,2){15}} 
   \put(75,30){\line(-1,2){15}}
  \put(60,60){\vector(1,2){15}} 
   \put(75,90){\line(1,2){15}}
  \put(95,107){$+$}
  \put(80,40){$\dtb$}
  \put(80,70){$\gamma$}

  \thinlines
  \multiput(30,60)(12,0){3}{\qbezier(0,0)(3,3)(6,0)}
   \multiput(36,60)(12,0){2}{\qbezier(0,0)(3,-3)(6,0)}

\end{picture}}

\put(320,265){\text{(c)}}
\put(350,150){\begin{picture}(150,60)

  \thicklines
  \put(-15,0){$-$}
  \put(30,30){\vector(-1,-1){18}} 
   \put(0,0){\line(1,1){12}}
   \put(15,5){\vector(1,1){10}}
  \put(60,0){\vector(-1,1){15}} 
   \put(45,15){\line(-1,1){15}}
  \put(0,120){\vector(1,-1){18}} 
   \put(18,102){\line(1,-1){12}}
   \put(25,105){\vector(-1,1){10}}
  \put(30,90){\vector(1,1){15}} 
   \put(45,105){\line(1,1){15}}
  \put(-15,115){$+$}

  \put(10,20){$\dta$}
  \put(45,20){$\b$}
  \put(45,90){$\gamma$}
  \put(10,90){$\dot{\delta}$}

  \thinlines
  \put(65,0){$+$}
  \multiput(30,30)(0,20){3}{\qbezier(0,0)(-5,5)(0,10)}
   \multiput(30,40)(0,20){3}{\qbezier(0,0)(5,5)(0,10)}
  \put(65,115){$+$}

\end{picture}}

\put(-10,120){\text{(d)}}
\put(55,0){\begin{picture}(150,60)

  \thicklines
  \put(-20,0){$-$}
  \put(30,30){\vector(-1,-1){18}} 
   \put(0,0){\line(1,1){12}}
   \put(15,5){\vector(1,1){10}}
  \put(60,0){\vector(-1,1){15}} 
   \put(45,15){\line(-1,1){15}}
  \put(0,120){\vector(1,-1){18}} 
   \put(18,102){\line(1,-1){12}}
   \put(25,105){\vector(-1,1){10}}
  \put(30,90){\vector(1,1){15}} 
   \put(45,105){\line(1,1){15}}
  \put(-20,115){$+$}

  \put(10,20){$\a$}
  \put(45,20){$\dtb$}
  \put(45,90){$\gamma$}
  \put(10,90){$\dot{\delta}$}

  \thinlines
  \multiput(30,30)(0,20){3}{\qbezier(0,0)(-5,5)(0,10)}
   \multiput(30,40)(0,20){3}{\qbezier(0,0)(5,5)(0,10)}
  \put(65,115){$+$}
  \put(65,0){$+$}

\end{picture}}

\put(175,120){\text{(e)}}
\put(200,0){\begin{picture}(150,60)

  \thicklines
  \put(-5,7){$+$}
  \put(0,0){\vector(1,2){15}} 
   \put(15,30){\line(1,2){15}}
  \put(30,60){\vector(-1,2){15}} 
   \put(15,90){\line(-1,2){15}}
  \put(-5,107){$-$}
  \put(10,70){$\dot{\delta}$}
  \put(10,40){$\dot{\alpha}$}

  \thicklines
  \put(95,7){$+$}
  \put(90,0){\vector(-1,2){15}} 
   \put(75,30){\line(-1,2){15}}
  \put(60,60){\vector(1,2){15}} 
   \put(75,90){\line(1,2){15}}
  \put(95,107){$+$}
  \put(80,40){$\beta$}
  \put(80,70){$\gamma$}

  \thinlines
  \put(30,60){\line(1,0){5}}
  \put(40,60){\line(1,0){10}}
  \put(55,60){\line(1,0){5}}

\end{picture}}

\put(320,120){\text{(f)}}
\put(350,0){\begin{picture}(150,60)

  \thicklines
  \put(-15,0){$-$}
  \put(30,30){\vector(-1,-1){18}} 
   \put(0,0){\line(1,1){12}}
   \put(15,5){\vector(1,1){10}}
  \put(60,0){\vector(-1,1){15}} 
   \put(45,15){\line(-1,1){15}}
  \put(0,120){\vector(1,-1){18}} 
   \put(18,102){\line(1,-1){12}}
   \put(25,105){\vector(-1,1){10}}
  \put(30,90){\vector(1,1){15}} 
   \put(45,105){\line(1,1){15}}
  \put(-15,115){$+$}
  \put(65,0){$+$}
  \put(65,115){$+$}

  \put(10,20){$\dta$}
  \put(45,20){$\dtb$}
  \put(45,90){$\gamma$}
  \put(10,90){$\delta$}

  \thinlines
  \put(30,30){\line(0,1){5}}
  \put(30,40){\line(0,1){10}}
  \put(30,55){\line(0,1){10}}
  \put(30,70){\line(0,1){10}}
  \put(30,85){\line(0,1){5}}

\end{picture}}

\end{picture}
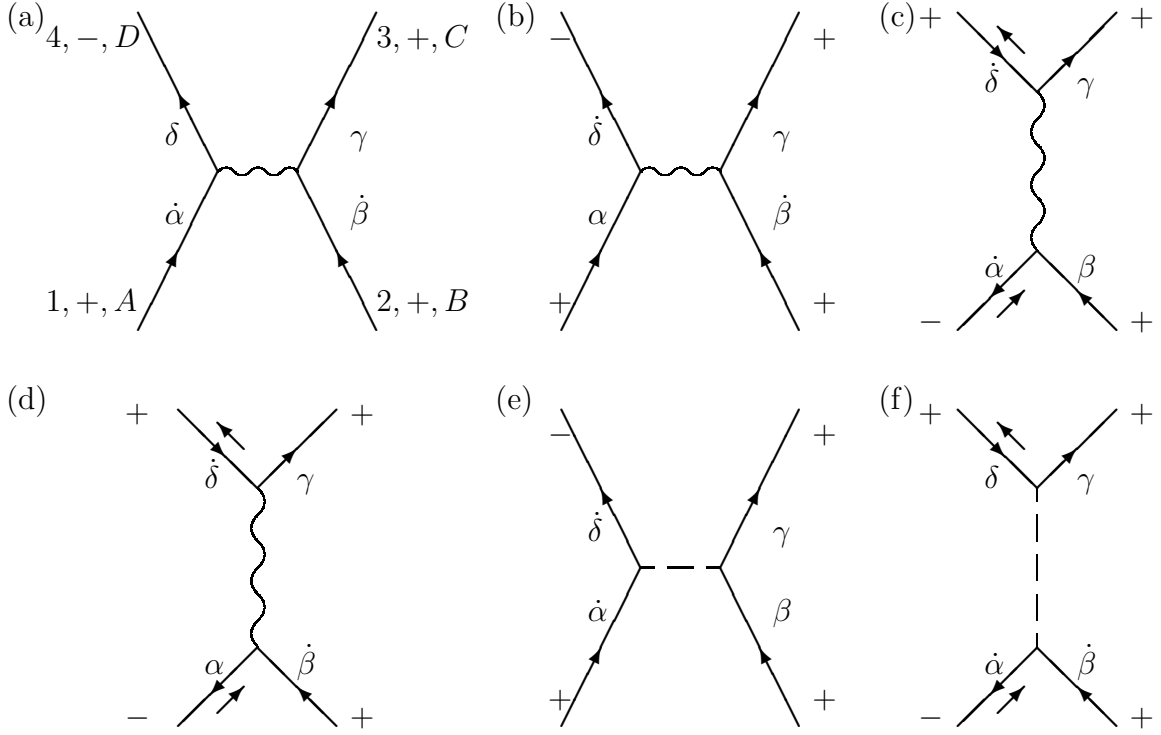
\caption{ Planar Feynman diagrams with four external fermions corresponding to
the extended MHV amplitude
$A_{\mathcal{O}(M)}(+1/2,+1/2,-1/2,+1/2)$. }\label{fig:ffffppmp}
}

and
\bear
A_d
&=&
(\wvrho_{4D}\tlam_{4}^{\dtd})(\varrho_3^C\lam_{3}^{\g})
\left(ig\sigma^{\mu}_{\g\dtd}{\delta^D}_C\right)
\left[\frac{-i\eta_{\mu\nu}}{(p_1+p_2)^2}\right]
\left(ig\sigma^{\nu}_{\a\dtb}{\delta^B}_A\right)
(-M^{AE}\wvrho_{1E}\eta_{1}^{\a})(\wvrho_{2B}\tlam_2^{\dtb})\nn\\
&=&
2ig^2(M^{AB}\wvrho_{1A}\wvrho_{2B})(\varrho_3^C\wvrho_{4C})
\frac{\langle3,\eta_1\rangle[2,4]}{\langle1,2\rangle[1,2]}\nn\\
&=&-2ig^2(M^{AB}\wvrho_{1A}\wvrho_{2B})(\varrho_3^C\wvrho_{4C})
\frac{\langle2,3\rangle\langle3,1\rangle\langle3,\eta_1\rangle\langle4,1\rangle}{\prod_{i=1}^4 \langle i,i+1\rangle}.
\eear

The Feynman rules in \figref{fig:scalarvertex} give:
\bear
A_e &=& (\wvrho_{4D}\lam_{4\dtd})(\wvrho_{1A}\lam_{1}^{\dta})
\left(2ig\Gamma^{\mathcal{I}DA} {\delta_{\dta}}^{\dtd}\right)
\left[\frac{-i\delta_{\mathcal{I}\mathcal{J}}}{(p_1-p_4)^2}\right]
\left(2ig \Gamma^\mathcal{J}_{CB} {\delta^{\b}}_{\g}\right)
(\varrho_3^C\lam_3^{\g})(M^{BE}\wvrho_{2E}\eta_{2\b})\nn\\
&=& -2ig^2 \frac{\langle3,\eta_2\rangle}{\langle1,4\rangle}
\Big(M^{AB}\wvrho_{1A}\wvrho_{2B}\varrho_3^C\wvrho_{4C}-M^{BD}\wvrho_{2B}\wvrho_{4D}\varrho_3^C\wvrho_{1C} \Big),
\eear
and
\bear
A_f
&=&
(\varrho_{3}^{C}\lam_{3}^{\g})(-M^{DH}\wvrho_{4H}\eta_{4\d})
\left(2ig\Gamma^\mathcal{I}_{CD} {\d^{\d}}_{\g}\right)
\left[\frac{-i\d_{\mathcal{I}\mathcal{J}}}{(p_1+p_2)^2}\right]
\left(2ig \Gamma^{\mathcal{J}AB} {\delta_{\dtb}}^{\dta}\right)
(-\wvrho_{1A}\tlam_{1\dta})(\wvrho_{2B}\tlam_{2}^{\dtb})\nn\\
&=& -2ig^2 \frac{\langle3,\eta_4\rangle}{\langle1,2\rangle}
\Big(M^{BD}\wvrho_{4D}\wvrho_{2B}\varrho_3^C\wvrho_{1C}-M^{AD}\wvrho_{1A}\wvrho_{4D}\varrho_3^C\wvrho_{2C} \Big)
\eear
by ithe dentity \eqref{eqn:phi2}.

Altogether,
the extended MHV amplitude for the four external fermions is
\bear
&&A_{\mathcal{O}(M)}(+1/2,+1/2,-1/2,+1/2) = A_a+\cdots +A_f
\nn \\
&\longrightarrow& \frac{2ig^2}{\prod_{1}^{4} \langle i, i+1\rangle}
\Big\{ \wvrho_{1A}M^{AB}\wvrho_{2B}\varrho_3^C\wvrho_{4C} \langle1,2\rangle\langle2,3\rangle\langle3,1\rangle
\nn \\
&& \qquad +  \wvrho_{1A}\varrho_3^A \wvrho_{2B} M^{BD}\wvrho_{4D} \langle2,4\rangle\langle2,3\rangle\langle3,4\rangle
\nn\\
&&\qquad
+ \wvrho_{2B}\varrho_3^B \wvrho_{1A} M^{AD}
\wvrho_{4D}  \langle4,1\rangle\langle1,3\rangle\langle3,4\rangle \Big\}.\label{eqn:AmpA(+1/2,+1/2,-1/2,+1/2) app}
\eear


\subsubsection{$A_{\mathcal{O}(M)}(+1/2,0,0,+1/2)$}

The relevant Feynman diagrams are given in \figref{fig:fssfp00p}.
Diagram (e) involves a 3-scalar vertex,
which is due to the presence of the chiral mass term (as discussed in \secref{sec:susy} and \appref{app:3phi}).


\FIGURE{
\begin{picture}(420,270)


\put(0,265){\text{(a)}}
\put(50,150){\begin{picture}(150,60)

  \thicklines
  \put(-30,0){$+,A$}
  \put(-6,-6){\vector(1,1){18}} 
   \put(12,12){\line(1,1){18}}
  \put(30,30){\vector(0,1){30}} 
   \put(30,60){\line(0,1){30}}
  \put(30,90){\vector(-1,1){18}} 
   \put(12,108){\line(-1,1){18}}
  \put(-30,125){$-,B$}

  \put(10,20){$\dta$}
  \put(20,40){$\dtd$}
  \put(20,75){$\g$}
  \put(10,90){$\b$}

  \thinlines
  \multiput(30,30)(13,-13){3}{\line(1,-1){10}}
  \multiput(30,90)(13,13){3}{\line(1,1){10}}
  \put(70,0){$\mathcal{I}$}
  \put(70,125){$\mathcal{J}$}

\end{picture}}

\put(150,265){\text{(b)}}
\put(200,150){\begin{picture}(150,60)

  \thicklines
  \put(-20,0){$+$}
  \put(-6,-6){\vector(1,1){18}} 
   \put(12,12){\line(1,1){18}}
  \put(30,30){\vector(0,1){30}} 
   \put(30,60){\line(0,1){30}}
  \put(30,90){\vector(-1,1){18}} 
   \put(12,108){\line(-1,1){18}}
  \put(-20,125){$-$}

  \put(10,20){$\a$}
  \put(20,40){$\d$}
  \put(20,75){$\dtg$}
  \put(10,90){$\dtb$}

  \thinlines
  \multiput(30,30)(13,-13){3}{\line(1,-1){10}}
  \multiput(30,90)(13,13){3}{\line(1,1){10}}

\end{picture}}

\put(300,265){\text{(c)}}
\put(320,145){\begin{picture}(150,60)

  \thicklines
  \put(-5,7){$+$}
  \put(0,0){\vector(1,2){15}} 
   \put(15,30){\line(1,2){15}}
  \put(30,60){\vector(-1,2){15}} 
   \put(15,90){\line(-1,2){15}}
  \put(-5,107){$-$}
  \put(10,40){$\a$}
  \put(10,70){$\dtb$}

  \thinlines
  \multiput(60,60)(8,-16){4}{\line(1,-2){6}}
  \multiput(60,60)(8,16){4}{\line(1,2){6}}

  \thinlines
  \multiput(30,60)(12,0){3}{\qbezier(0,0)(3,3)(6,0)}
   \multiput(36,60)(12,0){2}{\qbezier(0,0)(3,-3)(6,0)}

\end{picture}}

\put(60,120){\text{(d)}}
\put(90,0){\begin{picture}(150,60)

  \thicklines
  \put(-5,7){$+$}
  \put(0,0){\vector(1,2){15}} 
   \put(15,30){\line(1,2){15}}
  \put(30,60){\vector(-1,2){15}} 
   \put(15,90){\line(-1,2){15}}
  \put(-5,107){$-$}
  \put(10,40){$\dta$}
  \put(10,70){$\b$}

  \thinlines
  \multiput(60,60)(8,-16){4}{\line(1,-2){6}}
  \multiput(60,60)(8,16){4}{\line(1,2){6}}

  \thinlines
  \multiput(30,60)(12,0){3}{\qbezier(0,0)(3,3)(6,0)}
   \multiput(36,60)(12,0){2}{\qbezier(0,0)(3,-3)(6,0)}

\end{picture}}

\put(240,120){\text{(e)}}
\put(270,0){\begin{picture}(150,60)

  \thicklines
  \put(-5,7){$+$}
  \put(0,0){\vector(1,2){15}} 
   \put(15,30){\line(1,2){15}}
  \put(30,60){\vector(-1,2){15}} 
   \put(15,90){\line(-1,2){15}}
  \put(-5,107){$-$}
  \put(10,40){$\dta$}
  \put(10,70){$\dtb$}

  \thinlines
  \multiput(60,60)(8,-16){4}{\line(1,-2){6}}
  \multiput(60,60)(8,16){4}{\line(1,2){6}}

  \thinlines
  \put(30,60){\line(1,0){5}}
  \put(40,60){\line(1,0){10}}
  \put(55,60){\line(1,0){5}}

\end{picture}}

\end{picture}
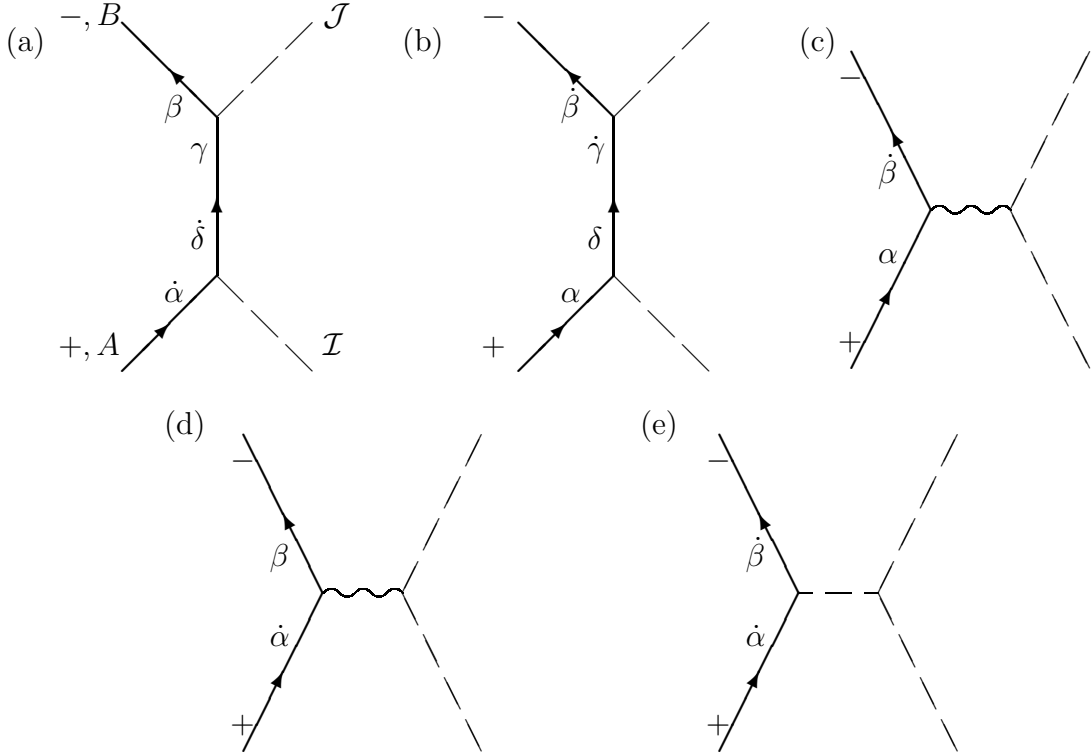
\caption{ Planar Feynman diagrams with two external fermions and two
external scalars corresponding to the extended MHV amplitude
$A_{\mathcal{O}(M)}(+1/2,0,0,+1/2)$. }\label{fig:fssfp00p}
}


The fermion-exchanging diagrams are calculated as usual:
\bear
A_a &=&
\varphi_{3\mathcal{J}}M^{BM}\wvrho_{4M}\eta_4^\b
\left(2ig\Gamma^{\mathcal{J}}_{BC}\right)
\left[\frac{i(p_1+p_2)_{\b\dta}\delta_D^C}{(p_1+p_2)^2}\right]
\left(2ig\Gamma^{\mathcal{I}DA}\right)
\wvrho_{1A}\tlam^\dta_1\,\varphi_{2\mathcal{I}}\nn\\
&=&-2ig^2\wvrho_{4C}M^{CB}\varphi_{3BD}\varphi_2^{DA}\wvrho_{1A}
\frac{\langle2,3\rangle\langle3,4\rangle\langle4,1\rangle\langle\eta_4,2\rangle}{2\prod_{i=1}^4 \langle
i,i+1\rangle} \nn \\
&\longrightarrow&-2ig^2\wvrho_{4C}M^{CB}\varphi_{3BD}\varphi_2^{DA}\wvrho_{1A}
\frac{\langle2,3\rangle\langle3,4\rangle\langle4,1\rangle}{2\prod_{i=1}^4 \langle
i,i+1\rangle},
\eear
and
\bear
A_b &=&
\varphi_{3\mathcal{J}}\wvrho_{4B}\tlam_{4\dtb}
\left(2ig\Gamma^{\mathcal{J}BC}\right)
\left[\frac{i(p_1+p_2)^{\dtb\a}\delta_C^D}{(p_1+p_2)^2}\right]
\left(2ig\Gamma^{\mathcal{I}{DA}}\right)
M^{AD}\wvrho_{1D}\eta_{1\a}\,\varphi_{2\mathcal{I}}\nn\\
&=&-2ig^2\wvrho_{4B}\varphi_{3}^{BD}\varphi_{2DA}M^{AD}\wvrho_{1D}
\frac{\langle1,2\rangle\langle2,3\rangle\langle4,1\rangle\langle3,\eta_1\rangle}{2\prod_{i=1}^4 \langle
i,i+1\rangle} \nn \\
&\longrightarrow&-2ig^2\wvrho_{4B}\varphi_{3}^{BD}\varphi_{2DA}M^{AD}\wvrho_{1D}
\frac{\langle1,2\rangle\langle2,3\rangle\langle4,1\rangle}{2\prod_{i=1}^4 \langle
i,i+1\rangle}.
\eear
The gluon-exchanging diagrams are given by
\bear
A_c &=&
\varphi_{3\mathcal{J}}\wvrho_{4B}\tlam_{4\dtb}
\left(ig\bar{\sigma}^{\mu\dtb\a}\d^{B}_{A}\right) M^{AC}\wvrho_{1C}\eta_{1\a}
\left[\frac{-i\eta_{\mu\nu}}{(p_4-p_1)^2}\right]
\left(\frac{ig}{2}(p_2+p_3)^{\nu}\delta^\mathcal{IJ}\right)
\,\varphi_{2\mathcal{I}} \nn \\
&=& ig^2\wvrho_{4B} M^{BC} \wvrho_{1C} \varphi_{1A'B'}\varphi_2^{A'B'}
\frac{\langle1,2\rangle\langle3,4\rangle(\langle1,3\rangle\langle \eta_1,2\rangle+\langle1,2\rangle\langle\eta_1,3\rangle)}{2\prod_{i=1}^4 \langle i,i+1\rangle} \nn \\
&\longrightarrow& -ig^2\wvrho_{4B} M^{BC} \wvrho_{1C} \varphi_{1A'B'}\varphi_2^{A'B'}
\frac{\langle1,2\rangle\langle3,4\rangle(\langle1,3\rangle+\langle1,2\rangle)}{2\prod_{i=1}^4 \langle i,i+1\rangle}
\eear
and
\bear
A_d &=&
\varphi_{3\mathcal{J}}M^{BD}\wvrho_{4D}\eta_4^\b
\left(ig\sigma^{\mu}_{\b\dta}\delta_{B}^{A}\right)\wvrho_{1A}\tlam^\dta_1
\left[\frac{-i\eta_{\mu\nu}}{(p_4+p_1)^2}\right]
\left(\frac{ig}{2}(p_2+p_3)^{\nu}\delta^\mathcal{IJ}\right)
\,\varphi_{2\mathcal{I}}\nn\\
&=& ig^2\wvrho_{4B} M^{BC} \wvrho_{1C} \varphi_{1A'B'}\varphi_2^{A'B'}
\frac{\langle1,2\rangle\langle3,4\rangle(\langle2,4\rangle\langle3,\eta_4\rangle - \langle2,\eta_4\rangle\langle3,4\rangle)}{2\prod_{i=1}^4 \langle i,i+1\rangle} \nn \\
&\longrightarrow& -ig^2\wvrho_{4B} M^{BC} \wvrho_{1C} \varphi_{1A'B'}\varphi_2^{A'B'}
\frac{\langle1,2\rangle\langle3,4\rangle(\langle2,4\rangle+\langle3,4\rangle)}{2\prod_{i=1}^4 \langle i,i+1\rangle}.
\eear

Finally, the Feynman rule for the 3-scalar vertex as depicted in \figref{fig:scalar-vertex} gives
\bear
A_e &=&
\wvrho_{4B}\tlam_{4\dtb}\wvrho_{1A}\tlam_{1}^{\dta}
\left(2ig\Gamma^{\mathcal{L}BA}\d_{\dta}^{\dtb}\right)
\left[
\frac{-i\delta_{\mathcal{LK}}}{(p_1-p_4)^2}
\right]
\Big(-\frac{ig}{2}M^{CD}\Gamma^{\mathcal{K}}_{CE}
\Gamma^{\mathcal{I}}_{DF}\Gamma^{\mathcal{J}EF}\Big)
\,\varphi_{3\mathcal{J}}\,\varphi_{2\mathcal{I}}
\nn\\
&=&-2ig^2\frac{\wvrho_{1A}\wvrho_{4B}}{\langle1,4\rangle[1,4]}
[4,1]\Big(
\delta_\mathcal{LK}\Gamma^{\mathcal{L}BA}\Gamma^\mathcal{K}_{CE}\Big)M^{CD}\varphi_{2DF}\,\varphi_3^{EF}
\nn\\
&=&-ig^2
\frac{\langle1,2\rangle\langle2,3\rangle\langle3,4\rangle}{
\prod_{i=1}^4 \langle
i,i+1\rangle}\,\wvrho_{1A}\wvrho_{4B}
\Big(
M^{BC}\varphi_{2CD}\,\varphi_3^{AD}
-M^{AC}\varphi_{2CD}\,\varphi_3^{BD}
\Big)
\nn \\
&=& -ig^2
\frac{
\langle1,2\rangle\langle2,3\rangle\langle3,4\rangle}{
\prod_{i=1}^4 \langle i,i+1\rangle}
\Big\{
\wvrho_{4C}M^{CB}\varphi_{3BD}\varphi_2^{DA}\wvrho_{1A}
+\wvrho_{4B}\varphi_{3}^{BD}\varphi_{2DA}M^{AD}\wvrho_{1D}
\nn \\
&& \qquad\qquad\qquad\qquad\qquad\qquad
+\frac{1}{2}\wvrho_{4C}M^{CB}
\wvrho_{1B}\varphi_{3}^{B'D'}\varphi_{2D'B'}
\Big\},
\eear
where the identity \eqref{eqn:phi7} is used in the last line.

Put all together, we have
\bear\label{eqn:AmpA(+1/2,0,0,+1/2) app}
&&A_{\mathcal{O}(M)}(+1/2,0,0,+1/2)=A_a+A_b+A_c+A_d+A_e\nn\\
&\longrightarrow&\frac{ig^2}{\prod_{i=1}^4 \langle i,i+1\rangle}
\Big\{(\langle2,3\rangle\langle3,4\rangle\langle1,4\rangle + \langle2,3\rangle\langle3,4\rangle\langle2,1\rangle)\wvrho_{4C}M^{CB}\varphi_{3BD}\varphi_2^{DA}\wvrho_{1A} \nn \\
&& \qquad\qquad\qquad
+ (\langle1,2\rangle\langle2,3\rangle\langle1,4\rangle + \langle2,3\rangle\langle3,4\rangle\langle2,1\rangle) \wvrho_{4B}\varphi_{3}^{BD}\varphi_{2DA}M^{AD}\wvrho_{1D}
\nn\\
&& \qquad\qquad\qquad \frac{1}{2}(\langle1,2\rangle\langle1,4\rangle\langle3,4\rangle + \langle2,3\rangle\langle3,4\rangle\langle2,1\rangle )\wvrho_{4B}M^{BC}\wvrho_{1C}\varphi_{3}^{B'D'}\varphi_{2B'D'} \Big\} \nn \\
&=&\frac{ig^2}{\prod_{i=1}^4 \langle i,i+1\rangle}
\Big\{\langle2,3\rangle\langle3,4\rangle\langle4,2\rangle\wvrho_{4C}M^{CB}\varphi_{3BD}\varphi_2^{DA}\wvrho_{1A}+ \langle1,2\rangle\langle2,3\rangle\langle3,1\rangle \wvrho_{4B}\varphi_{3}^{BD}\varphi_{2DA}M^{AD}\wvrho_{1D}
\nn\\
&& \qquad\qquad\qquad \frac{1}{2}
(\langle1,2\rangle\langle2,3\rangle\langle3,4\rangle + \langle1,2\rangle\langle1,4\rangle\langle3,4\rangle)
\wvrho_{4B}M^{BC}\wvrho_{1C}\varphi_{3}^{B'D'}\varphi_{2B'D'} \Big\}.
\eear


\end{document}